\documentclass[12pt]{article}
\usepackage[left=0.5in,top=1in,right=2in,bottom=1in,headheight=3ex,headsep=3ex]{geometry}
\usepackage{fullpage}
\usepackage{amsfonts}
\usepackage{amsmath}
\usepackage{amssymb}
\usepackage{bm}

\usepackage[T1]{fontenc}

\newcommand{\indep}{\perp \!\!\! \perp}

\usepackage[authordate, backend=biber, uniquename=false, noibid]{biblatex-chicago}

\DeclareFieldFormat{citehyperref}{%
  \DeclareFieldAlias{bibhyperref}{noformat}
  \bibhyperref{#1}}

\DeclareFieldFormat{textcitehyperref}{%
  \DeclareFieldAlias{bibhyperref}{noformat}
  \bibhyperref{%
    #1%
    \ifbool{cbx:parens}
      {\bibcloseparen\global\boolfalse{cbx:parens}}
      {}}}

\savebibmacro{cite}
\savebibmacro{textcite}

\renewbibmacro*{cite}{%
  \printtext[citehyperref]{%
    \restorebibmacro{cite}%
    \usebibmacro{cite}}}

\renewbibmacro*{textcite}{%
  \ifboolexpr{
    ( not test {\iffieldundef{prenote}} and
      test {\ifnumequal{\value{citecount}}{1}} )
    or
    ( not test {\iffieldundef{postnote}} and
      test {\ifnumequal{\value{citecount}}{\value{citetotal}}} )
  }
    {\DeclareFieldAlias{textcitehyperref}{noformat}}
    {}%
  \printtext[textcitehyperref]{%
    \restorebibmacro{textcite}%
    \usebibmacro{textcite}}}

\AtBeginBibliography{

}

\usepackage{framed}
\usepackage[dvipsnames]{xcolor}
\usepackage{setspace}
\usepackage{marginnote}

\usepackage{hyperref}
\hypersetup{
  colorlinks   = true, 
  urlcolor     = blue, 
  linkcolor    = blue, 
  citecolor   = blue 
}

\DeclareCiteCommand{\citeyear}
    {\usebibmacro{prenote}}
    {\bibhyperref{\printfield{year}}\bibhyperref{\printfield{extrayear}}}
    {\multicitedelim}
    {\usebibmacro{cite:postnote}}

\title{\textbf{Dynamic Games \\in Empirical Industrial Organization}\thanks{We are grateful to the editors and to four anonymous referees for constructive comments and suggestions. We would like to thank Seohee Kim for research assistance, and Peter Arcidiacono, John Asker, Nick Buchholz, Yanyou Chen, Jan De Loecker, Gautam Gowrisankaran, Robin Lee, Yao Luo, Mathieu Marcoux, Eduardo Souza-Rodrigues, Adam Rosen, and Daniel Xu for productive conversations about the manuscript.}}
\author{Victor Aguirregabiria \\ University of Toronto and CEPR\and Allan Collard-Wexler \\ Duke University and NBER \and Stephen P. Ryan \\ Washington University in St. Louis, NBER, and CESIfo}
\date{September 15, 2021}

\begin{document}

\maketitle

\thispagestyle{empty}

\begin{abstract}

This survey is organized around three main topics: models, econometrics, and empirical applications. Section \ref{Section: Models} presents the theoretical framework, introduces the concept of Markov Perfect Nash Equilibrium, discusses existence and multiplicity, and describes the representation of this equilibrium in terms of conditional choice probabilities. We also discuss extensions of the basic framework, including models in continuous time, the concepts of oblivious equilibrium and experience-based equilibrium, and dynamic games where firms have non-equilibrium beliefs. In section \ref{Section: Identification Estimation}, we first provide an overview of the types of data used in this literature, before turning to a discussion of identification issues and results, and estimation methods. We review different methods to deal with multiple equilibria and large state spaces. We also describe recent developments for estimating games in continuous time and incorporating serially correlated unobservables, and discuss the use of machine learning methods to solving and estimating dynamic games. Section \ref{Section: Empirical Applications} discusses empirical applications of dynamic games in IO. We start describing the first empirical applications in this literature during the early 2000s. Then, we review recent applications dealing with innovation, antitrust and mergers, dynamic pricing, regulation, product repositioning, advertising, uncertainty and investment, airline network competition, dynamic matching, and natural resources. We conclude with our view of the progress made in this literature and the remaining challenges.

\vspace{0.4cm}

\vspace{0.4cm}
\end{abstract}

\tableofcontents

\newpage 

\section{Introduction}

\subsection{Role of dynamic games in empirical industrial organization}
A central focus in industrial organization (IO) is understanding the role of market structure on equilibrium outcomes such as prices, product quality and variety, and market shares, and how those outcomes influence producer profits and consumer welfare. Market structure encompasses all of the features of the supply side of a market: the number of competitors in a market, their cost structure, where firms are located, the size of installed base in network industries, productivity advantages due to learning and managerial expertise, types and capacities of capital, the state of technology, vertical and horizontal relationships between firms, and everything else that is critical to understand competition. Dynamics is key to understanding the endogenous evolution of market structure. There are many questions in IO that revolve around the dynamic aspects faced by firms. Investment, production, or pricing decisions can affect firms' future profits, as well as future profits of their competitors in the industry. Supply-side dynamics can arise from different sources, including sunk costs of entry, partially irreversible investments, product repositioning costs, price adjustment costs, or learning-by-doing. Demand is dynamic when consumers have switching costs, or when products are durable or storable. Accounting for dynamics can change our view of the impact of competition in industries and our evaluation of public policies.

An early attempt to understand these relationships, the structure-conduct-performance paradigm \parencite{bain1951relation,bain1956} looked at empirical associations between market structure and outcomes across industries. However, the endogeneity of market structure---that market structure is itself the result of many firm decisions in equilibrium---was the core reason why the structure-conduct-performance paradigm ended up generating more questions than answers in empirical IO. Thus, the so-called New Empirical IO movement (see the survey by \cite{bresnahan1989empirical}) emphasized the need to build models of endogenous market structure. To that end, \cite{bresnahan1990entry, bresnahan1991entry} and \cite{berry1992estimation} proposed entry models that used observed market structure to infer aspects of the underlying profit function. While it is possible to include predetermined variables in the payoff function (e.g., firm size, capacity, incumbent status) and to interpret the payoff function as an intertemporal value function (see \cite{Bresnahan1994}), these models are essentially static in nature and have important limitations. The observed structure of an industry or market is the result of years or decades of cumulative firm decisions, and these static models leave no room for the history of demand or technology, or the process by which firms enter the industry and then shake out, to fundamentally alter the predictions for market structure. In addition, empirical questions in IO that have to do with the effects of uncertainty on firm behavior and competition, or that try to distinguish between short-run and long-run effects of exogenous shocks, typically require the specification and estimation of dynamic structural models that explicitly take into account firms' forward-looking behavior. For these reasons, most of the recent work in IO dealing with industry dynamics has relied on a more explicit modeling of dynamics.

Advances in econometric methods and modeling techniques, in conjunction with the increased availability of data and computing power, have led to a large body of empirical papers that study the dynamics of competition in oligopoly markets, especially over the last decade. The history of applications in the dynamic games literature in industrial organization can be delineated by two defining innovations. The first innovation was the introduction of Markov Perfect Nash Equilibrium (henceforth MPNE) by \cite{maskin1988theory1,maskin1988theory2} and \cite{Ericson1995}. The concept of MPNE restricts players’ strategies to depend only on payoff-relevant state variables, which reduces substantially the set of equilibria and facilitates their computation. At the same time, it provides a flexible framework that allows for rich dynamics and can be applied to a wide variety of settings. A first generation of applications focused on the calibration of stylized computational models that illustrated general economic principles, such as \cite{Pakes1994} and  \cite{Gowrisankaran1999}.\footnote{See \cite{besanko2014economics},  \cite{Doraszelski2012}, and \cite{Borkovsky2012} for more recent examples of this computational theory papers in IO. These computational models have been covered in a previous Handbook Chapter (\cite{Doraszelski2007}).} At the same time, a handful of empirical applications directly applied those theoretical frameworks, such as \cite{Benkard2004} and \cite{gowrisankaran1997dynamic}. These papers were highly influential for both the substantive research questions they addressed and for highlighting the need for an econometric approach to estimation that sidestepped the computational burden of repeatedly solving the theoretical model. The second innovation was the subsequent development of conditional choice probability (henceforth CCP) based methods inspired by the dynamic single agent work of \cite{hotz1993conditional} and \cite{hotz1994simulation} adapted to dynamic games by a collection of papers (\cite{Aguirregabiria2007,Bajari2007,Pesendorfer2008,Pakes2007}) that has directly led to the current era of empirical applications. These CCP based approaches allowed very complex dynamic games to be estimated---indeed much more complex than the types of dynamic games that can usually be solved for. 

\subsection{Organization of this chapter}

This survey is organized around three main topics: Models, in section \ref{Section: Models}; Econometrics, in section \ref{Section: Identification Estimation}; and Empirical Applications, in section \ref{Section: Empirical Applications}.

In section \ref{Section: Models}, we present the theoretical framework, introduce the concept of Markov Perfect Nash Equilibrium, discuss existence and multiplicity, describe the representation of this equilibrium in terms of conditional choice probabilities, and illustrate it with some examples. We also discuss some extensions of the basic framework, including models in continuous time, the concepts of oblivious equilibrium and experience-based equilibrium, and dynamic games where firms have non-equilibrium beliefs.

In section \ref{Section: Identification Estimation}, we first provide a brief overview of the types of data used in this literature, in section \ref{sec:data}, before turning to a discussion of identification issues and results in section \ref{sec:identification}. As a dynamic model, a key issue is the separate identification of true dynamics from spurious dynamics due to serially correlated unobservables. As a game, the presence of multiple equilibria in the data and unobservables correlated across players introduce relevant identification issues. The identification of the discount factor and the biases from normalization restrictions of the payoff function are issues that dynamic games share with other structural models where agents are forward-looking. For discrete choice models, the misspecification of the distribution of the unobservables is an important issue as it affects average marginal effects and counterfactuals using the estimated model. We discuss these identification issues and present positive and negative identification results for dynamic games.

Section \ref{Section: Estimation methods} deals with estimation methods. Empirical applications of dynamic games in IO need to deal with two main computational issues. First, for a given value of the structural parameters, the model typically has multiple equilibria. This introduces important challenges in the implementation of standard estimation methods such as maximum likelihood or generalized method of moments. Second, the curse of dimensionality in the solution of dynamic programming problems is particularly important in dynamic games with heterogeneous agents. We review different methods to deal with these issues. We also describe recent developments for estimating games in continuous time and incorporating serially correlated unobservables. We finish \ref{Section: Identification Estimation} with a discussion on the use of machine learning methods to solving and estimating dynamic games.

Section \ref{Section: Empirical Applications} discusses empirical applications of dynamic games in IO. We start in section \ref{sec:early_papers} describing the main features -- data, model, estimation, computation, and research questions -- in the first empirical applications in this literature during the early 2000s. Then, we review more recent applied papers. We have organized these applications around the following topics: innovation; antitrust and mergers; dynamic pricing; regulation; retail; product repositioning; advertising; uncertainty and investment; airline network competition; dynamic matching ; and natural resources. Admittedly, this classification is based in multiple criteria (e.g., empirical question, firms’ decisions, industry), but we prefer that each section covers clearly related papers, perhaps at the expense of missing some relevant connections between papers at different sections.

Previous survey papers on dynamic games in IO include \cite{Doraszelski2007}, \cite{arcidiacono_ellickson_2011}, \cite{Aguirregabiria2013}, and \cite{berry_compiani_2021}. \cite{Doraszelski2007} deal with algorithms for computing MPNE in dynamic games. Estimation methods are the main focus in \cite{arcidiacono_ellickson_2011} and \cite{Aguirregabiria2013}. Our chapter also covers these topics, though our treatment of solution methods is quite limited. Instead, we provide a more extensive coverage of recent developments on identification of dynamic games, and on empirical applications. Other chapters in this new volume of the Handbook of IO include sections on dynamics, e.g., the chapters on demand, productivity, collusion, or innovation, among others. In this chapter, we focus on methodological issues (specification, computation, identification, inference), which are particularly critical for dynamic games, as well as empirical applications where the dynamics of strategic interactions plays a key role for the empirical results.

\section{Models\label{Section: Models}}

\subsection{Basic framework}

We start presenting the Ericson-Pakes model (\cite{Ericson1995}), which is a general framework that includes as particular cases most empirical applications of dynamic games in IO.

The game is played by $N$ firms that we index by $i\in\mathcal{I} =\{1,2,...,N\}$. Time is discrete and indexed by $t$. Firms maximize their expected and discounted flow of profits in the market, 
$\mathbb{E}_{t}\left({\textstyle\sum\nolimits_{s=0}^{\infty}} \beta_{i}^{s}\text{ }\pi_{i,t+s}\right)$ where $\beta_{i}\in(0,1)$ is the firm's discount factor, and $\pi_{it}$ is its profit at period $t$. Every period, each firm $i$ makes two strategic decisions: a static decision, that affects current profits but not future profits; and an investment (dynamic) decision, that has implications on future profits. For instance, in a differentiated product industry, incumbent firms choose their prices (static decision) and make investments to improve the quality of their products (dynamic decision). See, for instance, \cite{Pakes1994}). These decisions correspond to a static and a dynamic game, respectively, that firms play in this market.

In the differentiated product example, given demand and marginal costs at period $t$, active firms compete in prices, and a static Bertrand equilibrium determines the current profit of each incumbent firm at period $t$ -- up to investment costs. We represent this component of the profit function as $r_{i}(\mathbf{x}_{t})$, where $\mathbf{x}_{t}$ is a vector of state variables affecting demand and costs at period $t$. We represent the investment decision using variable $a_{it}$. This dynamic action involves an investment cost $c_{i}(a_{it},\mathbf{x}_{t})$. The total profit function is $\pi_{i}(a_{it},\mathbf{x}_{t})=r_{i}(\mathbf{x}_{t})-c_{i}(a_{it},\mathbf{x}_{t})$. The set of feasible investment decisions may depend on the state variables, and it is represented as $\mathcal{A}_{i}(\mathbf{x}_{t})$. For instance, investments may be restricted to be positive, or not larger than a borrowing constraint that may evolve endogenously.

The vector of state variables $\mathbf{x}_{t}$ follows a first order controlled Markov process with transition CDF $F_{x}(\mathbf{x}_{t+1}|\mathbf{x}_{t},\boldsymbol{a}_{t})$, where $\boldsymbol{a}_{t}\equiv(a_{it}:i\in\mathcal{I})$ is the vector with the investment decisions of the $N$ firms. Vector $\mathbf{x}_{t}$ includes endogenous state variables -- such as capital stock, capacity, or product quality -- with transition rules that depend on firms' investments. For instance, a firm's stock of physical capital, $k_{it}$, may evolve according to equation $k_{i,t+1}= \delta_{i,t+1} \text{ } k_{it}+a_{it}$, where $a_{it}$ in this example is investment, and $\delta_{i,t+1}$ is a depreciation rate that may be stochastic or not. The vector $\mathbf{x}_{t}$ may also include exogenous state variables with transition probabilities that do not depend on firms' investment decisions such as demand shifters (e.g., market population and demographics) and input prices.

In our description of the model, we have implicitly assumed that a firm's investment affects its own profit -- other than the investment cost --
and other firms' profits one period after the investment decision is taken. This is the assumption of \textit{time-to-build} that has been used in many
models of firm investment such as \cite{kydland1982time}, but more specifically in \cite{Ericson1995}. Though time-to-build is a feature in many applications of dynamic games that we review in this chapter, there are also many studies that do not make this assumption. For instance, in a model of market entry, we may consider that entry decisions are made at at the beginning of period $t$ and are effective during the same period. For this alternative timing assumption, we need to modify the notation above and allow the variable profit and the total profit functions to depend on all the firms' current investment decisions. That is, these functions become $r_{i}(\boldsymbol{a}_{t},\mathbf{x}_{t})$ and $\pi_{i}(\boldsymbol{a}_{t},\mathbf{x}_{t})$, respectively. For the rest of the paper, unless we state otherwise, we adopt this notation without time-to-build.

\subsection{Markov Perfect Nash Equilibrium}

\subsubsection{Definition}

Building on the seminal work of \cite{maskin1988theory1,maskin1988theory2}, most of the IO literature studying industry dynamics has used the solution concept of Markov Perfect Nash Equilibrium (MPNE). A key assumption in MPNE is that players' strategies at period $t$ are functions only of payoff-relevant state variables at the same period. In this model, it means that firms' strategies are functions of the vector $\mathbf{x}_{t}$ only. Let $\boldsymbol{\alpha}=\{\alpha_{i}(\mathbf{x}_{t}):i\in\mathcal{I}\}$ be a set of strategy functions, one for each firm. A MPNE is a set of strategy functions such that every firm is maximizing its value given the strategies of the other players. 

For given strategies of the other firms, the decision problem of a firm is a single-agent dynamic programming (DP) problem. Let $V_{i}^{\boldsymbol{\alpha}}(\mathbf{x}_{t})$ be the value function of this DP problem. This value function is the unique solution to the Bellman equation:
\begin{equation}
    V_{i}^{\boldsymbol{\alpha}}(\mathbf{x}_{t})= \max_{a_{it}\in\mathcal{A}_{i}(\mathbf{x}_{t})}
    \left\{
    \begin{array}
    [c]{l}
        \pi_{i}^{\boldsymbol{\alpha}}(a_{it},\mathbf{x}_{t})
        +\beta_{i}
        {\displaystyle\int}
        V_{i}^{\boldsymbol{\alpha}}(\mathbf{x}_{t+1})
        \text{ }dF_{x,i}^{\boldsymbol{\alpha}}(\mathbf{x}_{t+1}|\mathbf{x}_{t},a_{it})
    \end{array}
    \right\}  
\label{Bellman equation}
\end{equation} 
where $\pi_{i}^{\boldsymbol{\alpha}}(a_{it},\mathbf{x}_{t})$ and $F_{x,i}^{\boldsymbol{\alpha}}(\mathbf{x}_{t+1}|\mathbf{x}_{t},a_{it})$ are the firm's profit and the transition CDF of the state variables given action $a_{it}$ for
firm $i$ and the strategy functions $\{\alpha_{j}(\mathbf{x}_{t}):j\neq i\}$ for firms other than $i$. That is, $\pi_{i}^{\boldsymbol{\alpha}}(a_{it},\mathbf{x}_{t})=$ $\pi_{i}(a_{it},\alpha_{j}(\mathbf{x}_{t}):j\neq
i,\mathbf{x}_{t})$ and $F_{x,i}^{\boldsymbol{\alpha}}(\mathbf{x}_{t+1} | \mathbf{x}_{t},a_{it})=$ $F_{x}(\mathbf{x}_{t+1}|\mathbf{x}_{t},a_{it}
,\alpha_{j}(\mathbf{x}_{t}):j\neq i)$. For the description of some results, it is convenient to define the expression in brackets $\{ \}$ in equation (\ref{Bellman equation}) as the \textit{conditional choice value function} $v_{i}^{\boldsymbol{\alpha}}(a_{it},\mathbf{x}_{t})$. That is,
\begin{equation}
    v_{i}^{\boldsymbol{\alpha}}
    (a_{it},\mathbf{x}_{t})
    \equiv 
    \pi_{i}^{\boldsymbol{\alpha}}
    (a_{it},\mathbf{x}_{t})
    +\beta_{i}
    {\displaystyle\int}
    V_{i}^{\boldsymbol{\alpha}}(\mathbf{x}_{t+1})
    \text{ }
    dF_{x,i}^{\boldsymbol{\alpha}}
    (\mathbf{x}_{t+1}|\mathbf{x}_{t},a_{it})
\label{eq: conditional choice value}
\end{equation}
The best response decision rule for firm $i$ is \textit{argmax} in $a_{it}\in\mathcal{A}_{i}(\mathbf{x}_{t})$ of 
$v_{i}^{\boldsymbol{\alpha}}(a_{it},\mathbf{x}_{t})$.

\subsubsection{Equilibrium existence}

\cite{doraszelski2010computable} show that existence of a MPNE in pure strategies is not guaranteed in this model under the conditions in \cite{Ericson1995}. They show that, when firms make discrete choices such as market entry and exit decisions,  the existence of an equilibrium cannot be ensured without allowing firms to randomize over these discrete actions. A possible approach to guarantee equilibrium existence is to allow for mixed strategies. However, computing a MPNE in mixed strategies poses important computational challenges. Instead, to establish equilibrium existence in this class of models, \cite{doraszelski2010computable} propose incorporating private information state variables.\footnote{This issue of existence is also discussed in \cite{gowrisankaran1995dynamic} for these models.} This approach is in the spirit of \cite{harsanyi1973games} technique for purifying mixed-strategy Nash equilibria of static games. This incomplete information version of the Ericson-Pakes model has been the one adopted in most empirical applications of dynamic games in IO.

\subsubsection{Incomplete information \label{sec:model incomplete info}}

\cite{rust_1994_worldcongress} was the first paper to present an incomplete information version of the Ericson-Pakes model (see his section 9 on discrete dynamic games). This is also the model in the first econometric papers in this literature, such as \cite{Jofre-Bonet2003}, \cite{Aguirregabiria2007}, and \cite{Pakes2007}.

In addition to the common knowledge state variables $\mathbf{x}_{t}$, a firm's profit depends on a private information shock (or vector of shocks) $\varepsilon_{it}$, such that the profit function is $\pi_{i}(\boldsymbol{a}_{t},\mathbf{x}_{t},\varepsilon_{it})$. This private information shock is independently distributed over time and across firms with a distribution function
$F_{\varepsilon}$ that has support over the real line. Similarly as in the complete information version of Ericson-Pakes model, a \textit{Markov Perfect Bayesian
Nash Equilibrium (MPBNE)} in this model is an N-tuple of strategy functions $\boldsymbol{\alpha}= \{\alpha_{i}(\mathbf{x}_{t},\varepsilon_{it}): i\in\mathcal{I}\}$ such that a firm's strategy maximizes its value taking as given other firms' strategies. 

For the computation of an equilibrium and for the estimation of the model, it is very convenient to represent firm's strategies as \textit{conditional choice probabilities} (CCP). For any strategy function $\alpha_{i}(\mathbf{x}_{t},\varepsilon_{it})$ we can define its corresponding
CCP function, $P_{i}(a_{it}|\mathbf{x}_{t})$, as the probability distribution of the firm's action conditional on common knowledge state variables that is induced by this strategy and the distribution of private information. That is,
\begin{equation}
    P_{i}(a_{it}|\mathbf{x}_{t})
    \equiv
    \int 
    1\left\{
    \alpha_{i}(\mathbf{x}_{t},\varepsilon_{it})=a_{it}
    \right\}
    \text{ } dF_{\varepsilon}(\varepsilon_{it}), 
\end{equation}
where $1\left\{\cdot\right\}$ is the indicator function. This CCP function represents the expected behavior of a firm from the point of view of other firms (or the researcher) who do not know this firm's private information. 

We can represent firms' best responses and equilibrium conditions as a fixed point mapping in the space of these CCPs. Let $\mathbf{P} \equiv \{P_{i}(a_{it}|\mathbf{x}_{t})\}$ be a vector of CCPs for every firm $i\in\mathcal{I}$, every action $a_{it}\in\mathcal{A}$, and every state $\mathbf{x}_{t}\in\mathcal{X}$. Define $\pi_{i}^{\mathbf{P}}(a_{it},\mathbf{x}_{t},\varepsilon_{it})$ as firm $i$'s expected profit given that the other firm behave according to their respective CCPs in $\mathbf{P}$. That is,
\begin{equation}
    \pi_{i}^{\mathbf{P}}
    (a_{i},\mathbf{x}_{t},\varepsilon_{it}) 
    \equiv
    {\displaystyle\sum\limits_{\boldsymbol{a}_{-it}\in\mathcal{A}_{-i}
    (\mathbf{x}_{t})}}
    \left[
    {\displaystyle\prod\limits_{j\neq i}}
    P_{j}(a_{jt}|\mathbf{x}_{t})\right]  \pi_{i}(a_{it},\boldsymbol{a}_{-it},\mathbf{x}_{t},\varepsilon_{it}).
\label{eq:expected_profit_function}
\end{equation}
Similarly, let $F_{i}^{\mathbf{P}}(\mathbf{x}_{t+1}|\mathbf{x}_{t},a_{it})$ be the transition probability of the state variables from the perspective of firm $i$' and given that the other firms behave according their CCPs in $\mathbf{P}$:
\begin{equation}
    F_{i}^{\mathbf{P}}
    (\mathbf{x}_{t+1}|\mathbf{x}_{t},a_{it}) 
    \equiv
    {\displaystyle\sum\limits_{\boldsymbol{a}_{-it}\in\mathcal{A}_{-i}
    (\mathbf{x}_{t})}}
    \left[
    {\displaystyle\prod\limits_{j\neq i}}
    P_{j}(a_{jt}|\mathbf{x}_{t})\right]  F_{x}(\mathbf{x}_{t+1}|\mathbf{x}_{t},a_{it},\boldsymbol{a}_{-it})\text{.}
\end{equation}
Then, for every firm $i$, action $a_{it}$, and state $\mathbf{x}_{t}$, we have that CCPs satisfy the following equilibrium condition:
\begin{equation}
    P_{i}(a_{it}|\mathbf{x}_{t})
    =
    \int 1\left[  a_{it}=\arg\max_{a_{i}\in
    \mathcal{A}_{i}(\mathbf{x}_{t})}
    \left\{
    \begin{array}
    [c]{l}
        \pi_{i}^{\mathbf{P}}
        (a_{i},\mathbf{x}_{t},\varepsilon_{it})
        +\beta_{i}
        {\displaystyle\int}
        V_{i}^{\mathbf{P}}
        (\mathbf{x}_{t+1})
        \text{ } dF_{x,i}^{\mathbf{P}}
        (\mathbf{x}_{t+1}|
        \mathbf{x}_{t},a_{i})
    \end{array}
    \right\}  
    \right]  dF_{\varepsilon}
    (\varepsilon_{it})
\label{best reesponse probability}
\end{equation}
where $V_{i}^{\mathbf{P}}$ is the
(integrated) value function in 
firm $i$'s DP problem given that the other firms behave according their CCPs in $\mathbf{P}$. This value function uniquely solves the following \textit{integrated Bellman equation}:
\begin{equation}
    V_{i}^{\mathbf{P}}
    (\mathbf{x}_{t})=\int\max_{a_{i}\in\mathcal{A}_{i}(\mathbf{x}_{t})}
    \left\{
    \begin{array}
    [c]{l}
        \pi_{i}^{\mathbf{P}}
        (a_{i},\mathbf{x}_{t},\varepsilon_{it})+\beta_{i}
        {\displaystyle\int}
        V_{i}^{\mathbf{P}}
        (\mathbf{x}_{t+1})\text{ }
        dF_{x,i}^{\mathbf{P}}
        (\mathbf{x}_{t+1}|\mathbf{x}_{t},a_{i})
    \end{array}
    \right\}  dF_{\varepsilon}(\varepsilon_{it})\text{.}
\label{integrated Bellman equation}
\end{equation}

Using a more compact vector notation, a MPBNE is a vector of CCPs, $\mathbf{P} \in$ $[0,1]^
{|\mathcal{I}| |\mathcal{A}| |\mathcal{X}|}$ that solves the fixed point problem $\mathbf{P} =\Psi(\mathbf{P})$, where $\Psi(.)$
is a vector-valued function that is defined by stacking the function in the right-hand-side of equation (\ref{best reesponse probability}) over every  value $(i,a_{it},\mathbf{x}_{t})\in\mathcal{I}\times\mathcal{A}\times\mathcal{X}$. Vector
$\mathbf{P}$ lives in the compact simplex space, and under standard conditions on the distribution $F_{\varepsilon}$, the best response function $\Psi$ is
continuous. Therefore, Brower's fixed point theorem implies equilibrium existence (see footnote 45 in \cite{rust_1994_worldcongress}, and for more details, section 2.3 in \cite{Aguirregabiria2007}, and Proposition 2 in \cite{doraszelski2010computable}).

\subsubsection{Multiple equilibria \label{sec:model multiple equilibria}}

The model generically has multiple equilibria, as agents best respond to the strategies of other agents, and there are potentially many strategies that are consistent with this definition of equilibrium. There are conditions on the primitives of the model that guarantee equilibrium uniqueness, but they are typically strong restrictions. For instance, a set of sufficient conditions for uniqueness is: (i) the game has a finite time horizon; (ii) firms are (ex-ante) homogeneous in their profit functions and transition probabilities; and (iii) every period, only one firm can make an investment decision.\footnote{See \cite{Igami2017} for an empirical application that imposes this set of restrictions.} However, multiple equilibria are possible when we relax only one of the conditions (i), (ii), and (iii).  Alternatively, \cite{abbring2010} work through an entry model with identical firms, in which entry and exit decisions are assumed to follow a last-in first-out (LIFO) structure. With the addition of assumptions on the process for demand, they show that this model has a unique equilibrium in demand thresholds. An obstacle is how to extend this approach beyond identical firms. 

Dealing with multiple equilibria, in the estimation of the model and in counterfactual experiments using the estimated model, is a topic that has received substantial attention in this literature. We deal with this issue in section \ref{Section: Estimation methods}.

\subsection{Examples}

The framework presented above has been used in a wide range of empirical applications in IO, including market entry and exit, firms' adoption of new technologies or products, investment in physical capital or capacity, investment in R\&D and innovation, learning-by-doing, competition in product quality, product positioning, store geographic location, price competition with menu costs or/and with durable or storable products, search and matching, dynamic auction games, market networks, endogenous mergers, and exploitation of natural resources. We cover all these applications in section \ref{Section: Empirical Applications}. To illustrate some specific features and economic trade-offs in these models, we briefly describe here three examples.

\medskip

\noindent \textit{Market entry and exit}. The investment decision is binary, with $a_{it}=0$ if the firm is not active
in the market, and $a_{it}=1$ if active. The endogenous state variable is the lagged decision that determines if the firm is an incumbent ($a_{i,t-1}=1 $) or a potential entrant ($a_{i,t-1}=0$). Potential entrants pay an entry cost if they decide to enter. Incumbents do not pay an entry cost if they decide to be active, but pay exit costs (or receive a scrap value) if they choose to be inactive. A firm's number of years of experience in the market may have a positive effect on the firm's profit by increasing consumer demand or reducing costs (i.e., passive learning). A key parameter of interest in these applications is the sunk entry cost: the difference between the entry cost and the firm's scrap value upon exit. The magnitude of this sunk cost -- and its distribution across firms -- has important implications on market structure. Firms' uncertainty about future demand and costs also plays an important role in firms' entry and exit decisions. 

\medskip

\noindent \textit{Price competition with durable products}. In the market of a differentiated durable product, firms face a dynamic trade-off in their pricing decisions. A price reduction implies an increase in today's sales but also a reduction in future demand, as consumers buying the product today exit the market and will not be part of future demand. \cite{Goettler2011} study this type of dynamic pricing in the PC microchip industry, discussed in section \ref{section gg}. \cite{Esteban2007} study the effects of durability and secondary markets on dynamic price competition between automobile manufacturers.

\medskip

\noindent \textit{Exploitation of natural resources}. In industries where firms exploit a common-pool resource, a firm's amount of output implies a dynamic externality on other firms because of the depletion of the common stock. This is known popularly as \textit{the tragedy of the commons}. \cite{Huang2014} study this problem in the context of the shrimp fishery industry in North Carolina discussed in section \ref{sec natural resources}. They propose and estimate a dynamic game of fishermen's daily fishing decisions.

\subsection{Extensions of the basic framework}

\subsubsection{Continuous time}

\cite{Doraszelski2012} introduced continuous time methods to the dynamic games literature. A key property of continuous models is that, with probability one, only one firm moves at any time. This reduces the number of future states we need to integrate over to calculate expected continuation values, and  consequently it can generate computational savings when solving for an equilibrium. In a discrete time game, if there are $N$ agents with $A$ possible actions each, one has to integrate over $A^N$ states. By contrast, in continuous time, one has to sum over $(A-1)N$ terms; this is smaller than the previous quantity and also grows slower as either $A$ and $N$ increase. Furthermore, in continuous time models, transition matrices are typically more sparse than in discrete time, such that solution algorithms that exploit sparse matrices can be more effective when computing the solution to Bellman equations. A negative computational property of continuous time models is that the time discount factor becomes larger than in discrete time, such that a solution algorithm that iterates in the Bellman equation requires a larger number of iterations than in discrete time. However, when the number of players $N$ becomes larger, the savings from the smaller cost per iteration dominates the cost from the larger number of iterations.

The computational advantage of modelling dynamic games in continuous time comes with a cost in terms of flexibility and realism. In these models, a firm's ability to make a choice at time $t$ is exogenously determined and it does not depend on the value of the state variables. For instance, in an actual market, if a competitor cuts its price, a firm might want to respond almost instantaneously by cutting its own price. But in the continuous time model, the firm needs to wait until it has the opportunity to move, that has the same probability as if the competitor had charged a high price. The standard continuous time dynamic game can be modified to allow the hazard rate of a firm's move to depend on state variables. However, this extension eliminates the computational savings of the continuous time model, as it requires integrating over all possible states to calculate continuation values. Nevertheless, there may be  specifications that allow the hazard rate to depend on some of the state variables, allowing for more flexibility and realism than the standard model but still maintaining most of its computational advantages. \cite{Arcidiacono2016} developed methods for estimation and counterfactual experiments in dynamic discrete choice models in continuous time.

\subsubsection{Oblivious equilibrium}

In competitive models of industry dynamics, such as those explored in \cite{hopenhayn1992entry}, firms are atomistic. That is, any choice that they make has no bearing on the evolution of the aggregate state. This is very helpful for computation, since it means that firms can take the path of the aggregate state as given. If one assumes that there are no aggregate shocks, i.e., shocks that create uncertainty on the evolution of a market level state such as changes in demand, then the path of the aggregate state is also deterministic. In other words, firms have perfect foresight.\footnote{Of course, if there are aggregate shocks, such as shocks to demand, then firms will need consider the distribution of aggregate states in the future. This is the basis of many models in macroeconomics with heterogeneous agents. See \cite{krusell1998income} for a very influential example of this type of work.} 

In contrast, the work that builds on \cite{Ericson1995} takes as a fundamental that firms can, by themselves, affect the profits of their rivals. This brings up issue of strategic considerations, which requires firms to take into account the entire distribution of their rival's states. As a matter of computation, the size of the state space required for even very austere models of industry dynamics is quite large. For instance, a \cite{Pakes1994} model with 10 firms choosing 20 different quality levels, has over 10 trillion states, which raises an issue of how to store the value function in computer memory, let alone how to compute it.

\cite{weintraub2008markov} work out the consequences of a model where firms behave more closely to how they would in a competitive environment, with the goal of simplifying computation. More precisely, they propose a model where firms only keep track of their own state when making choices, that is they restrict strategies $\alpha_{i}(\cdot)$ to be functions of a firm-specific state $x_{it}$, which is a component of the vector $\mathbf{x}_{t}$. In addition, every firm believes that the state of the market (i.e., the average value of $x_{it}$ over all the firms in the market) is the long-run average of this variable in equilibrium. Given that the state space considered by oblivious firms is merely $x_{it}$, the state space is as large as the one considered by a single agent model, considerably alleviating computation. 

The restrictions imposed by the oblivious equilibrium concept on the variables that enter firms' strategies are somewhat ad hoc, much as the MPNE refinement restricts strategies to depend only on payoff relevant state variables and rules out the type of strategies typically used to describe tacit collusion such as \cite{green1984noncooperative}. However, \cite{weintraub2008markov} also show that, for a class of oligopoly models than include a specific differentiated product quality game, the oblivious equilibrium of this game converges to the MPNE as the market becomes large and the number of firms increases. Furthermore, under certain conditions, the quality of this approximation, measured by numerical experiments, can be quite reasonable even with a couple of firms in the market. Thus, one can think of oblivious equilibrium as a good approximation for MPNE in industries with many firms, none of which is particularly large, rather than a theoretical refinement which of interest by itself. Oblivious equilibria have been used in a number of applications, such as \cite{xu2008structural}. 

The immediate issue with oblivious equilbria is how to generalize its computational simplicity to environments where firms may need to keep track of more than their state variable $x_{it}$. For instance, many industries are characterized by a small number of leading firms and a large number of fringe firms. In this context it may make sense for firms to track the states of the dominant firms in addition to their own state variable. \cite{benkard2015oblivious} study this model. 

Even in an unconcentrated industry, there are some difficulties with oblivious strategies if there are aggregate state variables --- again think of a macro demand shifter. For instance, if we consider a market where demand has been declining, say Cleveland, there may be a large number of firms compared to a market where demand has been growing over time, like Austin, even if the two cities currently have comparable metro level populations. This means that current demand might not capture the number of firms in the market very well, and oblivious equilibria could differ substantially from MPNE. One way to get around this problem is the moment based Markov equilibrium (henceforth MME) proposed by \cite{Weintraub2017}. This equilibrium concept restricts firms' strategies to be functions of the firm's own state $x_{it}$ but also of as moments of the state $\mathbf{x}_t$, which are denoted as $\widehat{\mathbf{s}}_{t}$. For instance, in a dynamic game of investment in capacity where firms compete each period à la Cournot, relevant moments in $\widehat{\mathbf{s}}_{t}$ could be the number of active firms and total capacity of all the firms. These two moments are sufficient statistics for profits in a static Cournot game with ex-ante homogeneous firms. Note however, there is no guarantee that two states $\mathbf{x}_t$ that generate the same moments will have the same profits in the future in this game, so they are not usually sufficient statistics for a firm's value function.

An important contribution of \cite{Weintraub2017} is to structure how firms form expectations on current and future profits given their own state $x_{it}$ as well as the moment based state for the rest of the market given by $\widehat{\mathbf{s}}_{t}$. The issue here is that one needs to forecast current profits and future state transitions given the MME state $(x_{i,t},\widehat{\mathbf{s}}_{t})$. In the Cournot example discussed above, the total capacity and firm capacity are sufficient statistics to compute the profit function. This is not usually the case. Thus, one needs to associate the expected profit $\widehat{\pi}(x_{it},\widehat{\mathbf{s}}_{t})$ to the actual profit function $\pi(\mathbf{x}_{t})$ for the states $\mathbf{x}_{t}$ that generate moments $\widehat{\mathbf{s}}_{t}$, and this requires understanding what type of weighted sum will do this properly. Furthermore, the state transitions defined on the MME state denoted as $\widehat{F}[x_{i,t+1},\widehat{\mathbf{s}}_{t+1}|x_{it},\widehat{\mathbf{s}}_{t}]$ are also unknown, and these are essential for computing the firm's value function. \cite{Weintraub2017} propose to sample from the ergodic distribution. That is, given a set of MME strategies $\hat{\alpha}(x_{it},\widehat{\mathbf{s}}_{t})$, they use forward simulation to compute both the expected profit function $\pi(\mathbf{x}_{t})$ and expected state transition matrix $\widehat{F}[x_{i,t+1},\widehat{\mathbf{s}}_{t+1}|x_{it},\widehat{\mathbf{s}}_{t}]$. This allows them to solve for the value function and firm policies, and repeat this sampling procedure to compute $\widehat{\pi}$ and $\widehat{F}$, until this algorithm converges. Incidentally, this sampling idea is also found in the algorithm used in \cite{Fershtman2012}'s model with asymmetric information. Moreover, a natural specification of MME without any moments at all, reduces down to an oblivious equilibrium. A number of recent papers have used MME, such as \cite{jeon2020learning}, \cite{caoui2019estimating}, and \cite{vreugdenhil2020booms}, which all incorporate rich firm level heterogeneity making a reduction of the state space inescapable.\footnote{Notice than an attractive part of using MME's is that the reduction in the state space happens in the choice of the moments $\widehat{\mathbf{s}}_{t}$. Otherwise, this choice will be made earlier in the paper when choosing the richness of the underlying state space $\mathbf{x}_t$ to begin with.}

While the research can in principle choose any vector of moments $\widehat{\mathbf{s}}_{t}$, usually these moments are chosen so that the MME might be close to a MPNE of the game. This immediately poses the question of which moments to choose, which we discuss in more detail in the next section. 

\subsubsection{Large state spaces \label{large state space section}}

An alternative to using MME to reduce the state space is to approximate the value function with a basis approximation such as $V(\mathbf{x}_t) \approx \sum_{k} ^K \theta_k \phi^k(\mathbf{x}_t)$, where each $\phi^k (\cdot)$ is a basis function and $\theta_k$ is a coefficient. To make this more concrete, if two firms were competing in quality and $\mathbf{x}_t = (x_{1t},x_{2t})$ where $x_{it}$ is firm $i$'s quality at period $t$, then basis functions could be a second order polynomial such that value function $V(x_{1t},x_{2t})$ is approximated using $\theta_{1} + \theta_{2} x_{1t} + \theta_{3} x_{2t} + \theta_{4} x_{1t}^{2} + \theta_{5} x_{2t}^{2} + \theta_{6} x_{1t}x_{2t}$. A key feature of this approach is that to compute an approximation to the solution of the DP problem, one does not need to solve for a fixed point for the value function at each state $\mathbf{x}_t$, but for a fixed point in the space of the vector of coefficients $\boldsymbol{\theta} \equiv \{\theta_k\}_{k=1} ^K$. A good example of this approach using Chebyshev polynomials for a basis function is \cite{doraszelski2003r}'s work on an R\&D race in duopoly.\footnote{These \textit{approximate DP} methods are extensively covered in a book by \cite{powell2007approximate}.} \cite{farias2012approximate} explore the numerical implementation of these solutions in the context of a \cite{Pakes1994} model. Several empirical papers have also used this approach, such as 
\cite{Sweeting2013} for a dynamic game of competition between radio stations (that we describe in section \ref{sweet radio}), \cite{barwick2015costs} for the market for real estate agents, and \cite{Arcidiacono2016} for retail entry. 

In the context of dynamic games, the main difficulty in the application of this method is finding a suitable basis function, given the large dimension of the state $\mathbf{x}_t$. Clearly, polynomial basis  functions do not work when there are ten firms in the market since this means that the space of $\boldsymbol{\theta}$ has at least ten dimensions. Instead,  \cite{powell2007approximate} suggests including features of the states, which roughly translates to picking relevant moments of the state and have basis functions defined over them. This returns the question to how to properly pick moments of the state as in our discussion of MME. One approach is to think about the components of the state that help predict profits. In the Cournot example, total capacity of the industry is the relevant state variable. In a model with firms competing in quality with a logit demand system, the relevant aggregate is the inclusive value, such as in \cite{gowrisankaran2012dynamics} and \cite{aguirregabiria2012airline}. 

A promising approach for solving dynamic games is using tools from machine learning and artificial intelligence. For instance, in the context of \textit{approximate DP} methods described above, the relevant moments can be identified using an iterative algorithm that simultaneously solves for the solution of the Bellman equation. In this spirit, \cite{kalouptsidi2018detection} uses LASSO to pick out the basis functions in her application. More promising is the application to dynamic games of newly developed techniques in deep learning.

\subsubsection{Persistent asymmetric information}

The model considered so far assumes that the only form of firms' private information is an idiosyncratic shock that is independently distributed across firms and over time. This precludes interesting forms of asymmetric information. For instance, \cite{laffont1993theory}'s work on the regulation of a monopolist with unknown costs relies critically on persistent asymmetric information. There is a deep interest in IO theory on these types of models. 

In dynamic games, an important challenge of incorporating persistent private information is that the complete history of previous states and decisions becomes payoff relevant. For a firm trying to uncover its rival's private information, any action that they have taken in the past, and the context in which this action was taken, i.e., the state at that point, is relevant to form a posterior on their private information. Therefore, if we maintain the assumption of MPBNE without further restrictions, the dimension of the state space becomes intractable for solving or estimating even for simple versions of these games.

\cite{Fershtman2012} study this type of model and propose an alternative equilibrium concept to deal with the high dimensionality problem of MPBNE. In their model, every firm $i$ observes a public state $\widetilde{\mathbf{x}}_{t}$ and a private state $\varepsilon_{it}$, such that their information set at period $t$ is $(\widetilde{\mathbf{x}}_{t}, \varepsilon_{it})$ and its strategy function is $\alpha_{i}(\widetilde{\mathbf{x}}_{t}, \varepsilon_{it})$. This information set has two important differences with respect to the model considered so far. First, $\widetilde{\mathbf{x}}_{t}$  may contain lagged values $\{\mathbf{x}_{s}:s<t\}$. Second, $\varepsilon_{it}$ can be serially correlated or even time-invariant. To avoid the dimensionality problem of the MPBNE solution concept in this game, \cite{Fershtman2012} propose an equilibrium concept that they denote \textit{Experience Based Equilibrium} (EBE). An EBE imposes three types of restrictions on equilibrium strategies: (i) if a state is visited, then this state will be visited in the future repeatedly; (ii) strategies are optimal given the evaluations of outcomes (profits); and (iii) strategies generate expected discounted values of profits that are consistent with these evaluations in the recurrent subset of states. \cite{Fershtman2012} propose a reinforcement learning algorithm to compute EBE strategies, taking ideas from Q-learning implemented in \cite{Pakes2001}.

In dynamic games with persistent asymmetric information, an assumption that can reduce substantially the dimensionality problem is that firms' private information becomes common knowledge every $T$ periods. Under this condition, $\widetilde{\mathbf{x}}_{t} = \mathbf{x}_{t}$ when private information is revealed; $\widetilde{\mathbf{x}}_{t} = (\mathbf{x}_{t-1},\mathbf{x}_{t})$ one period after revealing private information; and so on, such that the maximum dimension of the state space is $\widetilde{\mathbf{x}}_{t} = (\mathbf{x}_{t-T+1}, ..., \mathbf{x}_{t})$.

\cite{asker2020computational} apply the EBE solution concept together with the assumption of information revelation every $T$ periods to
a dynamic auction game. Take the example of either timber or construction auctions, which have been studied extensively in empirical IO. In both contexts, firms are competing against rivals repeatedly, and there are important sources of persistent asymmetric information. For instance, firms have backlogs of construction projects or timber that affect their bidding behavior (i.e., a larger backlog lowers the value from winning an auction). A firm's backlog is not perfectly known by its rivals. However, when a firm wins an auction, its bid becomes publicly available and this yields information to rivals that is helpful for subsequent auctions. Furthermore, firms may benefit from a commitment to share their backlog information with each other. \cite{asker2020computational} compute EBE for different values of $T$ to evaluate the impact of information sharing among bidders. They show that information sharing, even of strategically important data, can be welfare increasing.

\cite{Fershtman2012}'s approach to deal with persistent asymmetric information can be critical for making empirical work on dynamic games less tied down to the MPNE solution concept. However, the issues with large state spaces become substantially more difficult since some history dependence needs to be tracked. Applied work will need to address these computational problems in order to make this approach more than a proof of concept.

\subsubsection{Firms' biased beliefs}

Firms' behavior depends on their beliefs about the actions of other firms in the same market. Managers and their firms have different abilities to collect and process information and, as a result, they are heterogeneous in their expectations. This heterogeneity in beliefs can have important implications on firms' performance and welfare. The importance of firms' heterogeneity in their ability to form expectations and the possibility of biased beliefs has been long recognized in economics, at least since the work of \cite{simon_1959_aer}. However, in most fields in economics, the status quo has been to assume rational expectations. In particular, as we described above, the assumption of Markov perfect equilibrium has been very common in empirical applications of dynamic games in IO.

Recent papers in empirical IO relax the assumption of rational expectations and present evidence of substantial heterogeneity and biases in firms' beliefs. As one would expect, biased beliefs are more likely in new markets and after regulatory changes. For instance, after deregulation of the US telecommunication industry (\cite{goldfarb_xiao_2011}), the UK electricity market (\cite{Doraszelski2018}), the Texas electricity spot market (\cite{hortaccsu_puller_2008}; \cite{hortaccsu_puller_2019}), or the Washington State liquor market (\cite{huang_ellickson_2020}), and in the early years of the fast-food restaurant industry in UK (\cite{aguirregabiria_magesan_2020}) or China (\cite{xie_2021_jbes}). 

Most of these applications consider static games of market competition and use the solution concepts of level-K rationality introduced by \cite{stahl_wilson_1995} and \cite{nagel_1995}, and the Cognitive Hierarchy (hereafter, CH) equilibrium introduced by \cite{camerer_ho_2004}. Let $B_{i}(\boldsymbol{a}_{-it} | \mathbf{x}_{t})$ be the probability distribution that represents firm $i$'s belief about other firms' actions given common knowledge state variables $\mathbf{x}_{t}$. Under Bayesian Nash Equilibrium (BNE), beliefs correspond to the actual probability distribution, as represented by players' CCPs, such that $B_{i}(\boldsymbol{a}_{-it} | \mathbf{x}_{t}) = \prod_{j \neq i} P_{j}(a_{jt} | \mathbf{x}_{t})$. In contrast, level-K rationality and CH are equilibrium concepts where firms have biased beliefs. In these models, firms are heterogeneous
in their beliefs and there is a finite number of belief types. That is, the probability distribution $B_{i}(\boldsymbol{a}_{-it} | \mathbf{x}_{t})$
belongs to a finite number $K$ of belief types. These types correspond to different levels of strategic sophistication and are determined by a hierarchical structure. A type-0 firm has an arbitrary belief function $B^{(0)}(\boldsymbol{a}_{-it} | \mathbf{x}_{t})$. In the level-k model, a type-k firm believes that all the other firms are type $k-1$. This recursive structure
defines the belief functions for every type k between 1 and K. The only unrestricted function is the beliefs function for
type-0: the rest of the belief functions are known functions of $B^{(0)}$ and the structural parameters of the model. The CH model is more flexible than the level-$K$ model. In the CH model, a type-k firm believes that the other firms come from a probability distribution over types $0$ to $k-1$. This is the model of firms' biased beliefs that has been most commonly used in IO applications, e.g., \cite{goldfarb_xiao_2011}, \cite{brown_camerer_2013}, and \cite{hortaccsu_puller_2019}. These models  
impose important restrictions: they do not include BNE or rational beliefs as a particular case, and there is a small number of belief types. However, they have the attractive feature of being equilibrium models where (biased) beliefs are determined endogenously. This feature makes them particularly attractive for counterfactual experiments.

In dynamic games, every period $t$ firms need to form probabilistic beliefs about
the actions of competitors not only at the current period but also at future periods. Let $B^{t}_{i,t+s}(\boldsymbol{a}_{-i,t+s} | \mathbf{x}_{t+s})$ be the probability distribution that represents firm $i$'s beliefs at period $t$ about the behaviour of
competitors at period $t+s$ if the state is $\mathbf{x}_{t+s}$. A firm can update its beliefs over time, and $B^{t+1}_{i,t+s}(.) - B^{t}_{i,t+s}(.)$ represents the updating from period $t$ to period $t+1$ in the beliefs that firm $i$ has about the behaviour of competitors at period $t+s$. This belief structure is very general and allows for general forms of firms' learning or forgetting. Given its beliefs at period $t$, a firm's best response is the solution of a single-agent dynamic programming problem. Under MPBNE, firms' beliefs are equal to the actual probability distribution of other firms' choices: $B^{t}_{i,t+s}(\boldsymbol{a}_{-i,t+s} | \mathbf{x}_{t+s}) = \prod_{j \neq i} P_{j,t+s}(a_{j,t+s} | \mathbf{x}_{t+s})$.\footnote{The notation here considers a non-stationary model. If the model is stationary, then CCP functions $P_{j}(.)$ are time invariant.} \textit{CH} and \textit{Level-K rationality} models have been extended to dynamic games (e.g., \cite{ho_su_2013_ms}). However, these models impose strong restrictions on the evolution of beliefs over time: a firm's belief type does not vary over time. As far as we know, there are not IO applications of these models in dynamic games. \cite{aguirregabiria_magesan_2020} consider empirical dynamic games where firms' belief functions have the general structure describe above. They study the nonparametric identification of belief and payoff functions in this model, and apply this model to study competition in number of stores between McDonalds and Burger King in UK.

\section{Identification and estimation \label{Section: Identification Estimation}} 

\subsection{Data} \label{sec:data}

The datasets used in most applications of dynamic games in IO can be described as panel data of $M$ markets (geographic locations, products), over $T$ periods of time, with information on actions and state variables for $N$ players (often firms). The order of magnitude of $M$, $T$, and $N$, and the structure of the panel (e.g., how unbalanced it is in the different dimensions) varies across applications. Since most applications study oligopoly markets, the number of firms $N$ is typically small, but there is also a literature considered in section \ref{dynamic matching} that handles cases with a large number of agents. The number of periods $T$ is often small too. In many applications, a substantial part of the sample variation comes from the number of markets $M$ that may include hundreds or thousands of locations. Nevertheless, there are applications where the global nature of the industry implies very few markets, or even a single national or world market. For instance, this is the case for PC microchips (\cite{Goettler2011}), or hard drives (\cite{Igami2017}) discussed in section \ref{innovation}. In these cases, enough sample variation is achieved by the joint combination of firms, time periods, and markets, where neither of the three dimensions is large but $MNT$ can be large enough.

More generally, we have a three dimensional panel dataset $\{a_{imt} ,\mathbf{x}_{imt}:i=1,2,...,N$; $t=1,2,...,T$; $m\in\mathcal{M}_{it}\}$, where $i$ indexes firms, $t$ time, and $m$ markets. We use $\mathcal{M}_{it}\subseteq\mathcal{M\equiv}\{1,2,...,M\}$ to represent the set of markets where firm $i$ is observed making decisions at period $t$. The structure of the sets $\mathcal{M}_{it}$ is important for the identification and estimation of the model. In some industries every firm is a player in all (or most of) the $M$ markets such that $\mathcal{M}_{it}=\mathcal{M}$ for every $(i,t)$. For instance, this is the case in a retail industry characterized by competition between large retail chains which are potential entrants in any of the geographic markets that constitute the industry, discussed in more detail in section \ref{retail}. With this type of data, the researcher can allow for rich forms of firm heterogeneity that is fixed across markets and time by estimating firm-specific structural parameters. In other industries, even if competition is local and $M$ is large, most of the firms specialize in operating in a few markets such that $\mathcal{M}_{it}$ is typically a small subset of $\mathcal{M}$. In these cases, allowing for rich forms of firm unobserved heterogeneity requires more sophisticated econometric methods, and sometimes restrictions.

In some applications, the data also includes information on prices and quantities, production functions and input costs, that can be used to estimate demand functions and variable costs. In fact, as we show in section \ref{Section: Empirical Applications}, empirical applications of dynamic games in IO are characterized by a wide variation in the types of data being used. The idea of looking a panel of independent markets using data on observed demand factors and market structure pioneered in the static setting by \cite{bresnahan1991entry} or \cite{berry1992estimation} has been carried over to dynamic settings. This starts with the incompletely dynamic models of \cite{Bresnahan1994} (used in \cite{collard2014mergers} as well), but adopted by both \cite{Collard-Wexler2013} and \cite{Dunne2013}. Indeed, the most popular methods for the estimation of dynamic games (the two-step CCP methods that we describe in sections \ref{sec: two step methods} and \ref{BBL method}) are particularly well suited to use panel data from many independent markets. However, this is not the only type of data that has been brought to bear on these questions. For many applications, such as those for industries with a single national of world market, this panel data with large $M$ approach is not feasible. 

Beyond the problem of not having a cross-section of markets, the original idea from \cite{bresnahan1991entry} of backing out markups purely from the pattern of market structure and market demand shifters, that is, without any information on prices or costs, is an admittedly heroic use of economic theory to structure estimation. A first alternative is to use more traditional static cost and demand estimation to fill in much of the period profit function. The second approach is to calibrate both the determinants of static profits, and dynamic costs, such as entry or scrap values, from more anecdotal accounting data or engineering estimates. Indeed, \cite{Benkard2004} is a particularly extreme case in that no dynamic choices are used to estimate the model, and there is only a single global market for aircraft. 

Many papers have used a more ragtag empirical approach in this literature. For instance, in \cite{Ryan2012}, demand and costs are estimated using traditional IO methods for these problems. Only a very parsimonious number of parameters are estimated using the dynamic structure of the industry, such as entry and exit costs and investment adjustment costs.  While there are fewer flagship methodological papers to illustrate how these methods work, most serious empirical applications of dynamic oligopoly use empirical support in the form of well researched calibrations, static demand and cost estimation, as well as estimation that leverages the dynamic choices in the dynamic oligopoly game.  

This makes the literature on dynamic games quite different from, say, the empirical literature on demand estimation and production function estimation that is organized around common data structures, such as data on prices and quantities, for demand, or firm level data on output and inputs, for production functions. 

\subsection{Identification} \label{sec:identification}

The structural parameters (or functions) of the model consist of the profit functions $\pi_{i}$, the discount factors $\beta_{i}$, the transition probability of the state variables $F_{x}$, and the distribution of private information shocks $F_{\varepsilon}$. We represent all these parameters in a compact form using $\boldsymbol{\theta} \equiv \{\pi_{i}$, $\beta_{i}$, $F_{x}$, $F_{\varepsilon}:i\in\mathcal{I}\}$. The researcher is interested in the identification of $\boldsymbol{\theta}$ using the data described above.

In this subsection, we present identification and non-identification results for dynamic games. We start with a well known non-identification result. Then, we present a set of sufficient conditions for identification that has been used, implicitly or explicitly, in most existing empirical applications of dynamic games in IO. Finally, we discuss recent studies showing identification of dynamic games under weaker conditions than some of these sufficient conditions. In particular, we focus on the
following identification issues that have received attention in the literature: (i) unobserved market heterogeneity; (ii) multiple equilibria in the data; (iii) normalization of the payoff of a choice alternative; (iv) time discount factor; (v) non-additive unobservables; (vi) nonparametric distribution of unobservables; and (vii) non-equilibrium beliefs. Most of these results have been developed in the context of discrete dynamic games (i.e., firms' actions and observed state variables are discrete), and this class of models is the focus of sections \ref{sec:no identification} to  \ref{sec:relaxingrestrictions}. In section \ref{sec:identification mixed}, we present results for mixed continuous-discrete choice models.

\subsubsection{Non-identification result \label{sec:no identification}}

In the context of single-agent dynamic discrete choice games, \cite{rust_1994_hbook} and \cite{magnac_thesmar_2002} present a non-identification result that is
well-known in the literature of dynamic structural models. Note that a single-agent dynamic model is a restricted version of the dynamic game in section \ref{Section: Models}: payoff functions $\pi_{i}(a_{it},\boldsymbol{a}_{-it},\mathbf{x}_{t})$ and transition probability functions $F_{x,i}(\mathbf{x}_{t+1}|\mathbf{x}_{t},a_{it},\boldsymbol{a}_{-it})$ do not depend on other firms' actions, $\boldsymbol{a}_{-it}$. Since this model is more restrictive than a dynamic game, and it is estimated using the same type of data, non-identification of single-agent models implies non-identification of the dynamic game version. For the same reason, a positive identification result for dynamic games implies identification of its restricted single-agent version.

The set of assumptions (ID.1) to (ID.4) below define a class of dynamic discrete choice models that has been used in many empirical applications. These assumptions were first introduced by 
\cite{rust1987optimal} and \cite{rust_1994_hbook} such that this model is often referred as \textit{Rust model}.

\medskip

\noindent\textbf{Assumption (ID.1). No common knowledge unobservables.} The researcher observes all the state variables that are common knowledge to firms, $\mathbf{x}_{t}$. The only unobservables for the researcher are the private information shocks $\varepsilon_{it}$.

\medskip

\noindent\textbf{Assumption (ID.2). Additive unobservables.} The private information
variables are additively separable in the payoff function. More specifically,
the profit function has the form $\pi_{i}(\boldsymbol{a}_{t},\mathbf{x}
_{t})+\varepsilon_{it}(a_{it})$, where $\{\varepsilon_{it}(a_{i}):a_{i}\in\mathcal{A}_{i}\}$ are the unobservable shocks.

\medskip

\noindent\textbf{Assumption (ID.3). Known distribution of the unobservables.} The
probability distribution $F_{\varepsilon}$ does not depend on any parameter that is unknown to the researcher. Furthermore, it is strictly increasing over the whole Euclidean
space $\mathbb{R}^{J+1}$.

\medskip

\noindent\textbf{Assumption (ID.4). Conditional independence.} Conditional only
$(\boldsymbol{a}_{t},\mathbf{x}_{t})$, the realization of $\mathbf{x}_{t+1}$ is independent of $\varepsilon_{t}$. Note that this assumption was already included in the description of the incomplete information dynamic game in section \ref{sec:model incomplete info}.

\medskip

Under conditions (ID.1) to (ID.4), the CCP functions $P_{i}(a_{it}| \mathbf{x}_{t})$ and the transition probability function $F_{x}(\mathbf{x}_{t+1}| \mathbf{x}_{t}, \boldsymbol{a}_{t})$ are nonparametrically identified. Furthermore, \cite{hotz1993conditional} show that there is a one-to-one relationship between CCPs and conditional choice value function (as defined in equation \ref{eq: conditional choice value}) relative to a baseline choice alternative, that we can represent as $\widetilde{v}_{i}(a_{it},\mathbf{x}_{t})\equiv v_{i}(a_{it},\mathbf{x}_{t})-v_{i}(0,\mathbf{x}_{t})$ (Proposition 1 in \cite{hotz1993conditional}). Therefore, under these assumptions, differences in conditional choice value functions $\widetilde{v}_{i}(a_{it},\mathbf{x}_{t})$ are uniquely identified. However, \cite{rust_1994_hbook} and \cite{magnac_thesmar_2002} show that knowledge of differences in conditional choice value functions is not sufficient to identify the payoff function $\pi_{i}$ and the discount factor $\beta_{i}$ (Proposition 2 in \cite{magnac_thesmar_2002}). 

There are two main identification issues involved in this non-identification result. First, as in any other revealed preference approach, we can identify payoff function $\pi_{i}$ only relative to the payoff of a baseline alternative. The typical approach is to normalize to zero all the payoffs of a baseline choice alternative, e.g., $\pi_{i}(a_{it}=0,\mathbf{x}_{t})=0$ for every value of $\mathbf{x}_{t}$. The problem is that, in contrast to static discrete choice models, this normalization condition is not innocuous in dynamic models (see \cite{Aguirregabiria2014}, and \cite{kalouptsidi_scott_2021}). In the equation describing value differences $\widetilde{v}_{i}$ as functions of the  structural parameters, the payoff for the baseline alternative interacts with the discount factor and with transition probabilities of the state variables. Therefore, the effect on value differences (and on CCPs) of a change in the discount factor or in transition probabilities depends on the level of the baseline payoffs, such that misspecification of baseline payoffs implies biases in the predictions about these effects.

A second problem comes from the identification of the time discount factor
$\beta_{i}$. The identified value difference $\widetilde{v}_{i}(a_{it},\mathbf{x}_{t})$ has two additive components: the difference in current payoffs, $\pi_{i}(a_{it},\mathbf{x}_{t})-\pi_{i}(0,\mathbf{x}_{t})$; and the difference in continuation values $\beta_{i}$ ${\textstyle\sum\nolimits_{\mathbf{x}_{t+1}}}
V_{i}(\mathbf{x}_{t+1})$ $[F_{x,i}(\mathbf{x}_{t+1}|\mathbf{x}_{t},a_{it})-F_{x,i}(\mathbf{x}_{t+1}|\mathbf{x}_{t},0)]$ where $F_{x,i}$ is identified and the value function $V_{i}$ only depends on this transition probability and on $\pi_{i}$,
and $\beta_{i}$. Knowledge of the value difference $\widetilde{v}_{i}(a_{it},\mathbf{x}_{t})$ is not enough to separately identify $\pi_{i}(a_{it},\mathbf{x}_{t})-\pi_{i}(0,\mathbf{x}_{t})$ and $\beta_{i}$. The
intuition is simple. Without further restrictions, the difference in
continuation values depends on the same variables as the difference in current
payoffs. Therefore, the model can explain the data -- i.e., the value differences --
equally well with any value of $\beta_{i}$ between $0$ and $1$.\footnote{Note that the same non-identification holds if the discount factor is restricted to be the same across firms.}

\subsubsection{A set of sufficient conditions for identification \label{sec:positive identification}}

\noindent \textit{(a) Identification of single-agent dynamic model.} Suppose that the researcher has data only on agents' choices and states, $\{a_{imt},\mathbf{x}_{mt}\}$. Consider the following additional assumptions.

\medskip

\noindent \textbf{Assumption (ID.5). Normalization of payoff of one choice alternative.} For one of the choice alternatives, say $a_{it}=0$, the profit
function is equal to zero (or to any other value known to the researcher): $\pi_{i}(a_{it}=0,\mathbf{x}_{t})=0$ for any value of $\mathbf{x}_{t}$.

\medskip

\noindent \textbf{Assumption (ID.6). Known time discount factors.} The discount factors
$\beta_{i}$ are known to the researcher.

\medskip

Under conditions (ID.1) to (ID.6) all the structural parameters in $\boldsymbol{\theta} \equiv \{\pi_{i}$, $\beta_{i}$, $F_{x}$, $F_{\varepsilon}:i\in\mathcal{I}\}$ are identified in the single-agent model.\footnote{This positive identification result is a corollary of Magnac and Thesmar's Proposition on non-identification (Proposition 2 in \cite{magnac_thesmar_2002}). See also Proposition 1 in \cite{aguirregabiria_2005_econletters} for an explicit statement and proof of this positive identification result.} This set of identification restrictions has been, so far, the most commonly used in empirical applications of single-agent dynamic structural models.

\medskip

\noindent \textit{(b) Identification of dynamic game.} The identification of the dynamic game needs to deal with two additional issues. First, the dynamic game can have multiple equilibria and, in principle, observations in the data may come from different equilibria. Second, under conditions (ID.1) to (ID.6), and ignoring for the moment the issue of multiple equilibria in the data, the expected profit function $\pi_{i}^{\mathbf{P}}(a_{it},\mathbf{x}_{t})$ -- as defined in equation (\ref{eq:expected_profit_function}) -- is identified. However, the profit function $\pi_{i}(a_{it},\boldsymbol{a}_{-it},\mathbf{x}_{t})$ depends on the actions of all the other players. That is, the dimension of the structural payoff function $\pi_{i}(a_{it},\boldsymbol{a}_{-it},\mathbf{x}_{t})$ is larger than the dimension of the identified expected payoff $\pi_{i}^{\mathbf{P}}(a_{it},\mathbf{x}_{t})$. As it is common in other empirical games or in econometric models with social interactions, solving this identification problem requires exclusion restrictions. Consider the following assumptions.

\medskip

\noindent \textbf{Assumption (ID.7). Single MPBNE in the data.} Every observation $(i,m,t)$ in the sample comes from the same Markov perfect equilibrium.

\medskip

\noindent \textbf{Assumption (ID.8). Exclusion restriction in payoff.} The vector of observable state variables $\mathbf{x}_{t}$ contains firm-specific state variables that enter in the profit function of a firm but not in the profit function of competitors. More specifically, $\mathbf{x}_{t}=$ $(\mathbf{x}_{t}^{c},z_{it}:i\in\mathcal{I)}$, and the profit function is $\pi_{i}(\boldsymbol{a}_{t},\mathbf{x}_{t}^{c},z_{it})$ that does not depend on $z_{jt}$ for $\mathit{j\neq i}$. Furthermore, the support of $z_{it}$ has at least as many points as the support of $a_{it}$.

\medskip

Under conditions (ID.1) to (ID.8) all the structural parameters $\boldsymbol{\theta} \equiv \{\pi_{i}$, $\beta_{i}$, $F_{x}$, $F_{\varepsilon}:i\in\mathcal{I}\}$ are identified in the dynamic game (see Proposition 3 in \cite{Pesendorfer2008}). Similarly as for the case of single-agent models, these have been the most commonly used identification restrictions in empirical applications of dynamic games in IO.

\subsubsection{Relaxing restrictions (ID.1) to (ID.8) \label{sec:relaxingrestrictions}}

Some of the identification restrictions (ID.1) to (ID.8) are strong, and they might not hold in some applications, such that imposing these restrictions may generate important biases in parameter estimates and in our understanding of firms' behavior and the determinants of market structure in those industries. During the last decade, there has been a substantial amount of research dealing with identification results relaxing some of the conditions (ID.1) to (ID.8). This subsection reviews this literature.

\medskip

\noindent \textbf{(i) Incorporating common-knowledge serially correlated unobservables}

Assumption (ID.1) establishes that the only unobservables for the researcher are the private information shocks, which are i.i.d. over firms, markets, and time. In most applications in IO, this assumption is not realistic and can be easily rejected by the data. Markets and firms differ in terms of characteristics that are payoff-relevant. Some of these differences can be captured by state variables that the researcher observes and puts into the model, the $\mathbf{x}_t$'s, but other variables are either tricky to measure properly, or their inclusion in the model would expand the size of the state space in infeasible ways. It is difficult to believe that the state variables that the researcher does not measure do not exhibit the same persistence over the time as the observed state variables that are in $\mathbf{x}_t$. As such, no serial correlation of unobservables is a strong assumption. 

Not accounting for this heterogeneity may generate significant biases in parameter estimates and in our understanding of competition in the industry. For instance, in the empirical applications in \cite{Aguirregabiria2007} and \cite{Collard-Wexler2013}, the estimation of a model without unobserved market heterogeneity implies estimates of competition effects (i.e., in this case, the effect on a firms' profit of other firms' market entry) that are strongly biased towards zero. In both applications, accounting for time-invariant unobserved market heterogeneity results in significantly larger estimates of competition effects.

\cite{Kasahara2009} study the identification of CCPs in dynamic discrete choice models -- either single-agent or games -- when the model includes time-invariant unobserved heterogeneity with finite support but with a nonparametric distribution. This unobserved heterogeneity may vary over firms, over markets, or both. They derive sufficient conditions for nonparametric identification of the CCP functions conditional on unobserved market type, and the distribution of the unobserved types. That is, if $\omega_{m}$ represents unobserved market heterogeneity, and $\{\omega^{(1)}, \omega^{(2)}, ..., \omega^{(L)}\}$ is the set of market types, they prove identification of the CCP function $P_{i}(a_{it}|\mathbf{x}_{t},\omega_{m})$ at every value in the support set of these variables, and of the probability distribution of market types, $\lambda(\omega_{m})$. The identification restrictions depend on the time-dimension of the panel data ($T$ should be large enough), and on the number of points in the support sets of the observable state variables $\mathbf{x}_{t}$ and of the unobservable $\omega_{m}$. Given these CCPs, and the other identification restrictions described above, all the structural parameters of the model are identified.

\cite{hu_shum_2012} (in a general Markov model) and \cite{hu_shum_2013} (more specifically in dynamic games) extend this result on identification of CCPs to a model where the unobservable $\omega$ can vary over time following a first order Markov process. They present identification results for different models depending on whether the decision variable and the serially correlated unobservable $\omega$ are discrete or continuous.

\medskip

\noindent \textbf{(ii) Multiple equilibria in the data}

In the context of discrete choice games of incomplete information, \cite{depaula_tang_2012} propose a test of the restriction of unique equilibrium in the
data based on the independence between players' actions conditional on observable state variables: a test $a_{it} \indep a_{jt} | \mathbf{x}_{t}$. They interpret failure of independence in terms of multiple equilibria across markets. They also show (in a similar spirit as \cite{sweeting_2009}) that the sample variation generated by multiple equilibria across markets provides identification of the sign of the parameters
that capture the strategic interactions between players (e.g., competition effects), without need of the exclusion restrictions in (ID.8). A key restriction for de Paula and Tang's identification results is that the model does not contain (payoff relevant) common knowledge unobservables.

\cite{otsu2016pooling} propose statistical tests of the null
hypothesis that panel data from a discrete dynamic game can be pooled over
multiple markets. This null hypothesis can be interpreted in terms of a
restriction of no unobserved market heterogeneity, either payoff relevant or
multiple equilibria. The asymptotics of the test is based on large $T$. The
authors apply their tests to data of the US\ Portland cement industry from \cite{Ryan2012} and reject the null hypothesis of homogeneity.

\cite{Aguirregabiria2019incomplete} study the identification of discrete games of
incomplete information when there are two forms of market heterogeneity unobservable to the researcher but common-knowledge to the players: payoff-relevant unobservables, and nonpayoff-relevant variables that determine the selection between multiple equilibria. The number of equilibria in this class of models is (generically) finite (see their Lemma 1, and also \cite{Doraszelski2010}) such that the unobservable that represents the selection of an equilibrium has discrete and finite support. Following \cite{Kasahara2009} and \cite{hu_shum_2013}, the authors assume that payoff-relevant unobserved market heterogeneity has discrete and finite support. The authors provide necessary
and sufficient conditions for the identification of all the primitives of the
model. Two types of conditions play a key role in their identification results:
independence between players' private information, and the exclusion
restriction in assumption (ID.8). This exclusion restriction identifies which part of the unobserved heterogeneity affects the payoff function and which part affects players' CCPs but not the payoff function (i.e., multiple equilibria across markets).

\medskip

\noindent \textbf{(iii) Identification without normalization restrictions}

In the context of a single-agent model of market entry and exit, \cite{Aguirregabiria2014} show that three components of a firm's profit function are not separately identified: the fixed cost of an incumbent firm, the entry cost of a new entrant, and the scrap value (or exit
cost) of an exiting firm. Empirical applications assume that one of these
three components is zero. \cite{Aguirregabiria2014} study the implications of this identification
problem and normalization restrictions on different comparative static
exercises and counterfactual public policies using the estimated model. They show that the normalization is innocuous (i.e., it does not introduce biases) for counterfactual experiments that consist of an additive change in the profit function, as long as the magnitude of the additive change is known to the researcher. They also show that the normalization restriction introduces important biases in the predictions from counterfactual experiments that change  transition probabilities of the state variables or the discount factor. The bias can modify even the sign of the estimated effects. \cite{kalouptsidi_scott_2021} extend this analysis to a general framework that covers virtually any counterfactual encountered in applied work in single-agent dynamic discrete choice models. 

For dynamic games, \cite{kalouptsidi_scott_2017_ijio} show that counterfactuals are not identified, even when analogous counterfactuals of single- agent models are identified, i.e., additive changes in players’ payoff functions. In dynamic games, a player’s best response function depends on other players’ CCPs in a similar way as it depends on exogenous transition probabilities. An additive change in player 1’s payoff function affects the CCP of this player, and in turn this change affects player 2’s best response. This second effect depends on the value of baseline payoffs. 

A possible approach to deal with this identification problem is to use partial identification. There are weak and plausible restrictions on the sign of entry cost, fixed cost, and scrap value that provide bounds on the estimation of all the structural parameters in the profit function. For instance, if the three components are always positive, then the model implies sharper lower bounds for the entry cost and the scrap value. \cite{Aguirregabiria2014} describe this approach in the context of the model of market entry/exit. \cite{kalouptsidi2020partial} present a general partial identification approach for the implementation of counterfactuals in dynamic discrete choice models.

In some industries, the acquisition of an incumbent firm is a common form of firm entry and exit: the owner of an incumbent firm sells all the firm's assets to a new entrant. In the shipping, hotel, and banking industries this is frequently the case. Sometimes, the researcher has data on firm acquisition prices, or else the firm's underlying valuation can be recovered from stock market data or other assessments of a firm's value such as accounting statements. Under some assumptions, these additional data can be used to deal with the identification problem that we describe in this subsection. This is exactly the identification strategy used by \cite{Kalouptsidi2014} for the bulk shipping industry. \cite{Aguirregabiria2014} discuss this approach and the economic restrictions on transaction costs that it requires.

\medskip

\noindent \textbf{(iv) Identification of discount factors}

The discount factor measures the strength of an agent's forward-looking behavior. For given payoff and transition probability functions, the discount factor plays a key role in the optimal decision rule in a DP problem. However, without restrictions on payoff or transition probability functions, the discount factor is not identified: see Lemma 3.3
in \cite{rust_1994_hbook}, and Proposition 2 in \cite{magnac_thesmar_2002}.

An exclusion restriction that has certain power in the identification of the discount factor is a state variable that affects the expectation of future payoffs but not current payoffs. Intuitively, if the agent's behavior does not respond to changes in this state variable, this is evidence that the agent is myopic; the stronger the observed response, the more forward-looking the agent is. This exclusion restriction has been used to identify the discount factor in different applications of single-agent dynamic models.\footnote{\cite{chevalier_goolsbee_2009} test whether textbook consumers are forward-looking. They consider that the resale price of a textbook in the second hand market is the special state variable that does not affect current utility of buying that textbook in the first-hand market but it affects future expected utility. They find strong evidence that students are forward-looking. \cite{fang_wang_2015_ier} study women's decisions to get a mammogram  . They assume that the mother's age at death -- truncated, if still alive -- affects a woman's expectations but not her current utility. \cite{bayer_mcmillan_2016} estimate a dynamic model of housing demand. Their identification of the discount factor exploits the assumption that the utility from housing depends on the current level of the neighborhood amenities, but the variables representing amenities follow a stochastic process with more than one-year memory such that lagged amenities shifts consumer expectations but not current utility. In a model of consumer stockpiling decisions, \cite{ching_osborne_2020} identify consumers’ discount factors under the assumption that current storage costs depend on the size of the package -- regardless of the level of inventory -- such that the actual inventory of the product affects expected future utilities but not current utility. \cite{degroote_verboven_2019} study households’ adoption of solar photovoltaic systems, and their response to a generous subsidy program in Belgium. Their identification of the discount factor exploits that the program mainly consisted of future production subsidies instead of upfront investment subsidies.} So far, this identification strategy has not been applied to dynamic games in IO, possibly because most applications are to decisions about large firms where it would be surprising that discount factors differ too much from the interest rates that firms pay for capital. 

These empirical applications also impose restrictions other than the exclusion restriction to identify the discount factor, such as parametric assumptions in the payoff function. Therefore, it is not obvious what is the actual identification power provided by the exclusion restriction, and how much of the identification comes from functional form restrictions. To answer this question, \cite{abbring_daljord_2020} study the identification of the discount factor in a dynamic discrete choice model under assumptions (ID.1) to (ID.5) but where the payoff function is nonparametrically specified. They show that the discount factor is partially identified, but it is not point identified. More specifically, there are multiple (but finite) values of the discount factor that are consistent with the moment conditions implied by the exclusion restriction. They also show that if the dynamic model exhibits finite dependence, as defined in \cite{Arcidiacono2011}, the identified set is smaller. In a model with $\tau$-periods finite dependence, the identified set contains at most $\tau$ values. Therefore, in a model of market entry-exit -- that has $\tau = 2$ periods finite dependence -- the identified set for the discount
factor contains only two values.

\cite{komarova_sanches_2018} study identification of the
discount factor in an scenario that is, somehow, the complement of \cite{abbring_daljord_2020}. They consider a model under assumptions (ID.1) to (ID.4) but without the exclusion restriction and with a linear in parameters specification of the payoff function. That is, the payoff function has the form $\pi_{i}(\boldsymbol{a}_{t},\mathbf{x}_{t})=h(\boldsymbol{a}_{t},\mathbf{x}_{t})^{\prime}\theta_{\pi i}$ where $h(\boldsymbol{a}_{t},\mathbf{x}_{t})$ is a vector of functions known to the researcher, and $\theta_{\pi i}$ is a vector of parameters. They show that the discount factor $\beta_{i}$ and the payoff parameters $\theta_{\pi i}$ are (generically) point identified. Their identification proof is constructive and provides a simple two-step estimator.

\medskip

\noindent \textbf{(v) Non-additive unobservables}

The Hotz-Miller inversion property -- i.e., the one-to-one mapping between CCPs and value differences -- is a key component in the proofs of identification of dynamic discrete choice structural models. This inversion property relies on the additive separability (and infinite support) of the unobservable shocks $\varepsilon_{it}$. This restriction, though convenient, is not always plausible or desirable. For instance, in a model of competition in a differentiated product market with a logit demand system, unobservable demand shocks (the so called $\xi$'s) do not enter additively in a firm's profit function. \cite{kristensen_nesheim_2015} show that additive separability is not necessary to obtain a one-to-one mapping between CCPs and value differences. The necessary and sufficient condition for the inversion property is that, for every value of $(\boldsymbol{a}_{-it},\mathbf{x}_{t})$,
the vector of $J+1$ payoffs $\{\pi_{i}(0,\boldsymbol{a}_{-it},\mathbf{x}
_{t},\varepsilon_{it})$, $\pi_{i}(1,\boldsymbol{a}_{-it},\mathbf{x}
_{t},\varepsilon_{it})$, ..., $\pi_{i}(J,\boldsymbol{a}_{-it},\mathbf{x}
_{t},\varepsilon_{it})\}$ has full support on the Euclidean space $\mathbb{R}^{J+1}$. This condition is satisfied by different models where the shocks $\varepsilon_{it}$ are not additively separable.

\medskip

\noindent \textbf{(vi) Non-parametric distribution of unobservable shocks}

Most empirical applications of dynamic discrete choice structural models have assumed a parametric specification for the distribution of the unobservables. However, it is well-known that, in discrete choice models, the misspecification of the distribution of unobservables $F_{\varepsilon}$ can generate substantial biases in the estimation of payoff parameters (e.g., \cite{horowitz_hbook_statistics_1993}). Relaxing this parametric assumption is quite relevant in this class of models. As in static discrete choice models, the shape of the distribution of the unobservables plays a key role in the effect on the choice probability of a marginal change of a state variable or a structural parameter. Furthermore, in dynamic discrete choice models, the distribution of the unobservables captures also agents' uncertainty about future payoffs and plays an important role in the magnitude and shape of the continuation values of the dynamic decision problem.

\cite{aguirregabiria_2010_jbes}, based on results by \cite{matzkin_1992_ecma}, shows the nonparametric identification of the distribution $F_{\varepsilon}$ in a binary choice dynamic structural model with finite horizon. A key condition in this identification result is the existence of an observable state variable that enters additively in the payoff function, i.e., a so-called \textit{special regressor}, using the term coined by Arthur Lewbel (see \cite{lewbel_1998} and \cite{lewbel_2000}). \cite{blevins_2014} extends this result to a more general class of dynamic models in which agents can make both discrete and continuous choices. \cite{norets_tang_2014} study partial identification when the model does not include
exclusion restrictions or "special" additive state variables, and the decision problem can have infinite horizon. They derive sharp bounds on the distribution function $F_{\varepsilon}$ and on per-period payoff functions $\pi_{i}$. \cite{buchholz_shum_2021} also consider an infinite horizon model and do not impose the restriction of a special regressor. Instead, they assume that the vector of state variables includes at least one continuous variable and the payoff function is linear in parameters. They establish the nonparametric point identification of $F_{\varepsilon}$. \cite{chen_2017_et} obtains point identification results under the restriction that a subset of the state variables affects only the current payoff function but not agents' beliefs about future utilities.

\medskip

\noindent \textbf{(vii) Non-equilibrium beliefs}

Models of belief formation that depart from Bayesian Nash Equilibria can be difficult since both payoffs and beliefs need to be identified.  \cite{aguirregabiria_magesan_2020} study the identification of biased beliefs in dynamic games. Their model allows payoff and belief functions to vary over time in an unrestricted way. First, the authors show that the exclusion restriction in (ID.8) provides testable non-parametric restrictions of the null hypothesis of equilibrium beliefs in dynamic games with either finite or infinite horizon. Second, they prove that this exclusion restriction, together with consistent estimates of beliefs at two points in the support of the variable involved in the exclusion restriction, is sufficient for non-parametric point-identification of players' belief functions and payoff functions. They apply these results to a dynamic game of competition in number of stores between McDonalds and Burger King in UK. They find significant evidence of biased beliefs by Burger King. Imposing the restriction of unbiased beliefs generates a substantial attenuation bias in the estimate of competition effects.

\cite{an_hu_2021_joe} study the identification of dynamic discrete choice models without assuming rational expectations. For finite horizon models, their key identification restriction is that payoff function and transition probabilities are time invariant. Under this restriction, all the variation over time in the CCP functions should be attributed to the proximity to the terminal period. This implies a recursive relationship between CCPs at two consecutive periods. The authors show that this relationship provides identification of subjective beliefs. For the identification of subjective beliefs in infinite horizon models, they impose the stronger restriction that for one of the state variables, agents' have rational expectations.

\subsubsection{Identification of mixed continuous-discrete choice models \label{sec:identification mixed}}

\cite{blevins_2014} studies the identification of dynamic structural models with a discrete decision, $d_{it} \in \{0,1,...,J\}$, and a continuous decision, $c_{it} \in  \mathbb{R}$. For instance, in the model of \cite{sweeting2015selective}, firms choose to enter auctions based on the valuation they have. Thus, the entry and bidding problem are linked. More generally, we often have some information on firms outcomes post-entry, and this creates a selection problem of what $\epsilon$'s are observed in the market. In \cite{blevins_2014}, the model includes two types of unobservable state variables: unobservables associated with the discrete choice, $(\varepsilon_{it}(d): d=0,1,...,J) \in \mathbb{R}^{J+1}$; and an unobservable associated with the continuous choice, $\eta_{it} \in \mathbb{R}$. For instance, consider a firm manager that every period decides whether to operate a production plant ($d_{it}=1$) or to keep it idle ($d_{it}=0$), and conditional on operating, the manager chooses the amount of output to produce, $c_{it}$. Unobservable $\varepsilon_{it}(1)-\varepsilon_{it}(0)$ represents a shock in the fixed cost of starting up the plant. Unobservable $\eta_{it}$ is a shock in the marginal cost of output.

\cite{blevins_2014}'s model maintains all the assumptions (ID.1) to (ID.8) presented above. More specifically, the unobservables $\varepsilon_{it}$ and $\eta_{it}$ satisfy the conditional independence assumption (ID.4), and the discrete unobservables are additively separable in the payoff function, as in assumption (ID.2). Firm $i$'s profit function is $\pi_{i}(d_{it},c_{it},\mathbf{x}_{t},\eta_{it})+\varepsilon_{it}(d_{it})$. Blevins includes three additional assumptions related to the continuous-choice part of the model. First, shocks $\varepsilon_{it}$ and $\eta_{it}$ are revealed to the firm sequentially within each period. The firm observes first the discrete-choice shocks $\varepsilon_{it}$ and makes its discrete choice at period $t$. Then, after making its discrete choice, the continuous-choice shock $\eta_{it}$ is revealed and the firm makes its continuous decision. A second assumption is that conditional on $\mathbf{x}_{t}$, the unobservables $\varepsilon_{it}$ and $\eta_{it}$ are independently distributed. These are arguably  strong assumptions that may not hold in some applications. However, these two assumptions facilitate substantially the analysis of this model. In particular, they imply that the conditional discrete-choice value functions at the beginning of period $t$ (conditional on optimal continuous decision and integrated over the distribution of $\eta_{it}$) have the standard structure in the literature where all the unobservables are additively separable. Finally, he assumes that the marginal profit function $\partial \pi_{i}/\partial c_{it}$ is strictly monotonic in $c_{it}$ and $\eta_{it}$. This is a standard condition in continuous decisions models.

Under these conditions, \cite{blevins_2014} proves the nonparametric identification of the profit function $\pi_{i}(.)$. Note that this function is nonparametric in all its arguments, including the the unobservable $\eta_{it}$. He presents identification results both when the distribution functions $F_{\varepsilon}$ and $F_{\eta}$ are known to the researcher and when these functions are nonparametrically specified.

\subsection{Estimation methods\label{Section: Estimation methods}}

The primitives of the model, $\{\pi_{i}, \beta_{i}, F_{x}, F_{\varepsilon}: i \in  \mathcal{I}\}$, can be described in terms of a vector of structural parameters $\boldsymbol{\theta}$ that is unknown to the researcher. In this section, we describe methods for the estimation of $\boldsymbol{\theta}$, as well as different econometric issues. It is convenient to distinguish three components in the vector of structural parameters: $\boldsymbol{\theta} = (\boldsymbol{\theta}_{\pi}, \boldsymbol{\theta}_{f},\boldsymbol{\beta})$, where $\boldsymbol{\theta}_{\pi}$ represents the parameters in payoff functions and in the distribuion of the unobservables (if any), $\boldsymbol{\theta}_{f}$ contains the parameters in the transition probabilities of state variables, and $\boldsymbol{\beta}$ is the vector of discount factors.

\subsubsection{Full solution methods \label{sec:full solution methods}}

\medskip

\noindent \textbf{(i) Nested fixed point algorithm with equilibrium uniqueness}

Rust's nested fixed point algorithm (NFXP; \cite{rust1987optimal}, \cite{rust_1994_hbook}) was a fundamental development in the estimation of dynamic structural models. NFXP is a gradient iterative search method to obtain the maximum likelihood estimator of the structural parameters. It was originally proposed for single-agent models, but it has been applied also to the estimation of games with unique equilibrium (e.g., \cite{seim_2006}, \cite{abbring2010}, and \cite{Igami2017}). The own concept of a likelihood function -- and not a correspondence -- seems to imply that the model has only one equilibrium for each value of the structural parameters. The condition of equilibrium uniqueness has been common in applications of NFXP to games. However, as we describe below, a conceptually simple modification of this algorithm can be applied to estimate dynamic games with multiple equilibria. The problem is not the conceptual definition of this algorithm but the computational cost of implementing it.

For the moment, suppose that the dynamic game has a unique MPBNE for every value of the structural parameters.\footnote{As we have mentioned in section \ref{sec:model multiple equilibria} above, equilibrium uniqueness in this class of dynamic games requires strong restrictions.} Let $\{ P_{i}(a_{i}|\mathbf{x},\boldsymbol{\theta}):i \in \mathcal{I} \}$ be the equilibrium CCPs associated with a value of the structural parameters $\boldsymbol{\theta}$. Under
assumptions (ID.1) to (ID.4) the full log-likelihood function of the data is
$\ell(\boldsymbol{\theta}) = {\textstyle\sum\nolimits_{m=1}^{M}} \ell_{m}(\boldsymbol{\theta})$, where $\ell_{m}(\boldsymbol{\theta})$ is the contribution of market $m$ and has the following form:
\begin{equation}
    \ell_{m}(\boldsymbol{\theta}) =
    {\displaystyle \sum\limits_{i=1}^{N}}
    {\displaystyle \sum\limits_{t=1}^{T}}
    \log P_{i}(a_{imt}|\mathbf{x}_{mt},
    \boldsymbol{\theta})
    + \log f_{x}(\mathbf{x}_{m,t+1}|\boldsymbol{a}_{mt},\mathbf{x}_{mt},\boldsymbol{\theta}_{f})+\log\Pr(\mathbf{x}_{m1}|\boldsymbol{\theta})
\label{Rust log-likelihood}
\end{equation}
where $f_{x}$ is the transition density function, and $\log\Pr(\mathbf{x}_{m1}|\boldsymbol{\theta})$ is the contribution of the initial conditions to the likelihood, e.g., the observed market structure at the first sample period. Most applications imposing the restriction of no serially correlated unobservables follow a conditional likelihood approach that ignores the term
$\log\Pr(\mathbf{x}_{m1}|\boldsymbol{\theta})$. Though this approach is consistent as long
as there are not serially correlated unobservables, it implies a loss of
efficiency that can be important in stationary dynamic games.\footnote{In applications with serially correlated unobservables, accounting for the endogeneity of the initial conditions is key
to generate consistent estimators. We describe this in section \ref{sec:unobserved heterogeneity} below.} Under the conditional independence assumption (ID.4), the subvector of structural parameters $\boldsymbol{\theta}_{f}$ can be estimated separately from the rest of the parameters without solving for an equilibrium of the game. To reduce the computational cost in the estimation of the model, most applications use a sequential approach where the parameters $\boldsymbol{\theta}_{f}$ are
estimated in a first step based on the partial likelihood ${\textstyle\sum\nolimits_{m,t}}
\log f_{x}(\mathbf{x}_{m,t+1}|\boldsymbol{a}_{mt},\mathbf{x}_{mt},\boldsymbol{\theta}_{f})$, and in a second step the rest of the parameters are estimated using the conditional partial likelihood $\ell^{c}(\boldsymbol{\theta}_{\pi},\boldsymbol{\beta})=$ ${\textstyle\sum\nolimits_{i,m,t}}
\log P_{i}(a_{imt}|\mathbf{x}_{mt},\boldsymbol{\theta}_{\pi},\boldsymbol{\beta},\widehat{\boldsymbol{\theta}}_{f})$.

The NFXP combines a \cite{berndt1974estimation} (BHHH hereafter) method for the outer algorithm, that searches for a root of the likelihood equations, with an solution algorithm (inner algorithm) that solves for the MPBNE of the game for each trial value of the structural parameters. The algorithm is initialized with an arbitrary vector of structural parameters, say $\widehat{\boldsymbol{\theta}}_{0}$. A BHHH iteration is defined
as:
\begin{equation}
    \widehat{\boldsymbol{\theta}}_{k+1}
    = 
    \widehat{\boldsymbol{\theta}}_{k}
    + \left(
        {\displaystyle\sum\nolimits_{m=1}^{M}}
        \frac{\partial\ell_{m}
        (\widehat{\boldsymbol{\theta}}_{k})}
        {\partial \boldsymbol{\theta}}
        \frac{\partial
        \ell_{m}
        (\widehat{\boldsymbol{\theta}}_{k})}
        {\partial \boldsymbol{\theta}^{\prime}}
    \right)  ^{-1}
    \left(
        {\displaystyle\sum\nolimits_{m=1}^{M}}
        \frac{\partial \ell_{m}
        (\widehat{\boldsymbol{\theta}}_{k})}
        {\partial \boldsymbol{\theta}}
    \right)
\label{BHHH NFXP}
\end{equation}
The score vector $\partial \ell_{m}(\widehat{\boldsymbol{\theta}}_{k})/ \partial \boldsymbol{\theta}$ depends on $\partial \log P_{i}(a_{imt}|\mathbf{x}_{mt},\widehat{\boldsymbol{\theta}}_{k})/\partial \boldsymbol{\theta}$. To obtain these derivatives, the inner algorithm of NFXP solves for the equilibrium CCPs given $\widehat{\boldsymbol{\theta}}_{k}$. This solution algorithm can be based on value function iterations, or policy function iterations, or a hybrid of the two. As any other gradient method, the NFXP algorithm returns a solution to the likelihood equations.\footnote{In general, the likelihood function of this class of models is not globally concave. Therefore, some global search is necessary to check whether the root of the likelihood equations that has been found is actually the global maximum and not just a local optimum.}

There is a long list of applications in IO which have used the NFXP algorithm to estimate single-agent Rust models. For instance, the machine replacement model used in \cite{rust1987optimal}, \cite{das_1992}, \cite{rust_rothwell_1995}. The list of applications for dynamic games is shorter but includes important recent contributions such as \cite{Igami2017} and \cite{IgamiConsolidation}.

\medskip

\noindent \textbf{(ii) Maximum likelihood estimation with multiple equilibria}

A modified version of NFXP can be applied to obtain the maximum likelihood estimator (MLE) in games with multiple equilibria. To define the MLE in a model with multiple equilibria, it is convenient to define an \textit{extended} or \textit{pseudo} likelihood function. For arbitrary values of the vector of structural parameters $\boldsymbol{\theta}$ and firms' CCPs $\mathbf{P}$, we define the following likelihood function of observed players' actions conditional on observed state variables:
\begin{equation}
    Q(\boldsymbol{\theta},\mathbf{P}) =
    {\displaystyle\sum\limits_{m=1}^{M}}
    {\displaystyle\sum\limits_{i=1}^{N}}
    {\displaystyle\sum\limits_{t=1}^{T}}
    \log\Psi_{i}(a_{imt}
    \text{ } | \text{ }
    \mathbf{x}_{mt}, \boldsymbol{\theta},
    \mathbf{P})
\label{pseudo likelihood}
\end{equation}
where $\Psi_{i}$ is the \textit{best response probability function} that we have defined in equation (\ref{best reesponse probability}).
We call $Q(\boldsymbol{\theta},\mathbf{P})$\ a pseudo likelihood function because players'
CCPs in $\mathbf{P}$\ are arbitrary and do not represent the equilibrium
probabilities associated with $\boldsymbol{\theta}$ implied by the model. An implication of using arbitrary CCPs, instead of equilibrium CCPs, is that likelihood
$Q$ is a function and not a correspondence.

The MLE is defined as the pair $(\widehat{\boldsymbol{\theta}}_{MLE},\widehat{\mathbf{P}}_{MLE})$ that maximizes the pseudo likelihood subject to the constraint that the CCPs in $\widehat{\mathbf{P}}_{MLE}$ are
equilibrium strategies associated with $\widehat{\boldsymbol{\theta}}_{MLE}$. This is a constrained MLE can be defined as the solution of the following Lagrangian
problem:
\begin{equation}
    (\widehat{\boldsymbol{\theta}}_{MLE},
    \widehat{\mathbf{P}}_{MLE},
    \widehat{\boldsymbol{\lambda}}_{MLE}) 
    = 
    \arg\max\limits_{
    (\boldsymbol{\theta},
    \mathbf{P}, \boldsymbol{\lambda})}
    \text{ } 
    Q(\boldsymbol{\theta},\mathbf{P})
    +\boldsymbol{\lambda}^{\prime}
    \left[  
        \mathbf{P} - 
        \Psi(\boldsymbol{\theta},\mathbf{P})
    \right]
\label{ML estimator as constrained}
\end{equation}
where $\boldsymbol{\lambda}$ is the vector of Lagrange multipliers, and
$\boldsymbol{\lambda}$, $\mathbf{P}$, and $\Psi(\boldsymbol{\theta},\mathbf{P})$ are vectors where each element corresponds to a value of $(i,a_{imt},\mathbf{x}_{mt})$. This constrained MLE satisfies the standard
regularity conditions for consistency, asymptotic normality, and efficiency of
maximum likelihood estimation.

In principle, this constrained MLE can be computed using Newton or
Quasi-Newton methods. The first order conditions of this problem imply the
following Lagrangian equations:
\begin{equation}
    \left\{
    \begin{array}
    [c]{rcc}
        \widehat{\mathbf{P}}_{MLE} - \Psi(\widehat{\boldsymbol{\theta}}_{MLE}, \widehat{\mathbf{P}}_{MLE}) 
        & = & 
        \mathbf{0} 
        \\
        \triangledown_{\boldsymbol{\theta}}
        Q(\widehat{\boldsymbol{\theta}}_{MLE}
        ,\widehat{\mathbf{P}}_{MLE})
        - \widehat{\boldsymbol{\lambda}}_{MLE}^{\prime} \text{ }
        \triangledown_{\boldsymbol{\theta}}
        \Psi(\widehat{\boldsymbol{\theta}}_{MLE}, \widehat{\mathbf{P}}_{MLE})
        & = & 
        \mathbf{0}
        \\
        \triangledown_{\mathbf{P}}
        Q(\widehat{\boldsymbol{\theta}}_{MLE},\widehat{\mathbf{P}}_{MLE})
        -\widehat{\boldsymbol{\lambda}}_{MLE}^{\prime} \text{ }
        \triangledown_{\mathbf{P}}
        \Psi(\widehat{\boldsymbol{\theta}}_{MLE}, \widehat{\mathbf{P}}_{MLE})
        & = &
        \mathbf{0}
    \end{array}
    \right.  
\label{Lagrange equations}
\end{equation}
A Newton method can be used to obtain a root of this system of Lagrangian equations. However, a key computational problem is the very high dimensionality of this system of equations. In the empirical applications of dynamic oligopoly games, the
vector of probabilities $\mathbf{P}$\ -- and the vector of Lagrange multipliers $\boldsymbol{\lambda}$ -- includes thousands, millions, or even more elements. In particular, the computationally most intensive part of this algorithm is in the calculation of the Jacobian matrix $\triangledown_{\mathbf{P}}
\Psi(\widehat{\boldsymbol{\theta}},\widehat{\mathbf{P}})$. In dynamic games, in general, this is not a sparse matrix, and can contain billions or trillions of elements. Furthermore, the evaluation of the best response mapping
$\Psi(\boldsymbol{\theta},\mathbf{P})$ for a new value of $\mathbf{P}$ requires solving for
a valuation operator and solving a system of equations with the same dimension as $\mathbf{P}$. Therefore, if $L = N |\mathcal{A}| |\mathcal{X}|$ is the dimension of the vector $\mathbf{P}$, the evaluation of the Jacobian matrix $\triangledown_{\mathbf{P}} \Psi(\widehat{\boldsymbol{\theta}},\widehat{\mathbf{P}})$ requires solving of the order of $L^{2}$ systems of linear equation with dimension $L$. Furthermore, this Jacobian matrix needs to be recomputed at each iteration of the Newton's method. Given the value of $L$ in empirical applications, this approach is impractical in most empirical applications. \cite{Su2012} have proposed using an MPEC algorithm, which is a general purpose algorithm for the numerical solution of constrained optimization problems. However, MPEC also requires the repeated computation of the high dimensional and non-sparse Jacobian matrix $\triangledown_{\mathbf{P}}\Psi(\widehat{\boldsymbol{\theta}},\widehat{\mathbf{P}})$. Due to serious computational issues, there are no empirical applications of dynamic games with multiple equilibria that compute the MLE, with either the NFXP or MPEC algorithms.

\medskip

\noindent \textbf{(iii) Nested pseudo maximum likelihood estimation}

Motivated by the computational challenges of implementing the MLE in dynamic structural models (and by limitations of the two-step methods that we describe below), \cite{aguirregabiria_mira_2002} (for single-agent models) and \cite{Aguirregabiria2007} (for dynamic games) propose an alternative estimation method that imposes the equilibrium restrictions but does requires neither repeatedly solving for equilibrium CCPs for each trial value of the structural parameters (as in the NFXP algorithm), nor computing Jacobian matrices $\triangledown_{\mathbf{P}}\Psi(\widehat{
\boldsymbol{\theta}},\widehat{\mathbf{P}})$ (as in the NFXP and MPEC algorithms). They denote their method the \textit{Nested pseudo likelihood} (NPL  hereafter) estimator. 

In the NPL method, the analogue to a root of the likelihood equations is a \textit{NPL root} (or \textit{NPL fixed point}). A NPL root is defined as a vector of structural parameters and a vector of CCPs, $(\widehat{\boldsymbol{\theta}}_{NPL},\widehat{\mathbf{P}}_{NPL})$, that satisfy two conditions: (1) given $\widehat{\mathbf{P}}_{NPL}$, the vector of structural parameters maximizes the pseudo likelihood function, $\widehat{\boldsymbol{\theta}}_{NPL}=$
$\arg\max\nolimits_{\boldsymbol{\theta}}$ $Q(\boldsymbol{\theta},\widehat{\mathbf{P}}_{NPL})$; and (2) given $\widehat{\boldsymbol{\theta}}_{NPL}$, the vector of CCPs satisfies the equilibrium
restrictions, $\widehat{\mathbf{P}}_{NPL}=\Psi(\widehat{\boldsymbol{\theta}}_{NPL},\widehat{\mathbf{P}}_{NPL})$. Define the \textit{NPL mapping} $\varphi: [0,1]^{N |\mathcal{A}| |\mathcal{X}|} \rightarrow [0,1]^{N |\mathcal{A}| |\mathcal{X}|}$  as $\varphi(\mathbf{P}) \equiv$ $\Psi(\widehat{\theta}(\mathbf{P)},\mathbf{P})$ where $\widehat{\theta}(\mathbf{P)}$ represents the value of $\boldsymbol{\theta}$ that maximizes $Q$ given $\mathbf{P}$. Using this mapping, we can define a NPL root as a fixed point of the NPL mapping:
\begin{equation}
    \left\{
    \begin{array}
    [c]{rcc}
    \widehat{\mathbf{P}}_{NPL}-\varphi(\widehat{\mathbf{P}}_{NPL}) 
    & = & 
    \mathbf{0}
    \\
    \widehat{\boldsymbol{\theta}}_{NPL} 
    - \widehat{\theta}(\widehat{\mathbf{P}}_{NPL})
    & = & 
    \mathbf{0}.
    \end{array}
    \right.  
\label{NPL first order conditions}
\end{equation}
The NPL estimator is defined as the NPL root with the largest value of the pseudo likelihood. The NPL estimator is consistent and asymptotically normal under the same regularity conditions as the MLE (Proposition 2 in \cite{Aguirregabiria2007}). For dynamic games, the NPL estimator has larger asymptotic variance than the MLE. In single-agent dynamic models, the two estimators are asymptotically equivalent (Proposition 4 in \cite{aguirregabiria_mira_2002}).

To compute a NPL root, Aguirregabiria and Mira propose a simple algorithm that consists of successive iterations in the NPL mapping $\varphi$. They denote it \textit{NPL fixed point algorithm}. Starting with an initial $\mathbf{P}_{0}$, at iteration $k\geq 1$
the vector of CCPs is updated using $\mathbf{P}_{k}=\varphi(\mathbf{P}_{k-1})$.
This updating or fixed point iteration involves two calculations: (1) obtaining the pseudo ML estimator $\widehat{\boldsymbol{\theta}}_{k}=\widehat{\theta
}(\mathbf{P}_{k-1}\mathbf{)}$ by solving in $\boldsymbol{\theta}$ the system
$\triangledown_{\boldsymbol{\theta}}Q(\boldsymbol{\theta},\mathbf{P}_{k-1})=0$; and (2)
given $\widehat{\boldsymbol{\theta}}_{k}$ and $\mathbf{P}_{k-1}$, obtain players' best
response CCPs if the other players behave according to $\mathbf{P}_{k-1}$ and the structural parameters are $\widehat{\boldsymbol{\theta}}_{k}$, i.e., $\mathbf{P}
_{k}=\Psi(\widehat{\boldsymbol{\theta}}_{k},\mathbf{P}_{k-1})$. Computation (1) is very simple in most applications, as it is equivalent to obtaining the MLE in a static single-agent discrete choice model. The main computational task in (2) comes from the calculation of present values that is equivalent to solving once a system of linear equations with the same dimension as $\mathbf{P}$. Therefore, one iteration of this algorithm is several orders of magnitude cheaper than one Newton or MPEC iteration for the solution of the MLE. This is because an iteration in the NPL mapping does not involve solving for an equilibrium (as in NFXP) or calculating the non-sparse Jacobian matrix $\triangledown_{\mathbf{P}}\Psi(\theta,\mathbf{P)}$ (as in MPEC).

The NPL estimator has been used in a good number of empirical applications in IO, for single-agent dynamic models (\cite{copeland_monnet_2009}, \cite{depinto_nelson_2009}, \cite{tomlin2014exchange}, \cite{aguirregabiria_alonso_2014}, \cite{huang_2015_ms}), dynamic games (\cite{Sweeting2013}, \cite{aguirregabiria2012airline}, \cite{Collard-Wexler2013},
\cite{kano2013menu}, \cite{Huang2014}, \cite{lin2015quality}, \cite{gayle_xie_2018}), static games (\cite{ellickson_misra_2008},
\cite{han_hong_2011}) and networks (\cite{lin_xu_2017}, \cite{liu_zhou_2017}).

An important limitation of the NPL is that, in games, the mapping $\varphi(\mathbf{P})$ is not a contraction, so that fixed point iterations do not guarantee convergence. In fact, mapping $\varphi$ may have multiple fixed points, and the fixed point algorithm may converge to a solution that is not the consistent NPL root. This issue has been pointed out and illustrated with numerical examples by \cite{Pesendorfer2010}, \cite{egesdal_lai_2015}, \cite{Kasahara2012}, and \cite{aguirregabiria_marcoux_2021_qe}. It has also motivated different authors to propose algorithms to compute the NPL estimator that share the low cost per iteration of fixed point NPL iterations but that have better convergence properties when the NPL mapping is not a contraction.

One way to resolve issues of convergence with the NPL is to modify the update rule for $\mathbf{P}$.  \cite{Kasahara2012} propose a relaxation method that modifies the NPL mapping so that $\mathbf{P}$ updated more slowly, where the speed of update is controlled by a tuning parameter $\alpha$. However, as shown in the numerical experiments in \cite{egesdal_lai_2015} and \cite{aguirregabiria_marcoux_2021_qe}, this approach comes at the cost of slower convergence. \cite{aguirregabiria_marcoux_2021_qe} propose instead to use a spectral algorithm. A key feature of this approach is that the stepsize is updated at each iteration, and no derivatives need to be computed. They apply this spectral algorithm to multiple data generating processes from dynamic games, including those considered by \cite{Pesendorfer2010} and  \cite{egesdal_lai_2015}, and find that it converges to the NPL estimator for every Monte Carlo simulated sample.

\subsubsection{Two-step CCP methods \label{sec: two step methods}}

To avoid the large computational cost of full-solution methods, simpler two-step methods have been proposed. \cite{hotz1993conditional} was a seminal
contribution on this class of methods. In a single-agent model, under assumptions (ID.1) to (ID.4), they show that the conditional
choice value function (as defined in equation (\ref{eq: conditional choice value}) above) can be written as known functions of CCPs, transition probabilities, and one-period payoffs $\pi_{i}$. If the flow payoff function is linear in parameters, $\pi_{i}(a_{it},\mathbf{x}_{t}\mathbf{)=}$
$h(a_{it},\mathbf{x}_{t})$ $\boldsymbol{\theta}_{\pi,i}$, this representation is particularly simple:
\begin{equation}
    v_{i}(a_{it},\mathbf{x}_{t}) 
    \text{ } = \text{ } \widetilde{h}_{i}^{\mathbf{P}}(a_{it},\mathbf{x}_{t})
    \text{ }
    \boldsymbol{\theta}_{\pi,i} 
    \text{ } + \text{ } 
    \widetilde{e}_{i}^{\mathbf{P}}
    (a_{it},\mathbf{x}_{t})
\label{HM representation values}
\end{equation}
where $\widetilde{h}_{i}^{\mathbf{P}}(a_{it},\mathbf{x}_{t})$ and $\widetilde{e}_{i}^{\mathbf{P}}(a_{it},\mathbf{x}_{t})$ are the expected present values of their untilded counterparts $h$ and $e$:
\begin{equation}
    \begin{array}
    [c]{rcc}
    \widetilde{h}_{i}^{\mathbf{P}}(a_{it},\mathbf{x}_{t}) 
    & = & 
    \mathbb{E}
    \left(
        {\displaystyle\sum_{j=0}^{\infty}}
        \beta_{i}^{j}
        \text{ } 
        h(a_{it+j},\mathbf{x}_{t+j}) 
        \text{ } | \text{ } 
        a_{it},\mathbf{x}_{t}
    \right)
    \\
    \\
    \widetilde{e}_{i}^{\mathbf{P}}(a_{it},\mathbf{x}_{t})
    & = & 
        \mathbb{E}
    \left(
        {\displaystyle\sum_{j=0}^{\infty}}
        \beta_{i}^{j}
        \text{ } 
        e_{i}^{\mathbf{P}}(a_{i,t+j},\mathbf{x}_{t+j}) 
        \text{ } | \text{ } 
        a_{it},\mathbf{x}_{t}
    \right)
    \end{array}
\label{eq: present values}
\end{equation}
where future actions are drawn from the CCPs in vector $\mathbf{P}$. Function $e_{i}^{\mathbf{P}}(j,\mathbf{x}_{t})$ represents the conditional expectation $\mathbb{E}(\varepsilon_{i}(j) | \mathbf{x}_{t}, a_{it}=j)$ and it is a known function of the CCPs at $\mathbf{x}_{t}$; i.e., the expectation of shocks conditional on firms behaving optimally with conditional choice probabilities $\mathbf{P}$. For instance, when $\varepsilon$'s are i.i.d. extreme value type I, we have that $e_{i}^{\mathbf{P}}(j,\mathbf{x}_{t}) = \gamma - \log P_{i}(j|\mathbf{x}_{t})$ where $\gamma$ is Euler's constant. The present values in equation (\ref{eq: present values}) can be represented as known functions of CCPs, transition probabilities, and discount factor. More
precisely, 
\begin{equation}
    \begin{array}
    [c]{rll}
    \widetilde{h}_{i}^{\mathbf{P}}(a_{it},\mathbf{x}_{t}) 
    & = & 
    h(a_{it},\mathbf{x}_{t})
    \text{ } + \text{ }
    \beta_{i} 
    \text{ }
    {\displaystyle\sum_{\mathbf{x}_{t+1}}}
    f_{x}(\mathbf{x}_{t+1}|a_{it},\mathbf{x}_{t})
    \text{ }
    W_{h,i}^{\mathbf{P}}(\mathbf{x}_{t+1})
    \\
    \\
    \widetilde{e}_{i}^{\mathbf{P}}(a_{it},\mathbf{x}_{t})
    & = & 
    \beta_{i}
    \text{ }
    {\displaystyle\sum_{\mathbf{x}_{t+1}}}
    f_{x}(\mathbf{x}_{t+1}|a_{it},\mathbf{x}_{t})
    \text{ }
    W_{e,i}^{\mathbf{P}}(\mathbf{x}_{t+1})
    \end{array}
\label{eq: present values 2}
\end{equation}
and the matrix of values $\mathbf{W}_{i}^{\mathbf{P}} = \{ [W_{h,i}^{\mathbf{P}}(\mathbf{x}_{t}),W_{e,i}^{\mathbf{P}}(\mathbf{x}_{t})]:\mathbf{x}_{t}\in\mathcal{X} \}$ can be
obtained solving the following systems of linear equations:%
\begin{equation}
    \mathbf{W}_{i}^{\mathbf{P}} 
    \text{ } =  \text{ }
    {\displaystyle\sum_{a_{t}=0}^{J}}
    \mathbf{P}_{i}(a_{i}) \circ 
    \left(  
        \text{ }
        \left[
            \mathbf{h}_{i}(a_{i}),
            \text{ }
            \mathbf{e}_{i}^{P}(a_{i})
        \right]  
        + \beta_{i}
        \text{ }
        \mathbf{F}_{x,i}(a_{i})
        \text{ }
        \mathbf{W}_{i}^{\mathbf{P}}
        \text{ }
    \right)
\label{W present values}
\end{equation}
where $\circ$ is the element-by-element or Hadamard product; $\mathbf{P}_{i}(a_{i})$ is the column vector of choice probabilities $(P_{i}(a_{i}|\mathbf{x}):\mathbf{x}\in\mathcal{X})$; $\mathbf{F}_{x,i}(a_{i})$ is the matrix of transition probabilities of $\mathbf{x}$ given choice $a_{i}$; $\mathbf{h}_{i}(a_{i})$ is the matrix $(h_{i}(a_{i},\mathbf{x}):\mathbf{x} \in \mathcal{X})$; and $\mathbf{e}_{i}^{P}(a_{i})$ is the vector $(e_{i}^{P}(a_{i},\mathbf{x}):\mathbf{x}\in\mathcal{X})$. One can compute $\mathbf{W}_{i}^{\mathbf{P}}$ using efficient methods for solving systems of linear equations, and exploit the possible sparsity of the transition matrices
$\mathbf{F}_{x,i}(a_{i})$. Solving this system of linear equations has a complexity of at most (worst case) $O(|\mathcal{X}|^{3})$ where $|\mathcal{X}|$ is the dimension of the state space. This complexity is of the same order as solving the DP problem once. From the point of view of estimation, the main advantage of this
representation is that -- combined with initial reduced form estimates of the CCPs -- can be used to estimate the structural parameters without having to solve repeatedly the DP problem.

\cite{hotz1993conditional} also show that for DP problems with an absorbing state, so called \textit{optimal stopping problems}, the representation of the conditional choice value functions in equation (\ref{HM representation values}) becomes extremely simple. To illustrate this assumption, the application in  \cite{hotz1993conditional} is the choice to have a vasectomy when families are choosing the number of children to have. Likewise, in many of the market entry and exit models considered in this chapter, exit is a permanent decision (e.g., \cite{Collard-Wexler2013,Dunne2013}).  In these models, $v_{i}(a_{it},\mathbf{x}_{t}\mathbf{)}$ can be represented using CCPs and transitions at only periods $t$ and $t+1$. More specifically, if $a_{it}=0$ represents the stopping decision and $a_{it}=j$ is any other choice alternative, we have that:
\begin{equation}
    v_{i}(j,\mathbf{x}_{t})-v_{i}(0,\mathbf{x}_{t}) 
    =
    \pi_{i}(j,\mathbf{x}_{t}) -
    \pi_{i}(0,\mathbf{x}_{t}) +
    \beta_{i} \text{ }
    \mathbb{E}_{t}
    \left[ 
        \pi_{i}(j,\mathbf{x}_{t+1}) -
        \pi_{i}(0,\mathbf{x}_{t+1}) +
        e_{i}^{P}(j,\mathbf{x}_{t+1})
    \right].
\label{oprimal stopping}
\end{equation}
It is clear that the representation in equation (\ref{oprimal stopping}) is computationally much simpler than the general representation in equation (\ref{HM representation values}): a complexity of $O(|\mathcal{X}|)$ instead of $O(|\mathcal{X}|^{3})$. \cite{Arcidiacono2011} generalize this result to DP models with \textit{finite dependence structure}, which is a substantially broader class than optimal stopping problems. We describe this extension in section \ref{sec:finite dependence} below.

Given either the general representation in (\ref{HM representation values}) or the finite dependence representation in (\ref{oprimal stopping}), the 
pseudo likelihood function $Q(\boldsymbol{\theta},\mathbf{P})$ has practically the same structure as in a static or reduced form discrete choice model. That is, the best response probabilities $\Psi_{i}(a_{imt}|\mathbf{x}_{mt},\mathbf{\theta,P)}$ that enter in the pseudo likelihood $Q(\boldsymbol{\theta},\mathbf{P})$ can be seen as the choice probabilities in a standard random utility model:
\begin{equation}
    \Psi_{i}(a_{imt}|\mathbf{x}_{mt},\mathbf{\theta,P})
    \text{ } = \text{ }
    \Pr 
    \left(  
        a_{imt} = 
        \arg\max_{j}
        \left\{ 
            \widetilde{h}_{i}^{\mathbf{P}}
            (j,\mathbf{x}_{mt})
            \text{ }
            \boldsymbol{\theta}_{i} +
            \widetilde{e}_{i}^{\mathbf{P}}
            (j,\mathbf{x}_{mt}) +
            \varepsilon_{it}(j)
        \right\}  
    \right).  
\label{best response as RUM}
\end{equation}
Given $\widetilde{h}_{i}^{\mathbf{P}}(.,\mathbf{x}_{mt})$ and $\widetilde{e}_{i}^{\mathbf{P}}(.,\mathbf{x}_{mt})$ and a parametric specification for the distribution of $\varepsilon$ (e.g., logit, probit), the vector of parameters $\boldsymbol{\theta}_{i}$ can be estimated as in a standard logit or probit model.

The method proceeds in two steps. Let $\mathbf{P}^{0}$\ be the vector with the population values of the CCPs. Under assumptions (ID.1) to (ID.4), these CCPs can be estimated consistently using standard nonparametric methods. Let $\widehat{\mathbf{P}}^{0}$ be a consistent nonparametric estimator of $\mathbf{P}^{0}$. The two-step estimator of $\boldsymbol{\theta}$\ is defined as:
\begin{equation}
    \widehat{\boldsymbol{\theta}}_{2S}
    \text{ } = \text{ }
    \arg\max_{\boldsymbol{\theta}}
    \text{ }
    Q(\boldsymbol{\theta},\widehat{\mathbf{P}}^{0})
\label{eq: two step estimator}.
\end{equation}
Under standard regularity conditions, this
two-step estimator is root-M consistent and asymptotically normal (see Proposition 2 in \cite{hotz1993conditional}, and more generally \cite{newey_1994_asymptotic}). \cite{aguirregabiria_mira_2002} show that, in single-agent models, this two-step estimator based on the maximization of the pseudo likelihood function $Q$ is asymptotically effecient due to the \textit{zero Jacobian property} in this class of models.\footnote{In single-agent dynamic discrete choice models, the Jacobian matrix $\triangledown_{\mathbf{P}} \Psi(\boldsymbol{\theta},\mathbf{P})$ evaluated at a fixed point $\mathbf{P}$ is zero (Proposition 2 in \cite{aguirregabiria_mira_2002}). Therefore, at the population parameters $(\boldsymbol{\theta}^{0},\mathbf{P}^{0})$ we have that $\triangledown_{\mathbf{P}} \Psi(\boldsymbol{\theta}^{0},\mathbf{P}^{0}) = \mathbf{0}$, and this implies asymptotic independence between the first step estimator of $\mathbf{P}^{0}$ and the second step pseudo maximum likelihood estimator of $\boldsymbol{\theta}^{0}$, and asymptotic efficiency of the later (Proposition 4 in \cite{aguirregabiria_mira_2002}).}

The first empirical applications of CCP methods in empirical IO were \cite{slade1998optimal} and \cite{aguirregabiria1999dynamics} on the estimation of dynamic models of firms' pricing and inventory decisions.\footnote{For the computation of $\widetilde{h}_{i}^{\mathbf{P}}(a_{it},\mathbf{x}_{t})$ and $\widetilde{e}_{i}^{\mathbf{P}}(a_{it},\mathbf{x}_{t})$, \cite{hotz1993conditional} considered only finite horizon models and optimal stopping models. For infinite horizon models, they suggest treating them similarly as finite horizon models by truncating the future stream of payoffs. \cite{aguirregabiria1999dynamics} was the first paper to consider the representation of the present values $\widetilde{h}_{i}^{\mathbf{P}}(a_{it},\mathbf{x}_{t})$ and $\widetilde{e}_{i}^{\mathbf{P}}(a_{it},\mathbf{x}_{t})$ in the infinite horizon stationary DP problems as presented above in equations (\ref{eq: present values 2}) and (\ref{W present values}). This representation has been used later in IO applications of CCP methods.} Different versions of this two-step method have been proposed and applied to the estimation of dynamic games by \cite{Aguirregabiria2007}, \cite{Bajari2007}, \cite{Pakes2007}, and \cite{Pesendorfer2008}. 

In dynamic games, the two-step pseudo likelihood estimator in equation (\ref{eq: two step estimator}) is not asymptotically efficient because the zero Jacobian property does not hold in dynamic games. \cite{Pesendorfer2008} propose a two-step estimator for dynamic games that is asymptotically efficient. Their estimator belongs to a general class of minimum distance (or asymptotic least squares) estimators described by the following expression:
\begin{equation}
    \widehat{\boldsymbol{\theta}}
    \text{ } = \text{ }
    \arg\min_{\boldsymbol{\theta}}
    \text{ }
    \left[
        \widehat{\mathbf{P}}^{0} -
        \Psi
        \left(  
           \widehat{\mathbf{P}}^{0},\boldsymbol{\theta}
        \right)
    \right] ^{\prime}
    \mathbf{A}_{M}
    \left[
        \widehat{\mathbf{P}}^{0} -
        \Psi
        \left(  
            \widehat{\mathbf{P}}^{0},\boldsymbol{\theta}
        \right)
    \right]
\label{md estimator}
\end{equation}
where $\mathbf{A}_{M}$ is a weighting matrix. Each estimator within this general class is associated with a particular choice of the weighting matrix. The asymptotically optimal estimator within this class has the following weighting matrix:
\begin{equation}
   \mathbf{A}_{M}^{\ast}
    \text{ } = \text{ }
    \left\{
        \left[
            \mathbf{I} - 
            \triangledown_{\mathbf{P}} \Psi(\boldsymbol{\theta}^{0},\mathbf{P}^{0})
        \right] ^{\prime}
        \text{ }
        \boldsymbol{\Sigma}_{\widehat{\mathbf{P}}^{0}}
        \text{ }
        \left[
            \mathbf{I} - 
            \triangledown_{\mathbf{P}} \Psi(\boldsymbol{\theta}^{0},\mathbf{P}^{0})
        \right] 
    \right\}^{-1}
\end{equation}
where $\boldsymbol{\Sigma}_{\widehat{\mathbf{P}}^{0}}$ is the variance matrix of the initial nonparametric estimator
$\widehat{\mathbf{P}}^{0}$. \cite{Pesendorfer2008} show that this estimator is asymptotically equivalent to the MLE defined in equation (\ref{ML estimator as constrained}). Therefore, there is no loss of asymptotic efficiency by using this two-step estimator of the structural parameters instead of the MLE. From a computational point of view, in contrast to the computation of the MLE, this two-step estimator requires computing the Jacobian matrix $\triangledown_{\mathbf{P}} \Psi$ only once. \cite{srisuma_linton_2012} generalize this method to dynamic games with continuous state variables.

This family of two-step estimation methods -- often referred as \textit{CCP estimators} -- are very attractive because they reduce substantially the computational cost of estimating dynamic models. However, they also have some limitations. A first limitation is the restrictions imposed by the assumption of no unobserved common knowledge variables. Ignoring persistent unobservables, if present, can generate important biases in the estimation of
structural parameters. In section \ref{sec:unobserved heterogeneity}, we review two-step methods that allow for persistent unobserved heterogeneity. Nevertheless, the type of unobserved heterogeneity that we can identify when using two-step methods is substantially more restrictive than when using full-solution methods.\footnote{This is because two-step methods require nonparametric identification of CCPs (conditional on unobserved types) in the first step. The conditions for nonparametric identification of CCPs with unobserved heterogeneity (i.e., nonparametric finite mixture models) are stronger than for the identification of the structural model imposing all its restrcitions, e.g., exclusion restrictions in profit functions. See the results on this point in section 5 in \cite{Aguirregabiria2019incomplete}.}

A second limitation of two-step methods is their finite sample bias. The initial nonparametric estimator can be very imprecise given the sample sizes and the dimension of the vector of state variables that we have in empirical applications in IO. In dynamic games with heterogeneous players, the number of observable state variables is proportional to the number of players and therefore the so called curse of dimensionality in nonparametric estimation can be particularly serious. The finite sample bias and variance of the first-step estimator can generate serious biases in the two-step estimator of structural parameters. To reduce this finite sample bias, \cite{aguirregabiria_mira_2002} have proposed a method that consists of fixed point iterations in the NPL mapping defined in section \ref{sec:full solution methods}. In single-agent models, this NPL mapping is a contraction, and this iterative procedure reduces higher order approximations to the bias (see \cite{Kasahara2009}). However, this is not the case in dynamic games, and this procedure may increase the bias, and even converge to an inconsistent estimator (\cite{Pesendorfer2010}). The development of an estimation procedure for dynamic games that guarantees bias reduction of two-step estimators -- but still is substantially cheaper to implement than full solution methods -- is an interesting topic of methodological research in this field that still needs further developments. 

\subsubsection{Bajari-Benkard-Levin (BBL) method \label{BBL method}}

\cite{Bajari2007} propose a two-step method for the estimation of dynamic games -- the so called \textit{BBL} method -- that has received substantial attention in empirical IO applications. This method has several distinguishing features with respect the two-step Hotz-Miller method described in section \ref{sec: two step methods} above. First, BBL uses Monte Carlo simulation to approximate the expected present values $W_{h,i}^{\mathbf{P}}(\mathbf{x}_{t})$ and $W_{e,i}^{\mathbf{P}}(\mathbf{x}_{t})$. Second, the estimator of the structural parameters in the second step is based on moment inequalities instead of pseudo maximum likelihood (as in \cite{Aguirregabiria2007}), GMM (as in \cite{Pakes2007}), or minimum distance (as in \cite{Pesendorfer2008}). Finally, the BBL method can be applied to models with discrete or/and continuous decision and state variables.

\medskip 

\textit{(i) Monte Carlo approximation of present values.} For the state spaces than we find in many applications of dynamic games (with millions or billions of states), the exact computation of the present values in equation (\ref{W present values}) is impractical, even if this evaluation needs to be done only once for the estimation of the model. An approach to deal with this issue consists in approximating expected present values using Monte Carlo simulation, an approach used early on by \cite{pakes1986patents} and called forward-simulation. In single-agent dynamic discrete choice models, \cite{hotz1994simulation} propose this simulation approach in combination with Hotz-Miller two-step method. Given $\mathbf{x}_{t}$ and $a_{it}$, we can use the estimated transition probability function $f_{x}(.| a_{it},\mathbf{x}_{t},\widehat{\boldsymbol{\theta}}_{f})$ to generate a random draw for $\mathbf{x}_{t+1}$. And given this simulated value of $\mathbf{x}_{t+1}$, we can use the estimated CCP function $\widehat{P}_{i}(.|\mathbf{x}_{t+1})$ to generate a random draw for the optimal choice $a_{i,t+1}$. We proceed sequentially to generate a simulated path of actions and states between periods $t+1$ and $t+T^{*}$ for some pre-specified time horizon $T^{*}$. We can generate many of these simulated paths. Let $\{a_{i,t+s}^{(r)}, \mathbf{x}_{t+s}^{(r)} :s=1,2,...,T^{*}; r=1,2,...,R \}$ be $R$ simulated paths, all of them starting from the sample observation $(a_{it},\mathbf{x}_{t})$. Then, the Monte Carlo approximation to the expected present value $\widetilde{h}_{i}^{\mathbf{P}}(a_{it},\mathbf{x}_{t})$ is:
\begin{equation}
    \widetilde{h}_{i}^{\mathbf{P},R}(a_{it},\mathbf{x}_{t}) 
    \text{ } = \text{ }
    h(a_{it},\mathbf{x}_{t}) 
    \text{ } + \text{ }
    \frac{1}{R}
    {\displaystyle\sum_{r=1}^{R}}
    \left(
        {\displaystyle\sum_{s=1}^{T^{*}}}
        \beta_{i}^{s}
        \text{ } 
        h\left( a_{i,t+s}^{(r)}, \mathbf{x}_{t+s}^{(r)} \right)
    \right).
\end{equation}
And we can use a similar expression to approximate the expected present value $\widetilde{e}_{i}^{\mathbf{P}}(a_{it},\mathbf{x}_{t})$.  \cite{Bajari2007} adapt this approach to approximate expected present values in dynamic games. 

This Monte Carlo approximation implies an approximation error, $u_{it}^{R} = \widetilde{h}_{it}^{\mathbf{P},R} - \widetilde{h}_{it}^{\mathbf{P}}$. This approximation error goes to zero as $R$ goes to infinity, but it can be substantial for the finite value that we use in an application. Simulation errors can increase the bias and variance of our estimators. However, for simulation-based GMM estimators where the simulation error enters additively in the moment conditions, this error does not generate (first order) asymptotic bias in the estimator and the estimator is consistent as the sample size goes to infinity but the number of simulations $R$ is fixed (\cite{mcfadden1989method}).\footnote{Of course, the asymptotic variance of the simulation-based GMM estimator, and higher order approximations to the bias, still depend on (decline with) the number of simulations $R$} This nice property of some simulation based GMM estimators is shared by the method proposed by \cite{hotz1994simulation}.

\medskip 

\textit{(ii) Moment inequalities.} The value of firm $i$ at state $\mathbf{x}_{t}$ when all the players behave according to their strategies in $\mathbf{P}$ can be written as:
\begin{equation}
    V_{i}^{\mathbf{P}}(\mathbf{x}_{t}) 
    \text{ } = \text{ }
    W_{i}^{\mathbf{P}}(\mathbf{x}_{t})
    \text{ }
    \boldsymbol{\theta }_{i}
\end{equation}
where $W_{i}^{\mathbf{P}}(\mathbf{x}_{t}) \equiv \left[ W_{h,i}^{\mathbf{P}}(\mathbf{x}_{t}), W_{e,i}^{\mathbf{P}}(\mathbf{x}_{t}) \right]$, and $\boldsymbol{\theta }_{i}^{\prime} \equiv (\boldsymbol{\theta }_{\pi,i}^{\prime},1)$. For notational simplicity, below we use $W_{it}^{\mathbf{P}}$ to represent $W_{i}^{\mathbf{P}}(\mathbf{x}_{t})$. We can split the vector of CCPs $\mathbf{P}$ into two sub-vectors: $\mathbf{P}_{i}$ with firm $i$'s CCPs, and $\mathbf{P}_{-i}$ containing the probabilities of firms other than $i$. Since $\mathbf{P}^{0}$ is an equilibrium associated to $\boldsymbol{\theta }^{0}$, we have that $\mathbf{P}_{i}^{0}$ is firm $i$'s best response to $\mathbf{P}_{-i}^{0}$. Therefore, for any vector $\mathbf{P}_{i}\neq \mathbf{P}_{i}^{0}$, the following inequality holds:
\begin{equation}
    W_{it}^{\left( \mathbf{P}_{i}^{0},\mathbf{P}_{-i}^{0}\right)}
    \text{ }
    \boldsymbol{\theta }_{i}^{0}
    \text{ }  \geq  \text{ }
    W_{it}^{\left( \mathbf{P}_{i},\mathbf{P}_{-i}^{0}\right)}
    \text{ }
    \boldsymbol{\theta }_{i}^{0}.
\end{equation}
We can define an estimator of $\boldsymbol{\theta }^{0}$ based on these (moment) inequalities. There are infinite alternative policies $\mathbf{P}_{i}$, and therefore there are infinite moment inequalities. For estimation, we should select a finite set of alternative policies. Indeed, a larger number of moments may lead to worse estimates in terms of larger variance, but tighter identified sets.\footnote{There is a large recent literature on moment selection and computing confidence set for models defined by moment inequalities. For instance,  \cite{andrews2010inference}, \cite{bugni2010bootstrap}, \cite{canay2010inference}, and \cite{romano2014practical} study selection of unconditional moment inequalities with varying procedures, while conditional moment inequalities are addressed in \cite{andrews2013inference}, \cite{chernozhukov2013intersection}. For work with a large number of moment inequalities, which is typical of applications such as BBL, work such as \cite{belloni2019subvector}, \cite{bai2021two}, and \cite{chernozhukov2019inference} is also more appropriate.} This is a very important choice for the researcher in the implementation of the BBL estimator (see our discussion below). Let $\mathcal{P}$ be a (finite) set of alternative CCPs selected by the researcher. Define the following criterion function:
\begin{equation}
    \widetilde{Q}
    \left( 
        \boldsymbol{\theta},\mathbf{P}^{0}
    \right) 
    \equiv
    \sum\limits_{m,i,t}
    \sum\limits_{\mathbf{P} \in \mathcal{P}}
    \left( 
        \min 
        \left\{ 
            0 \text{ } ; \text{ }
            \left[ 
               W_{imt}^
               { \left( \mathbf{P}_{i}^{0},
               \mathbf{P}_{-i}^{0} \right) }
                - W_{imt}^
                { \left(\mathbf{P}_{i}, 
                \mathbf{P}_{-i}^{0} \right) }
            \right] 
            \boldsymbol{\theta }_{i}
        \right\} 
    \right) ^{2},
\end{equation}
This criterion function, proposed by \cite{chernozhukov_hong_2007}, penalizes departures from the inequalities. Given an initial nonparametric estimator of $\mathbf{P}^{0}$, and replacing exact  present values $W_{imt}^{\mathbf{P}}$ with Monte Carlo approximations, \cite{Bajari2007} propose an estimator that minimizes in $\boldsymbol{\theta}$ the criterion function $\widetilde{Q}(\boldsymbol{\theta},\widehat{\mathbf{P}}^{0})$. 

In this model. the vector of structural parameters $\boldsymbol{\theta}$ is point identified. However, in most applications of the BBL method, the relatively small set of alternative CCPs, $\mathcal{P}$, selected by the researcher does not provide enough moment inequalities to achieve point identification such that the BBL method provides set estimation of the structural parameters.

This BBL estimator has been applied in a good number of important empirical applications of dynamic games in IO and marketing, such as \cite{Ryan2012}, \cite{ryan2012heterogeneity}, \cite{Suzuki2013}, \cite{Jeziorski2014}, \cite{fowlie2016market}, \cite{hashmi2016relationship}, and \cite{lim_yurukoglu_2018}, among others.

The distinguishing features of BBL method are key to explain its relative popularity. Monte Carlo approximation of present values can make the difference between being able to estimate a dynamic model or not. Nevertheless, this approximation method can be used along with any of the other estimation methods described above, either two-step or full solution methods. The applicability of the BBL method to models with either continuous or discrete variables is also very convenient, and it is a more substantial feature that distinguishes this method. Last but not least, the researcher's selection of the set of alternative CCPs to estimate the parameters (the set $\mathcal{P}$) can be quite attractive in some applications. Any model has its strengths and weaknesses, and sometimes a model provides a poor match for some aspects of the data that are not important to answer the main questions that motivate the paper. The freedom provided by the selection of the set $\mathcal{P}$ allows the researcher to focus on those predictions of the model that are key for the specific research questions. It can also provide a more clear and intuitive picture on the contribution of different features in observed firms' behavior and the identification of some parameters.

Nevertheless, these attractive features also have limitations. In some applications, the number of possible forward paths of length $T^{*}$ is greater than the number of atoms in the universe, but we use only a few million paths to approximate expected present values. These approximations can be seriously biased, but we do not have any practical way of knowing the order of magnitude of this bias in our specific application. Also, the selection of the set $\mathcal{P}$ can hide (intentionally or unintentionally) some sources of misspecification in the model which may be important for the purpose of the research.

\subsubsection{Large state space and finite dependence \label{sec:finite dependence}}

As we have mentioned above, in some empirical applications, the exact computation of present values is impractical as it would require months or years of computing time with even the most sophisticated computer equipment. We need to use approximations to these present values. We have already discussed Monte Carlo approximation methods, which have received substantial attention in this literature. Other approach that reduces this computational cost is exploiting the finite dependence property in some dynamic models.

As we have mentioned above, in \textit{optimal stopping problems} the difference between the conditional choice value functions of two choice alternatives are a simple closed-form expression of CCPs and profits at two consecutive periods, as illustrated in equation (\ref{oprimal stopping}). \cite{Arcidiacono2011} generalize this result to DP models with \textit{finite dependence structure}. For this class of models, two firms that make different choices at period $t$ have a positive probability of visiting the same state $\mathbf{x}$ after a finite number of periods. For instance, consider a \textit{multiple bandit} dynamic decision model where $\mathbf{x}_{t} = (a_{i,t-1}, \mathbf{z}_{t})$ where $\mathbf{z}_{t}$ is a vector of exogenous state variables. For this model, the finite dependence property implies that the difference between the conditional choice value functions of any two choice alternatives, say $j$ and $k$, has the following expression:
\begin{equation}
   \begin{array}
   [c]{l}
    v_{i}(j,\mathbf{x}_{t})-v_{i}(k,\mathbf{x}_{t}) 
    =
    \pi_{i}(j,\mathbf{x}_{t}) -
    \pi_{i}(k,\mathbf{x}_{t})
    \\
    \\
    + \beta_{i} \text{ }
    \mathbb{E}_{t}
    \left[ 
        \pi_{i}(0,j,\mathbf{z}_{t+1}) -
        \pi_{i}(0,k,\mathbf{z}_{t+1}) +
        e_{i}^{P}(0,j,\mathbf{z}_{t+1}) -
        e_{i}^{P}(0,k,\mathbf{z}_{t+1})
    \right].
    \end{array}
\label{mbandit finite dep}
\end{equation}
Furthermore, by Hotz-Miller inversion property, the difference between conditional choice value functions is also a known function of contemporaneous CCPs. For the sake of concreteness, suppose that the unobservables are i.i.d. extreme value type 1, and the profit function is linear in parameters. Then, equation (\ref{mbandit finite dep}) has the following form:
\begin{equation}
    \begin{array}
    [c]{l}
        \log P_{i}(j |\mathbf{x}_{t}) - \log P_{i}(0 |\mathbf{x}_{t}) 
        = 
        \left[ 
            h_{i}(j,\mathbf{x}_{t}) -
            h_{i}(k,\mathbf{x}_{t}) 
        \right]
        \boldsymbol{\theta}_{\pi,i}
        \\
        \\
        + 
        \beta_{i}
        \text{ }
        \mathbb{E}_{t}
        \left(
            \text{ }
            \left[ 
               h_{i}(0,j,\mathbf{z}_{t+1}) -
               h_{i}(0,k,\mathbf{z}_{t+1})
            \right]
            \boldsymbol{\theta}_{\pi,i} -
            \log P_{i}(0,j, |\mathbf{z}_{t+1}) +
            \log P_{i}(0,k, |\mathbf{z}_{t+1})
            \text{ }
        \right).
    \end{array}
\label{mbandit 2}
\end{equation}

This provides an optimality condition that does not include expected present values but only CCPs and profits at periods $t$ and $t+1$. This equation includes the conditional expectation at period $t$ of profits and CCPs at $t+1$, and therefore, it seems that it requires numerical integration over the state space. However, it is possible to use this equation to construct moment conditions that do not require any explicit integration over the space of state variables. The \textit{trick} has a long tradition in the estimation of continuous choice dynamic structural models using Euler equations (e.g., \cite{hansen_singleton_1982}). Under rational expectations, the conditional expectation at period $t$ of CCPs and profits at $t+1$ is equal to these variables minus an expectational error that is orthogonal to the state variables at period $t$. Therefore, for any vector of functions of $\mathbf{x}_{t}$, say $\mathbf{g}(\mathbf{x}_{t})$, we have the following moment conditions:
\begin{equation}
    \mathbb{E}
    \left(
        \mathbf{g}(\mathbf{x}_{t})
        \left[
        \begin{array}
            [c]{c}
            \log P_{i}(j |\mathbf{x}_{t}) - 
            \log P_{i}(k |\mathbf{x}_{t}) -
            \left[ 
                h_{i}(j,\mathbf{x}_{t}) -
                h_{i}(k,\mathbf{x}_{t}) 
            \right]
            \boldsymbol{\theta}_{\pi,i}
            \\
            \\
            - \beta_{i}
            \text{ }
            \left[ 
              h_{i}(0,j,\mathbf{z}_{t+1}) -
              h_{i}(0,k,\mathbf{z}_{t+1})
            \right]
            \boldsymbol{\theta}_{\pi,i} -
            \log P_{i}(0 |j,\mathbf{z}_{t+1}) +
            \log P_{i}(0 |k,\mathbf{z}_{t+1})
    \end{array}
    \right]
    \right)
    = 0. 
\label{moment conditions}
\end{equation}
Constructing sample counterparts of these moment conditions does not require integration over the space of state variables but only averaging over the sample observations. The computational cost of estimating the structural parameters using GMM based on these moment conditions does not depend on the dimension of the state space. The finite dependence property, and this estimation approach, also applies to dynamic games.

This estimation method has been used in IO applications of single-agent models (\cite{bishop2008dynamic} to locational choice; \cite{aguirregabiria2013Euler} to asset replacement; \cite{scott2014dynamic} to land use; \cite{murphy2018dynamic} to housing supply) and of dynamic games (\cite{ellickson2012repositioning}; \cite{Igami2016}).

\subsubsection{Unobserved market heterogeneity \label{sec:unobserved heterogeneity}}

So far, we have maintained the assumption that the only unobservables for the
researcher are the private information shocks that are i.i.d. over firms,
markets, and time. In most applications in IO, this assumption is not
realistic and it can be easily rejected by the data. Markets and firms are
heterogenous in terms of characteristics that are payoff-relevant for firms
but unobserved to the researcher. Not accounting for this heterogeneity may
generate significant biases in parameter estimates and in our understanding of
competition in the industry. For instance, in the empirical applications in
\cite{Aguirregabiria2007} and \cite{Collard-Wexler2013}, the estimation of a
model without unobserved market heterogeneity implies estimates of strategic
interaction between firms (i.e., competition effects) that are close to zero
or even have the opposite sign to the one expected under competition. In both
applications, including unobserved heterogeneity in the models results in
estimates that show significant and strong competition effects.

\cite{Aguirregabiria2007},  \cite{Collard-Wexler2013}, and \cite{Arcidiacono2011} have proposed methods for the estimation of dynamic games that
allow for persistent unobserved heterogeneity in players or markets. Here we concentrate on the case of permanent unobserved market heterogeneity in the profit function. \cite{Arcidiacono2011} propose a method that combines
the GMM-finite dependence method, that we hsve described in section \ref{sec:finite dependence}, with an EM algorithm that facilitates the estimation of the distribution of unobserved heterogeneity. 

Suppose that the payoff function $\pi_{i}$ depends on a time-invariant `random
effect' $\omega_{m}$ that is common knowledge to the players but unobserved to
the researcher. This unobservable is i.i.d across markets, with a distribution that has discrete and finite support. Each value in the support of
$\omega$ represents a `market type', we index market types by $\ell
\in\left\{  1,2,...,L\right\}$, and  $\lambda_{\ell}\equiv\Pr(\omega_{m}=\omega^{\ell})$. This unobservable does not enter into the transition probability of the observed state variables. Each market type $\ell$\ has its own equilibrium mapping (with a different level of profits given $\omega^{\ell}$) and its own
equilibrium. Let $\mathbf{P}_{\ell}$\ be a vector of strategies (CCPs) in
market-type $\ell$. The introduction of unobserved market heterogeneity also
implies that we can relax the assumption of only `a single equilibrium in the
data'\textit{\ }to allow for different market types to have different
equilibria. 

The pseudo log likelihood function of this model is
$Q(\boldsymbol{\theta},\boldsymbol{\lambda},\mathbf{P})=$ ${\textstyle\sum\nolimits_{m=1}^{M}}
\log q_{m}(\boldsymbol{\theta},\boldsymbol{\lambda},\mathbf{P})$, where $q_{m}(\boldsymbol{\theta},\boldsymbol{\lambda},\mathbf{P})$\ is the contribution of market $m$\ to the pseudo likelihood:
\begin{equation}
    q_{m}(\boldsymbol{\theta},
    \boldsymbol{\lambda},\mathbf{P)} 
    = 
    {\displaystyle \sum \limits_{\ell=1}^{L}}
    \lambda_{\ell|\mathbf{x}_{m1}}
    \left[
        {\displaystyle \prod \limits_{i,t}}
        \Psi_{i}(a_{imt} | \mathbf{x}_{mt},\omega{\ell},
        \boldsymbol{\theta},\mathbf{P}
    \right].
\label{likelihood unobserved het}
\end{equation}
where $\lambda_{\ell|\mathbf{x}}$\ is the conditional probability $\Pr
(\omega_{m}=\omega^{\ell}|\mathbf{x}_{m1}=\mathbf{x})$. The conditional
probability distribution $\lambda_{\ell|x}$\ is different from the
unconditional distribution $\lambda_{\ell}$. In particular, $\omega_{m}$\ is
not independent of the predetermined endogenous state variables that represent
market structure. For instance, if $\omega_{m}$ has a positive effect on
profits, we expect a positive correlation between firms' lagged entry
decisions and this unobservable. This is the so called initial conditions
problem (\cite{heckman_1981_incidental}). In short panels (for $T$\ relatively small), not taking into account this dependence between $\omega_{m}$\ and $\mathbf{x}_{m1}$\ can generate significant biases, similar to the biases associated to ignoring the existence of unobserved market heterogeneity. There are different ways to deal with the initial conditions problem in dynamic models. One possible approach is to derive the joint (ergodic) distribution of
$\mathbf{x}_{m1}$\ and $\omega_{m}$\ implied by the equilibrium of the model.
That is the approach proposed and applied in \cite{Aguirregabiria2007} and \cite{Collard-Wexler2013}. \cite{collard2014mergers} also models the initial conditions problem for a time varying market level unobserved state.

Let $p^{\mathbf{P}_{\ell}}\equiv\{p^{\mathbf{P}_{\ell}}(\mathbf{x}_{t}):\mathbf{x}_{t}\in\mathcal{X}\}$\ be the ergodic or steady-state distribution of $\mathbf{x}_{t}$\ induced by the equilibrium $\mathbf{P}_{\ell}$\ and the transition $F_{x}$. This stationary distribution can be simply obtained as the solution to the following system of linear equations:
for every value $\mathbf{x}_{t}\in\mathcal{X}$, $p^{\mathbf{P}_{\ell}}(\mathbf{x}_{t})={\textstyle\sum\nolimits_{\mathbf{x}_{t-1}\in\mathcal{X}}}
p^{\mathbf{P}_{\ell}}(\mathbf{x}_{t-1})$\ $[\sum_{\boldsymbol{a}_{t}}P_{\ell
}(\boldsymbol{a}_{t}|\mathbf{x}_{t})$ $f_{x}(\mathbf{x}_{t}$\ $|$%
\ $\boldsymbol{a}_{t},\mathbf{x}_{t-1})]$. Given the ergodic distributions for
the $L$\ market types, we can apply Bayes' rule to obtain:
\begin{equation}
    \lambda_{\ell|\mathbf{x}_{m1}}=\frac{\lambda_{\ell}
    \text{ }
    p^{\mathbf{P}_{\ell}}
    (\mathbf{x}_{m1})}
    {{\displaystyle \sum \limits_{\ell^{\prime}=1}^{L}}
    \lambda_{\ell^{\prime}}\text{ }
    p^{\mathbf{P}_{\ell^{\prime}}}
    (\mathbf{x}_{m1})}
\label{Bayes rule}
\end{equation}
Note that given the CCPs for each market type, this steady-state distribution does
not depend on the structural parameters $\boldsymbol{\theta}$.

For the estimators that we discuss here, we maximize $Q(\boldsymbol{\theta},\boldsymbol{\lambda},\mathbf{P})$\ with respect to $(\boldsymbol{\theta},\boldsymbol{\lambda})$\ for given $\mathbf{P}$. Therefore, the
ergodic distributions $p^{\mathbf{P}_{\ell}}$\ are fixed during this optimization. This implies a significant reduction in the computational cost associated with the initial conditions problem. Nevertheless, in the
literature of finite mixture models, it is well known that optimization of the
likelihood function with respect to the mixture probabilities $\lambda$\ is a
complicated task because the problem is plagued with many local maxima and
minima. To deal with this problem, \cite{Arcidiacono2011} propose using the EM algorithm.

The estimators of finite mixture models in \cite{Aguirregabiria2007},  \cite{Collard-Wexler2013}, and \cite{Arcidiacono2011} consider that the researcher cannot obtain consistent nonparametric estimates of market-type
CCPs $\{\mathbf{P}_{\ell}^{0}\}$. \cite{Kasahara2009} have derived
sufficient conditions for the nonparametric identification of market-type CCPs
$\{\mathbf{P}_{\ell}^{0}\}$\ and the probability distribution of market types,
$\{\lambda_{\ell}^{0}\}$. Given the nonparametric identification of
market-type CCPs, it is possible to estimate structural parameters using a
two-step approach similar to the one described above.

\cite{berry2020instrumental} (see also \cite{berry_compiani_2021}) advance a generalized instrumental variable approach, following the more abstract approach to this problem outlined in \cite{chesher2017generalized}, to estimating dynamic models with serially correlated unobservables allowed to change over time. Their instrumental variables approach relies on the existence of observable covariates that are uncorrelated with the unobservable component of the payoff function, do not directly enter the present period policy function, but are correlated with the present state variables. Shocks to investment costs in prior periods, changes in regulatory policies that limited or encouraged entry, and demographic changes across time are examples of external economic forces can be correlated with the present state of the market but are uncorrelated with unobservables. A similar approach is taken by \cite{kalouptsidi_scott_2021}. The focus in their work is on market-level unobserved heterogeneity, rather than the agent-level unobserved heterogeneity emphasized by \cite{berry2020instrumental}. The papers impose different assumptions on the nature of the unobservables, and thus are not nested within each other, but both illustrate two ways that the literature has moved forward regarding unobserved state variables.

\subsection{The promise of machine learning}

Machine learning, a term that covers a broad set of tools for statistical learning, has recently generated excitement for its potential to transform empirical and computational analysis. Generally speaking, machine learning methods are algorithmic approaches to solving problems where a minimum of guidance is provided by the researcher in guiding that algorithm to its goal.

There are numerous applications of machine learning in economics. In the context of functional approximation such as the value function approximations that we discussed in section \ref{large state space section}, neural networks, and their extension, deep learning, have shown remarkable promise in their ability to model nonlinear relationships between inputs and outputs. In econometrics, recent work has started to provide rigorous theoretical foundations to machine learning techniques and leverage them for model selection. See, for example, \cite{chernozhukov2018double} and \cite{nekipelov2018moment}. As applied to DP, reinforcement learning has investigated ways of solving for value functions and/or optimal controls using computational techniques based on trial-and-error while remaining agnostic about some aspects of the underlying theoretical machinery, such as the transitions across states, or exactly how rivals arrive at their strategies. For dynamic games, a fundamental question is: can techniques from the machine learning literature help researchers overcome the various computational challenges associated with solving DP problems with high-dimensional state spaces and complex action sets consisting of continuous and discrete decisions while accommodating a large number of potential agents? Our answer to this question, as of the time that this review is written, is, without a doubt, perhaps.

Machine learning has been used for solving dynamic games for decades. An early application is the reinforcement learning algorithm (also known as Q-learning) used in \cite{Pakes2001}, and the real-time algorithm in \cite{Weintraub2017} based on the work by \cite{bertsekas1996neuro}. Over time, under some regularity conditions, the learner traces out the ergodic set of states that are visited in equilibrium and act optimally at each state. One advantage of this approach is that states that are never visited in equilibrium do not need to be included in the solution, which may lead to a speed up in the computation of an equilibrium.\footnote{Note that this specific advantage of reinforcement learning algorithms does not apply to dynamic games that include private information shocks with unbounded support for every action, as suggested by \cite{doraszelski2010computable} to guarantee existence of an equilibrium in pure strategies. With this type of shocks, the probability of every action in every state is nonzero, and the ergodic set is the entire state space. The algorithm may still provide other advantages in the computation of an equilibrium. \cite{Collard-Wexler2013} is an example of reinforcement learning used along with full support shocks.}  Research in this area has continued at a rapid pace since the turn of the century; recent advancements include deep reinforcement learning (\cite{arulkumaran2017deep}). Deep reinforcement learning also encompasses many different techniques, but the basic aim of all of them is to utilize deep neural networks to approximate the optimal policy function. The deep neural network may be augmented with convolutional neural networks that effectively reduce the dimensionality of the inputs. The canonical applications for these techniques are in teaching computers to play video games.\footnote{\cite{shao2019survey} surveys the literature.}. Highlighting the minimal amount of modeling in some reinforcement learning approaches, the basic inputs are simply pictures of the video game screen, while outputs are controller actions (such as up, down, left, or right). The underlying neural network learns the optimal policy function through a trial and error association between states and long-run outcomes.

Early applications of these algorithms focused on simple single-player video games (i.e., the opponent is a computer that does not learn), such as Space Invaders or Snake. It is perhaps not surprising that it is possible to learn optimal policies in such environments, where the best response of the game is relatively straightforward. What is surprising is how well these systems have been adapted to play much more open-ended multi-agent games where the state space is extremely complex, there are a huge number of actions available to players, and your opponent optimizes back against your strategy. A very high-profile example of this was the headline-generating defeat of the (human) world-champion team playing the game Defense of the Ancients 2 (DOTA2) in 2018 and 2019 by OpenAI Five \cite{1912.06680}.\footnote{See, for example, a popular press discussion in \url{https://www.vox.com/2019/4/13/18309418/open-ai-dota-triumph-og}, accessed August 26, 2021.} This example is notable for several reasons. First, the human team was composed of the very best players in the world who have very high-powered incentives to become experts in playing the game---the prize money at the world championships in 2021 is \$40 million. Second, the pace of advancement on the OpenAI side is astounding. In 2016, the AI could only play limited versions of the game with single opponents. Three years later it was roundly and repeatedly defeating the best human players.

The experience of OpenAI Five suggests some important lessons for the promise of machine learning in DP problems.  OpenAI makes admirable progress on all of the criteria: it was able to develop successful policies in an environment with high-dimensional state spaces, complex action sets, and multiple strategic actors. However, there are some caveats. 

First, much of the computer science work on machine learning often focuses on providing improved, rather than exact, solution to decision problems. A machine learning approach which provides a better quality translation of a text, or a more competitive player in DOTA2, is clearly useful. However, in economics, we usually assess algorithm for solving games by how closely they compute equilibrium strategies. Thus, a better machine learning algorithm for solving a game might nevertheless be quite far away from the equilibrium policy. Indeed, in 2019, the OpenAI algorithm was still learning how to play DOTA2. While there is some work providing worse case bounds for these dynamic decision problems, the evidence here is much less clear. Indeed, reading some of the computer science literature reminds us of what econometrics would look like if estimators were judged exclusively based on Monte-Carlo's without reference to any theorem proving consistency or asymptotic distributions. It can be difficult to assess how well these method can be extended to somewhat different dynamic games.

Second, if the goal of machine learning is providing tools that ease computational burden in both time and effort, OpenAI is not a particularly compelling test case. To achieve its current level of sophistication, it has played millions of games, both against itself and against human opponents. This has taken years of computer time, and significant amounts of hardware.

Third, there is clearly some adaption that the research team had to make to bring ideas of deep learning to playing DOTA2; this is not an off-the-shelf endeavor. For instance, the AI was initially restricted to play reduced-complexity versions of DOTA2. This suggests that even the cutting-edge machine learning techniques still require some hand-tuning in defining and restricting the scope of the underlying context that it is trying to learn. As a practical example in economics, when using reinforcement learning approaches for dynamic games it is critical to make sure that the algorithm explores a large enough part of the state space to ensure it is not confined to a locally optimally policy.  

Some of these outstanding issues are driven by a fundamental result proved by \cite{chow1989complexity} thirty years ago: in general, no algorithm can solve the DP problem, for some level of error, without suffering from the curse of dimensionality when the state space is continuous. \cite{iskhakov_rust_2020_ej} have an engaging discussion about the prospects for machine learning techniques that help reduce the state space:
\begin{quote}
Even though machine learning can potentially address the curse of dimensionality by employing model selection when estimating high dimensional choice probabilities, data still limits what we can learn about underlying model structure. But even in the ideal case where machine learning can recover a precise, sparse representation of $P(d|x)$ that allow us to estimate the structural parameters, we cannot rely on this approach for counterfactual simulations. If choice probabilities $P(d|x)$ fundamentally change shape in the counterfactual and require a different set of variables and basis functions, it is still necessary to solve high dimensional DP problems whose sparsity structure is unknown.
\end{quote}
This emphasizes a fundamental difference between some machine learning contexts like computer vision, where dimensionality reduction and neural networks have combined to produce high-performance classification systems for identifying objects in photographs, and dynamic games: in the latter, the value functions and optimal policies are not objects to be identified from a static snapshot, but rather are endogenous, fluid objects that change in response to decisions made in other parts of the state space. Any solution method using machine learning techniques for complexity reduction in the state space has to be adaptive, as states that one might group together as homogeneous at the beginning of the solution process may turn out to be ultimately very different from each other at the final solution. In this sense, the consistent classification of states for the purposes of reducing state space complexity generates yet another fixed point that needs to be solved in parallel to those governing the value functions.

Circling back to the original question of whether machine learning techniques will be beneficial for dynamic games estimation and counterfactual computation, our assessment is a cautious one. Fundamental challenges remain: many machine learning techniques that are marketed as solving the world's problems are nothing more than effective ways to approximate functions. The econometrics literature has already delivered a library of nonparametric techniques that are capable of consistently recovering nearly-arbitrary functions, some of which are much easier to use than the very nonlinear neural networks currently in vogue. No algorithm can ever escape the curse of dimensionality when the state space is continuous, as many are. Finally, dimensionality reduction techniques, like those from computer vision, are promising but still need to confront the problem that the underlying object they are approximating changes while computing the solution. On the other hand, the practical performance of specific implementations like OpenAI Five raise the possibility that future advancements will bring us much closer to the promise of an easy-to-use, accurate, and quick off-the-shelf methodology for estimating and computing dynamic games.

\section{Empirical applications \label{Section: Empirical Applications}}

There are, at present, a large number of applications of dynamics games in IO. This literature is eclectic, motivated by specific applications. To organize our venture in this field, we start by tracing a chronology of the major methodological innovations in the field and how they were applied, then move out to different topics of substantive interest to IO economists, such as innovation, antitrust, asymmetric information, regulation, uncertainty, natural resources, and dynamic matching. 

\subsection{Earlier empirical work on dynamic games\label{sec:early_papers}}

The history of empirical applications of dynamic games in industrial organization can be split into two distinct periods. Early on, a handful of empirical applications directly applied the \cite{Pakes1994} algorithm, mainly  Yale graduate students such as \cite{Benkard2004} and \cite{gowrisankaran1997dynamic}. These papers addressed substantive research questions and highlighted the need for an econometric approach to estimation that sidestepped the computational burden of repeatedly solving the theoretical model. The main innovation was the subsequent development of the estimation methods described above based on CCPs that has directly led to the current era of empirical applications. We organize our discussion of this literature roughly chronologically, beginning with a discussion of \cite{Benkard2004} and \cite{gowrisankaran1997dynamic} before turning to four papers that bridged the two eras: \cite{Jofre-Bonet2003}, \cite{Ryan2012}, \cite{Collard-Wexler2013}, and \cite{Dunne2013}. Much like the econometric methodology upon which they are based, these papers co-evolved contemporaneously; they are important not only as they are among the first examples of applications using these methods, but because they also shed light on challenges to implementation. 

\subsubsection{Competition in the hospital market}

\cite{gowrisankaran1997dynamic} is one of the very first applications using the MPNE framework in an empirical context. Based on the theoretical quality ladder model of \cite{Pakes1994}, the authors examine competition in the US hospital market. This market is an economically important part of the US economy, both in terms of direct expenditures (approximately 5 percent of GDP) and its role in ensuring the health of the population. It is also an industry with heavy government involvement (via service requirements and regulated payments for certain types of consumers), rapid technological advancement, and, in later years, consolidation. The authors build a dynamic model of competition in this industry that aims to capture many of these salient institutional details. 

On the supply side, there are two types of firms: for-profit and non-profit hospitals, each with different objective functions, taxation treatment, and investment costs. Non-profit hospitals differ from for-profit hospitals by maximizing a weighted sum of profits and consumer surplus in their objective function, whereas for-profits care only about the former. For-profit hospitals also have to pay additional taxes that non-profits often do not. Non-profits cannot engage in the same range of financial transactions as for-profits, which may influence their investment costs. 

Hospitals are differentiated by location and quality and may improve their quality through investment as the \cite{Pakes1994} quality ladder model. Hospitals may enter, exit, and set prices for the private market, but are required to accept Medicare patients at a government-imposed regulated price. On the demand side, consumers differ by income and insurance coverage and decide which hospital to attend; the authors use a tailored logit model to estimate demand. 

They use this model to evaluate three different counterfactuals: a change to the Medicare reimbursement rate mechanism, the introduction of universal health-care coverage, and an adjustment to the taxes of non-profit hospitals. Understanding how these policies affect the provision of hospitals in the United States is important. Given how slow the entry and exit process for hospitals is likely to be, it is hard to find good sources of identification for a strictly empirical approach to these questions.  

Their empirical strategy is a modified version of the nested fixed point from \cite{rust1987optimal} or \cite{pakes1986patents} using a simulated method of moments \parencite{mcfadden1989method,pakes1989simulation} approach. In an inner loop, for each guess of the parameter vector they solve the MPNE of their model. They then simulate data from the ergodic distribution of this equilibrium, and construct simulated moments of demand and supply that aggregate data over hospitals and time. In an outer loop, a nonlinear optimizer searches over the parameter space to minimize a distance metric between simulated moments and their empirical counterparts. 

There are several drawbacks to using this aggregated moment approach. The first is that it is statistically inefficient in two ways: aggregating information loses granularity in the underlying data-generating process that would be captured by a full-information estimation method, and the choice of which moments of the data to use ignores some of the empirical restrictions of the model. Second, this approach presumes that all markets are mature enough that they have reached their ergodic distributions. If an industry is still growing to maturity, this approach cannot be used since one effectively is matching the long-run distribution of states to a non-ergodic transition path. Third, the nested fixed-point approach is also computationally burdensome, as it requires repeatedly solving the entire dynamic game, for each market, for each guess of the parameter vector. Finally, the dynamic game has multiple equilibria but their implementation of the NFXP algorithm does not account for this issue. 

The authors make several simplifying assumptions to reduce the dimension of the state space. First, they aggregate a rich set of observable hospital characteristics into a single quality index, and assume that this index has a discrete and finite support with a relatively small number of points. Second, the stochastic process for quality is restricted to move up or down by at most one unit. Third, the authors only consider a small number of firms as potentially being active. All these restrictions are made because of computational convenience, but they may have important impact on the estimation of the model and its predictions on the evolution of market structure. For instance, it is well known that the number of potential entrants in a market can have important effects on incumbent firms' incentives to guard against entry, which in turn changes the evolution of market structure. 

Despite all these limitations, this paper was highly innovative and was among the very first papers to take the MPNE framework to an empirical setting. As such, the authors had to confront an entire host of issues that had never been dealt with previously in the literature. It is also important to note that many of these problems are still present (and potentially acute) at the frontier today. 

\subsubsection{Dynamic output competition with learning by doing}

\cite{Benkard2004} considers the production decisions of wide-body aircraft manufacturers: Boeing, Airbus, and for the time period considered, Lockheed as well. The dynamics here are driven by learning by doing, where aircraft production costs fall with accumulated experience. This mechanism produces intertemporal strategic considerations when pricing an aircraft, as lower prices not only increase sales, but also speed the learning process, while also potentially robbing rival firms of additional experience. 

A manufacturer $i$ produces different varieties of aircraft which are indexed by $\ell$. The production technology includes an equation that represents the causal relationship between a manufacturer's labor requirement for producing one aircraft of type $\ell$, $L_{\ell it}$, and the manufacturer's experience as measured by the number of planes of that type produced in the past, $E_{\ell it}$. This log-linear equation: $\log L_{\ell it} = \theta \log E_{\ell it} + \gamma \log S_{\ell it} + u_{\ell it}$, where $S_{\ell it}$ represents observable characteristics of the aircraft type such as size and speed, and $u_{\ell it}$ is an unobservable productivity shock.\footnote{This labor requirement equation is estimated in the companion paper \cite{Benkard2000}.} Experience evolves based on cumulative production, discounted by a forgetting rate, $E_{\ell i,t+1}=\delta E_{\ell it} + q_{\ell it}$, where $q_{\ell it}$ represents the number of planes produced at time $t$. The demand side of the model is a nested logit demand system, with unobserved product level quality $\xi_{\ell it}$ that evolves exogenously over time. The model includes potential entrants who may choose to enter the market after paying a sunk entry cost. Benkard calibrates the entry cost based on accounting data on development costs of aircraft released by Lockheed.

In this dynamic oligopoly model, every period $t$ firms decide how much to produce of each product. The vector of state variables $\mathbf{x}_{t}$ includes the firm-product specific variables $x_{\ell it} = (E_{\ell it}, \xi_{\ell it})$ for every firm and product, and a time-varying market size $M_t$. This model is solved using a version of the \cite{Pakes1994} algorithm that exploits symmetry in firms' strategy functions. As in the work of \cite{gowrisankaran1997dynamic}, the state space needs to be reduced for computation, and this reduction is achieved in the base case by ruling out multi-product firms, that is, assuming that Boeing's 777 and 747 are produced by two separate firms. 

The model is used for counterfactuals, but also to see if a dynamic oligopoly model can rationalize, quantitatively, some intriguing characteristics of the industry. Aircraft are frequently sold below marginal cost, especially at the early years of a product line. This below-cost pricing may be interpreted as predation (which often triggers anti-dumping sanctions), but it could be partly explained by learning by doing motives which also affect the pricing strategy of a monopolist not concerned about potential entrants. In an oligopoly industry, learning by doing can also exacerbate predatory motives, since lower prices at the early years of a product can induce the exit of rivals. Benkard's estimates show that prices can be up to 50 percent below cost when an aircraft is first introduced, and these prices are even lower when facing competitors that have more experience producing aircraft. This matches observed data on price/cost margins earned by Lockheed. The model is also used to predict concentration in the industry. Learning by doing functions as an additional entry barrier above development costs, and this leads to more concentrated market structure. 

An attentive reader will notice that the dynamic decisions made by firms are not used for estimation. Instead, Benkard uses static techniques to estimate the firm's cost function and demand. So one can think of this line of research trying to uncover what are the dynamic implications of a model, and whether these are quantitatively accurate. While this type of quantitative exercise is common in macroeconomics, in IO, this is the most successful exercise of this type. 

\subsubsection{Dynamics in auctions}

An early paper that presages the following explosion in CCP-based applications is \cite{Jofre-Bonet2003}. This transitional paper sits at the crossroads of the methodological innovations discussed above and the empirical applications that followed. The authors leverage the insights from \cite{elyakime1994first} and \cite{guerre2000optimal} (hereinafter GPV) to estimate bidder valuations in a series of repeated first price procurement auctions for highway paving services. 

There are two potential sources of dynamics: capacity constraints and learning by doing. Firms only have a limited amount of paving capacity in the short run, so winning a large paving contract today may preclude the firm from being able to compete for future contracts. On the other hand, winning a contract today gives the firm additional experience and expertise which may lower costs for future projects. The optimal bidding strategy for a firm has to account for these two economic forces in addition to the standard trade-off between the probability of winning and rents. The dynamic incentives change the standard first-order condition considered in GPV through the inclusion of an extra term that accounts for changes in future costs that may accrue as the result of winning the auction 
today.

The estimation approach proceeds in two steps, in the same spirit as the CCP methods described in section \ref{sec: two step methods}. The first step consists of nonparametric estimation of the reduced form bidding functions relating a firm's bid with the observable state variables. In the second step, these reduced forms are plugged into the first order condition characterizing a firm's best response, and then structural parameters are estimated.

In this model, time is discrete with an infinite horizon. Firms share a common discount factor $\beta$. There are two types of firms: regular firms and fringe firms. Fringe firms are differentiated from regular firms in that they only exist for one period, while regular firms are infinitely-lived. The authors classify firms into these two categories on the basis of how frequently they bid in the data; the largest ten firms are considered to be regular firms and everyone else is a fringe firm. In every period, a sequence of events occur. First, the government presents a single paving contract with idiosyncratic characteristics that the firms may bid on. Second, bidders obtain a draw of private costs, $c_{it}$, for performing the job that comes from a commonly-known distribution that depends on the vector of state variables. Crucially, costs are assumed to be conditionally independent of the contract characteristics. Third, the auction runs and the seller awards the contract to the lowest bidder, subject to a reservation price.

The vector of common knowledge observable state variables $\mathbf{x}_{t}$ is $(\mathbf{s}_{0t}, \mathbf{s}_{it}: i \in \mathcal{I} )$ where $\mathbf{s}_{it}$ is a list of projects, each with an associated size and time left to completion, that firm $i$ has won in the past, and $\mathbf{s}_{0t}$ contains the characteristics of the current contract being auctioned off. This state vector is updated in two ways: the backlog increases (endogenously) when a firm wins a contract, declines (exogenously) each period as the firm works on finishing off existing paving contracts. The authors assume that contracts are completed at a fixed rate each period.

Let $b_{it}$ be firm $i$'s bid at auction $t$, and let $\alpha_{i}(\mathbf{x}_{t},c_{it})$ be firm $i$'s bidding strategy function. The firm's expected profit at period $t$ is equal to its rent if winning, $b_{it} - c_{it}$, times the probability of winning, $W_{i}^{\boldsymbol{\alpha}}(b_{it},\mathbf{x}_{t}) \equiv \mathbb{E}( 1\{ b_{it} < \alpha_{j}(\mathbf{x}_{t},c_{jt})$ for any $j \neq i\}  \text{ }|  \text{ } b_{it},\mathbf{x}_{t})$. Given other firms' bidding strategies, the value function for bidder $i$ is the solution to the following Bellman's equation:
\begin{equation}
    V_{i}^{\boldsymbol{\alpha}}
    (\mathbf{x}_{t},c_{it})
     = 
    \max_{b_{it}} 
    \left[ 
    \begin{array}
    [c]{l}
        \left( b_{it} - c_{it} \right) 
        W_{i}^{\boldsymbol{\alpha}}(b_{it},\mathbf{x}_{t})
        \\
        + \beta  \text{ } 
        W_{i}^{\boldsymbol{\alpha}}(b_{it},\mathbf{x}_{t})
        \text{ }
        \mathbb{E} 
        \left(
            V_{i}^{\boldsymbol{\alpha}}
            (\mathbf{x}_{t+1},c_{i,t+1}) 
            \text{ } | \text{ }
            \mathbf{x}_{t},c_{it}, 
            i \text{ } wins
        \right)
        \\
        + \beta \text{ }
        \left(
            1 -
            W_{i}^{\boldsymbol{\alpha}}(b_{it},\mathbf{x}_{t})
        \right)
        \text{ }
        \mathbb{E} 
        \left(
            V_{i}^{\boldsymbol{\alpha}}
            (\mathbf{x}_{t+1},c_{i,t+1}) 
            \text{ } | \text{ }
            \mathbf{x}_{t},c_{it}, 
            i \text{ } loses
        \right)
    \end{array}
    \right].
 \label{eqn:valueW}
\end{equation}
In the right hand side, the first term is the familiar static payoff from a first-price auction. The second and third terms are the continuation values if winning and if losing the auction, respectively. Each firm forms expectations about the value of the world in the next period for each of the possible winners of the contract today, including itself. Note that, once we account for the probability of winning, the continuation values do not depend on the current bid $b_{it}$. This property plays an important role in the structure of the first order conditions of the model. This formulation is very general, as firms in principle are carrying around a huge state space. As we explain below, the authors introduce important simplifying assumptions in this general framework.

A profit-maximizing firm sets marginal cost equal to marginal revenue. In GPV, one can solve for cost as a function of the observed bid and a markup term. Here, the cost equation has an extra term that comes from the continuation values. Let $h_{i}^{\boldsymbol{\alpha}}(\mathbf{x}_{t})$ be the hazard function of firm $i$'s bids: $h_{i}^{\boldsymbol{\alpha}}(\mathbf{x}_{t}) = g_{i}^{\boldsymbol{\alpha}}(\mathbf{x}_{t}) [1-G_{i}^{\boldsymbol{\alpha}}(\mathbf{x}_{t})]^{-1}$, where $g$ and $G$ are the density and cumulative functions in the distribution of firm $i$'s bids. The authors show the following expression for the first order condition of optimality in firm $i$'s best response:
\begin{equation}
    c_{it} = b_{it} - 
    \frac{1}
    {
    \sum \limits_{j \neq i}
    h_{j}(b_{it}|\mathbf{x}_{t})
    }
    + \beta 
    {\displaystyle \sum \limits_{j \neq i}}    
    \frac{h_{j}(b_{it}|\mathbf{x}_{t})}
    {
    \sum \limits_{\ell \neq i}
    h_{\ell}(b_{it}|\mathbf{x}_{t})
    }
    \left[
        EV_{it}^{\boldsymbol{\alpha}}
        (i \text{ } wins) -
        EV_{it}^{\boldsymbol{\alpha}}
        (i \text{ } loses)
    \right]
\label{eqn:costsJBP}
\end{equation}
where $EV_{it}^{\boldsymbol{\alpha}}(i \text{ } wins)$ is firm $i$'s continuation value if it wins the current auction (that is, $\mathbb{E}( V_{i}^{\boldsymbol{\alpha}}$ $(\mathbf{x}_{t+1},c_{i,t+1}) \text{ } | \text{ } \mathbf{x}_{t},c_{it}, i \text{ } loses)$), and similarly, $EV_{it}^{\boldsymbol{\alpha}}(i \text{ } loses)$ is the continuation value if it loses. The third term in the right hand side represents the dynamic marginal value of winning the auction.

Equation (\ref{eqn:costsJBP}) is the key condition for the estimation of the structural parameters. The econometric object of interest is the cost in the left hand side of this equation. In the right hand side, the hazard functions can be estimated using data on bids and state variables. As in many other applications, the authors consider that the discount factor is known. The authors show that the continuation values $EV_{it}^{\boldsymbol{\alpha}}(i \text{ } wins)$ and $EV_{it}^{\boldsymbol{\alpha}}(i \text{ } loses)$ can be represented as a recursive equation involving the bid distribution function.

Even when the specification of the cost function (that relates a firm's cost $c_{it}$ with the firm's backlog in vector $\mathbf{x}_{t}$) is parametric, the first step in the sequential method to estimate the parameters in this cost function is nonparametric. That is, for consistency of the estimator, the estimation of the hazard functions $h_{i}(b_{it}|\mathbf{x}_{t})$ should be nonparametric because these functions are endogenous equilibrium objects such that a parametric specification of these functions is, in general, incompatible with the equilibrium outcome. However, in this model, the dimension of the space of $\mathbf{x}_{t}$ is very large, such that nonparametric estimates of hazard rates $h_{i}(b_{it}|\mathbf{x}_{t})$ can be extremely imprecise given the curse of dimensionality in nonparametric estimation. Therefore, the authors end up estimating hazard functions under strong  exclusion, aggregation, and parametric restrictions on how the vector $\mathbf{x}_{t}$ enters in these functions. This is a common issue in this literature when using two-step estimation methods. 

\cite{Jofre-Bonet2003} estimate the bid distribution function as a parameterized Weibull distribution for regular bidders and as a beta distribution function for fringe bidders. These choices have substantive restrictions, as the likelihood is only well-specified for a range of parameter values. To capture the dependence of these distributions on state variables, they impose the symmetry condition that all bidders behave identically conditional on equal states. This is a strong assumption that would be violated if the identity of the firm matters beyond what is captured in the state variables, e.g., if there is persistent difference in firm types. These restrictions are imposed through a parameterization of several of the arguments of the bid distribution.

This approach illustrates a common trade-off that practitioners face when using two-step estimation methods in dynamic structural models. Even for unidimensional distributions, the nature of dynamic games may require knowledge of functions evaluated at states that are visited rarely. Having enough observations at every point in the state space is a very high burden outside of the simplest dynamic models, and as a result practitioners have resorted to using parametric approaches. However, this also comes with a cost. 

In addition, while the GPV techniques used in the paper make estimation possible, there remains an issue of how to compute a solution to the dynamic bidding model in the paper, which is, currently, a topic of ongoing research. Needless to say, this limits the scope of counterfactuals from their model. 

\cite{jeziorski2016dynamic} extend Jofre-Bonet and Pesendorfer model to allow for subcontracting in response to capacity constraints. \cite{groeger2014study} studies dynamics generated by sunk entry costs that involve multiple sequential auctions. \cite{hopenhayn2016bidding} develop a dynamic model of bidding in second price online auctions agents can revise their bids, and bidding opportunities and values follow a joint Markov process. They estimate the model using data from eBay auctions. \cite{dee2020} proposes and estimates a dynamic model for pay-per-bid auctions -- a type of auction where bidders incur a cost each time they place a bid.

\subsubsection{Environmental regulations in concentrated industries}

\cite{Ryan2012} studies the cost of environmental regulation in concentrated industries, where the effects of long-run changes to market structure can dwarf the direct costs associated with regulatory compliance. Specifically, he measures the welfare costs associated with the 1990 amendments to the Clean Air Act in the US Portland cement industry, that is the upstream industry from the concrete market.\footnote{See section \ref{sec:demand_shocks} below for our description of \cite{Collard-Wexler2013} study of the US concrete industry.} The amendments introduced a variety of new regulations that applied to the cement industry, including new environmental assessment standards for greenfield cement plants and additional classes of regulated pollutants. In principle, these regulations could have changed the cost structure of the industry in several ways and, by extension, led to a different evolution of market structure. Ryan constructs a model of the cement industry, which has several features which make it well-suited for analysis in the BBL framework, estimates a change in the cost structure of firms before and after regulation using a panel on every cement plant in the United States from 1980 to 1998, and compares realized outcomes against a simulated counterfactual where the regulation did not exist. His primary finding is that entry costs increased, leading to fewer firms in equilibrium and a loss of between \$810 million and \$3.2 billion in surplus. A static analysis would miss the change in entry costs and find the wrong sign of costs to incumbent firms, who actually benefit from reduced competition under the amendments.

The cement industry has several institutional features that make it an attractive setting for two-step estimation. The first is that cement is a largely homogeneous commodity due to its use as a construction material. Cement is also hygroscopic (i.e., it absorbs water from the air), making storage expensive, and has a relatively low value to volume ratio. The combination of these two factors leads it to be shipped overland only relatively short distances, which means that most cement markets are quasi-independent geographically-differentiated regional markets. This is useful for modeling and estimation purposes, as it both reduces the number of firms that need to be considered in each market and generates a cross-section of observations. The spot market for cement is also highly seasonal, due to construction demand peaking in the warm weather months. Given that storage is expensive, most firms do not hold significant amounts of stock from year to year, which also simplifies the state space.

The technology of cement production also lends itself to parsimonious modeling. Technological progress in the industry is very slow; the basic process of making cement has changed very little since the late 1800s. To produce cement, firms mine limestone (often co-located with the plant), grind it up into small pieces, and then heat it through a large rotating kiln that is fired to very high temperatures at one end. It is then ground up and gypsum is added to create cement. Cement plants typically produce nonstop at a constant rate for most of the year before shutting down in the winter to perform maintenance on the kiln. This is important to note for several reasons: first, marginal costs should be reasonably flat until that maintenance period is reached; at that point, the opportunity cost of production increases as the firm eats into the maintenance period. Second, firms are primarily differentiated through their location and their productive capacity. Third, emissions are a key component of cement production, both through the pyrochemical process of converting limestone into clinker and through the burning of fossil fuels to produce that heat. Fourth, fixed costs are a first-order feature of the industry. The typical cement market has only a handful of firms active, and the average size of cement plants is large and has increased steadily over the twentieth century. A typical cost for a greenfield plant is half a billion dollars, and plant lifetimes approach one hundred years. Finally, most plants in most years are capacity constrained and produce right up to their boilerplate ratings. This suggests that long-run changes to market structure may be the dominant margin for assessing the costs of regulation, as firms may have relatively little margin for adjustment in the output market.

The theoretical model has three primary components: a state space, induced transitions over those states in response to firm actions, and per-period payoff functions which depend on firm actions, market demand, and the state vector. As with the prior work on dynamic games, there is a relatively simple state space consisting of the productive capacities of each firm in a regional market. Potential entrants are encoded with a zero capacity. In contrast to prior literature, the state space is continuous, as capacity is not naturally discrete. The industry has been in a long period of sustained consolidation, as a smaller number of larger firms become the dominant firm type. This is also useful for bounding the number of potential firms in the industry, as it is very unlikely that any market would see more than one entrant, especially after the passage of the 1990 amendments.

In each period, firms compete in Cournot competition subject to their capacity constraints. Capacity constraints are modeled through a ``hockey-stick'' specification for cost that generates constant marginal costs before increasing as a function of the firm's capacity:
\begin{equation}
    c(q_{it};\delta) = 
    \delta_0 + \delta_1 \text{ } q_{it} + 
    \delta_2 \text{ } 
    1(q_{it} > \nu \text{ } s_{it}) 
    \text{ } (q_{it} - \nu \text{ } s_{it})^2,
\end{equation}
where $q_{it}$ is firm $i$'s output at year $t$, $s_{it}$ is its production capacity, $\delta$s are cost parameters, and $\nu$ is a parameter that determines the output/capacity ratio at which the additional cost kicks in. The lack of meaningful dynamics in production, due to high storage costs and seasonal demand, is particularly helpful in this setting to pin down the range of admissible dynamic parameters. 

The vector of common knowledge state variables is $\mathbf{x}_{t} = (z_{t}, s_{it}: i \in \mathcal{I})$, where $z_{t}$ is a demand shock. Every period, firms make dynamic decisions: incumbent firms decide investment (or divestment) in capacity and whether to stay or exit from the market, and potential entrants decide whether to enter in the market. Let $a_{it}$ represent firm $i$'s dynamic decision.  Firms' capacities change endogenously as a result of decision $a_{it}$. All transitions are assumed to take one period to happen. There is an adjustment cost function that captures investment, divestment, entry, and exit costs associated with decision $a_{it}$. Since all these actions are discrete, i.i.d. private information shocks in adjustment costs are introduced to ensure the existence of a pure strategy equilibrium \parencite{doraszelski2010computable}. All firms, including potential entrants, receive a draw from a distribution of fixed adjustment costs and decide whether to engage in costly capacity investment/divestment. Additionally, incumbent firms receive a draw of exit costs from a common distribution and decide whether to exit or continue. Potential entrants receive a draw of entry costs from a common distribution and decide whether to enter the industry (and at what capacity level), or remain outside the industry. As discussed in Section \ref{sec:identification} above, fixed costs of production are not jointly separately identified from entry and exit costs, and are assumed to be zero.

In this dynamic game, as described in section \ref{sec:model incomplete info}, we can use a CCP function $\mathbf{P}_{i} \equiv \{P_{i}(a_{it}| \mathbf{x}_{t}: (a_{it},\mathbf{x}_{t}) \in \mathcal{A} \times \mathcal{X})\}$ to represent a firm's strategy. The key equilibrium requirement is that a firm's strategy $\mathbf{P}_{i}$ should be optimal given the strategies of its competitors. Optimality requires that the expected present value from following that strategy is at least as good as from using any other feasible strategy, $\mathbf{P}_{i}^{\prime}$:
\begin{equation}
    \mathbb{E}_{\mathbf{P}_{i}, \mathbf{P}_{-i}} \left[ 
        \sum_{t=0}^{\infty} 
        \beta^{t} \text{ }
        \pi(a_{it},\mathbf{x}_{t},
        \boldsymbol{\theta}_{\pi})
    \right] >
    \mathbb{E}_{\mathbf{P}_{i}^{\prime}, \mathbf{P}_{-i}} 
    \left[ 
        \sum_{t=0}^{\infty} 
        \beta^{t}  \text{ }
        \pi(a_{it},\mathbf{x}_{t},
        \boldsymbol{\theta}_{\pi})
    \right] 
\label{eqn:RyanInequality}
\end{equation}
where strategy-dependent expectations are taken over future actions, states, and shocks, and $\pi(a_{it},\mathbf{x}_{t},\boldsymbol{\theta}_{\pi})$ is the profit function, where $\boldsymbol{\theta}_{\pi}$ is the vector of structural parameters in profits.

The empirical approach closely follows the BBL method described in section \ref{BBL method} above: a first step where the econometrician estimates reduced form equilibrium policy functions (i.e., CCPs) directly from the data, followed by a second step where those reduced forms are projected onto an underlying theoretical model. The inequality in equation (\ref{eqn:RyanInequality}) is the heart of that second step projection. Ryan assumes that all of the markets play the same equilibrium, allowing him to pool across markets when estimating policy functions.\footnote{See \cite{otsu2016pooling} for a statistical test and evaluation of the pooling assumption.} BBL requires high-quality, flexible reduced-form estimates of the policy functions for each element of the state space: the probability of entry and exit, the probability of investment/divestment, and the level of investment/divestment if it occurs. Without solving for an equilibrium, these reduced form functions are nonparametric objects for the econometrician. However, there is a huge curse of dimensionality in the nonparametric estimation of these reduced form functions. Using panel data from a few hundred markets over two decades, achieving the maximum possible precision in the estimation of structural parameters in the second step requires imposing substantial smoothing / aggregation restrictions in the nonparametric estimation in the first step (\cite{ackerberg2014asymptotic}, \cite{chernozhukov2016locally}). To deal with this problem, Ryan estimates parsimonious parametric policy functions in the first step of the method. He uses probits to estimate the probability of entry and exit, where the arguments of the probit are a constant, the sum of capacity of competitors, a dummy variable for post-1990, and the firm's own capacity for the exit policy function. The investment/divestment probabilities and levels are estimated using an adaptation of the (S, s) rule from \cite{attanasio2000consumer}, where two latent bands around the current state define when firms adjust and to what level. The critical aspect of this specification is that it allows for lumpy adjustment, where firms do nothing for long periods of time and then abruptly make a large change to their capacity. The arguments of these band functions are b-spline basis functions of the firm's own capacity and the sum of competitors capacity.

After projecting the reduced forms down onto the underlying dynamic structural model, Ryan finds that the distribution of fixed entry costs both increased in mean and decreased in variance after the 1990 policy change. Both factors lead to potential entrants facing much higher draws of entry costs. In contrast, the distributions of exit costs, investment costs, and divestment costs did not have statistically significant differences before and after 1990.

Ryan performs a counterfactual experiment where the cost structure of firms is held constant at pre-1990 levels, and compares that outcome to that with the actual post-1990 parameters. For computational reasons, Ryan  restricts the experiment to two different initial conditions in a four-firm market, which was chosen to be close to the representative cement market in the US. Starting without any firms, the regulation severely restricts entry, lowering profits and consumer surplus. The distribution of active firms is compressed downward (by about one firm on average), although this is partially offset by firms choosing larger capacities when they do enter. Prices go up very modestly, but it is really the lack of entry (and associated capacity) that drives the total surplus declines, leading both firms and consumers to be worse off.

In a second experiment, Ryan starts the market with two mature firms, one large and one small, that are endowed with a combined capacity similar to the average US market. In this setting, the incumbent firms actually do better under the regulations, as higher costs effectively prevent entry while not harming existing firms directly. While pre-1990 entry costs has two firms active only 4 percent of the time, after the amendments that proportion increases to 11 percent.

These two counterfactuals are intended to put very rough bounds on the costs of the amendments. While there are no markets that have zero firms, the estimated cost in this setting should be a conservative upper bound, conditional on the market size. On the other hand, many markets do look more like the second setting, with mature firms and low turnover. The weakness of both experiments is clearly that they do not actually model the cement market in the US directly. This was completely driven by computational restrictions, as it proved impossible to compute equilibria for markets with five or more firms. 

\subsubsection{Demand shocks and market structure \label{sec:demand_shocks}}

An important limitation of the static models of entry of \cite{bresnahan1990entry}, \cite{bresnahan1991entry}, and \cite{berry1992estimation} is the inability to look at how uncertainty affects market structure in oligopolies. \cite{Collard-Wexler2013} directly address the question of how volatility of demand affects market structure in the market for ready-mix concrete, the downstream industry of Portland cement studied by \cite{Ryan2012} and described in the previous section. In section \ref{Section: uncertainty} we discuss other papers that study how uncertainty influences the organization of production, that also relates to long-standing debates in macroeconomics on the role of uncertainty in investment. 

Collard-Wexler studies the market for ready-mix concrete, which is a combination of water, gravel, sand, and cement, and is used for construction projects such as basements, sidewalks, and roads. This industry is even more geographically differentiated than the market for cement studied by \cite{Ryan2012}. Because ready-mix concrete is heavy and starts to set once cement and water have been mixed in, transportation is quite limited, with the average load of ready mix concrete being delivered no more than a half and hour away by truck. This means that one can think of the industry as a collection of hundreds of geographically segmented local markets. It is this geographical segmentation combined with the production of a reasonably homogeneous good that makes ready-mix concrete a good setting for looking at the empirical consequences of differences in competition. It has  been studied first by \cite{syverson2004market}, but also in \cite{foster2008reallocation} and \cite{backus2020productivity}. In addition, because ready-mix concrete is part of the manufacturing sector, in contrast to other locally-segmented markets considered in \cite{bresnahan1991entry}, like dentists or tire dealers, it is included in Census of Manufacturers with data on all plants in the industry going back to the early 1960's. So there is data on the choices of thousands of plants over decades in terms of entry and exit decisions, as well as investment choices. This combination of variation in market structure and many plant level decisions allows the paper to rely less on parametric assumptions to estimate conditional choice probabilities. 

A distinguishing feature of the ready-mix concrete industry is that demand is very volatile, with year to year demand changes averaging 30 percent. This demand volatility is usually caused by variation in government spending on local construction projects. To evaluate the effect of removing this demand volatility, Collard-Wexler estimates a structural model of entry and exit and discrete investment. In this model, the state of the market, represented by vector $\mathbf{x}_{t}$, includes the size distribution of firms, $(s_{it}: i \in \mathcal{I})$ (where $s_{it}=0$ means that the firm is not active in the market), and an exogenous state variable $z_{t}$ that measures the state of the demand for construction in a local market. Every period (year), firms choose to be active or not in the market, as well as three discrete levels of plant size. That is, a firm's decision at period $t$ is its size at period $t+1$, i.e., $a_{it} = s_{i,t+1}$. There is an assumption of one year time-to-build.

The profit function, $\pi_{i}(a_{it},\mathbf{x}_{t},\boldsymbol{\theta}_{\pi})$ is equal to $r_{i}(\mathbf{x}_{t},\boldsymbol{\theta}_{r}) - \tau(a_{it},s_{it},\boldsymbol{\theta}_{\tau})$, where $r_{i}(.)$ is a variable payoff function (revenue minus variable cost), and $\tau(.)$ is an adjustment cost function that captures the costs of market entry and exit, and the cost of growing and shrinking firm size. The dataset in this paper does not include information on firms' output. Therefore, in contrast to the modelling and estimation approach in \cite{Ryan2012}, the payoff function $r_{i}(.)$ is not based on an explicit specification of demand, variable costs, and the form of (static) competition. Instead, following a common approach in static models of market entry based on \cite{bresnahan1991entry}, the payoff function $r_{i}(.)$ is a semi-reduced-form linear-in-parameters function of the firm's own size ($s_{it}$), competitors' sizes ($\mathbf{s}_{-it}$), and the state of demand ($z_{t}$). Finally, there are private information shocks, $\varepsilon_(a_{it})$, to the value of taking an action, which are assumed to be i.i.d. extreme value type 1.

In this paper, all the parameters in the profit function $\pi_{i}$, both $\boldsymbol{\theta}_{\tau}$ and $\boldsymbol{\theta}_{r}$, are estimated from the equilibrium conditions in the dynamic game, based on firms' entry, exit, and investment decisions. This approach is not feasible without a large amount of data on entry and exit decisions of firms in markets with differing demand and market structure. This explains why this modeling approach is relatively unusual in the broader literature. Moreover, the adjustment cost function $\tau(.)$ has many parameters to estimate, as it measures the cost of moving between any two discrete size categories. 

More than two dozen parameters are estimated using a two-step CCP method similar to the ones in \cite{Aguirregabiria2007} and \cite{Bajari2007}. As we have discussed in section \ref{sec:unobserved heterogeneity}, a major issue with standard CCP methods is the presence of persistent unobserved market heterogeneity. A common effect of of ignoring this type of unobservables when present is that the response of entry to the number of firms is biased. Indeed it can be positive. As a diagnostic of this issue, Collard-Wexler finds far more negative coefficients of competition on entry when market fixed effects are included, suggesting that there is indeed the presence of persistent unobserved profit shocks in these markets.

The ``hack'' used in \cite{Collard-Wexler2013} is to group markets into a couple of categories based on their market fixed effect. This group becomes an additional observed state that can simply be added to the rest of the state space. This grouping does well at replicating the results from market fixed effects regressions, without having to estimate different market fixed effects in the structural model. Of course, this approach is problematic since endogenous variables are being used to create this grouping, and the estimated fixed effects suffer the \textit{incidental parameters problem} (\cite{heckman_1981_incidental}). Thus, a more holistic approach to classification, such as the one in \cite{Arcidiacono2011}, seems more appropriate. This approach has been used by \cite{Igami2016} for the estimation of a dynamic game of market entry/exit in the Canadian fast food industry.

Collard-Wexler uses the estimated structural parameters to simulate out the effect of shutting off demand shocks associated to local government projects. To evaluate this effect, he needs to solve for firms' equilibrium strategies under the counterfactual scenario where demand shocks are eliminated. Given that the state space in this model has around 50 million points, standard methods to compute a MPNE, such as \cite{Pakes1994}, are not feasible.  Instead, the stochastic algorithm of \cite{Pakes2001}, a machine learning algorithm, is adapted to solve this dynamic programming problem. 

Because of high sunk costs of entry, there is no effect of demand volatility on plant shutdown and new plant entry. It is worthwhile for plants to wait out periods of low demand, even if they lose money for several years. However, demand fluctuations do change the size distribution in the industry, as firms would build larger plants in the absence of demand volatility. This effect opens up interesting avenues by which macroeconomic policy that reduces swings in demand may permanently alter market structure, which is not attainable with static models or entry. Later in this section, we will discuss the work of \cite{Kalouptsidi2014} which further investigates the role of adjustment frictions, such as time to build, in an volatile demand environment. 

\subsubsection{Subsidizing entry}

\cite{Dunne2013} examine the determinants of market structure in two service industries using the empirical framework of \cite{Pakes2007}. This paper is of interest both for substantive reason, they assess an important entry subsidy for helping locate health care providers into underserved geographic areas, and because it directly connects back to two of the most influential early papers on structural entry models: \cite{bresnahan1990entry} and \cite{bresnahan1991entry}. Those papers advanced a two-stage model of entry, and used the relationship between the total number of active firms and market size (population) to indirectly infer the nature of competition. For example, suppose that we observe only one firm active in all markets below a population threshold of $20,000$ people, and only two firms for populations above that threshold. If that is the case, we can infer that, in markets with more than $20,000$ people, competition must be near Bertrand-levels of intensity, as no additional amount of demand, as proxied for by population, can induce additional entry. That could only be true if the firms are minimally differentiated and pricing near marginal cost. On the other hand, if we observe a steady increase in the number of active firms as population increases, we can infer that competition is less intense. At the extreme, a linear relationship between population and active firms would be consistent with collusion, where prices do not fall with entry and firms only have enough demand to cover their fixed costs. A data innovation that Bresnahan and Reiss use, focusing on small, isolated markets to obtain a cross section of independent markets, is carried over to this paper.

\cite{Dunne2013} extend the static two-stage framework to a dynamic game. This is necessary to understand the effects of different types of subsidies (e.g., subsidies on entry costs versus subsidies on fixed operating costs) and their differential impact on potential entrants and incumbents in the short-run and long-run. In their extended model, there are two types of firms: potential entrants and incumbents. Potential entrants take a draw from a distribution of entry costs before deciding to enter. Incumbent firms earn product market profits and receive a draw from the fixed costs of operation. If the fixed costs are sufficiently high, that firm exits. The vector of state variables $\mathbf{x}_{t}$ consists of the number of incumbent firms, $n_{t}$, and a vector of exogenous profit-shifters, $\mathbf{z}_{t}$, that evolves as a finite-order Markov process. Following the tradition in Bresnahan and Reiss's entry models, the flow profit of an incumbent firm, $\pi(n_{t},\mathbf{z}_{t})$, is modeled as a reduced form: it is a linear-in-parameters function of variables $n_{t}$ and $\mathbf{z}_{t}$.\footnote{For the purpose of this paper which is interested in the effects of entry subsidies, a drawback of a reduced form specification of the profit function is that it is not possible to measure consumer surplus. This limits the content in the counterfactual evaluations.} In addition to this flow profit, there are fixed costs, $\theta^{FC} + \varepsilon_{t}^{FC}$, paid by any incumbent firm, and entry costs, $\theta^{EC} + \varepsilon_{t}^{EC}$, paid by potential entrants that choose to enter in the market. The authors assume that $\varepsilon_{t}^{FC}$ is i.i.d. Expontential, and $\varepsilon_{t}^{EC}$ is i.i.d. chi-square.

The authors study two different health care industries: dentists and chiropractors. They argue that balance sheet data from the US Census Bureau provides good measures of flow profits $\pi_{mt}$ in the geographic markets included in their sample. Given they observe profits, they estimate the parameters in the profit function $\pi(n_{t},\mathbf{z}_{t})$ by estimating the following linear regression model:
\begin{equation}
    \pi_{mt} = 
    \theta_{0} + 
    \sum_{n=0}^{5} \theta_{n} \text{ } 1\{n_{mt}=n\} + 
    \theta_{6} \text{ } n_{mt} + 
    \theta_{7} \text{ } n_{mt}^2 +
    h(\mathbf{z}_{mt},\boldsymbol{\theta}_{z}) +
    \omega_{m} +
    u_{mt}
\label{eq:dunne_regression}
\end{equation}
where vector $\mathbf{z}_{mt}$ includes socioeconomic variables at the local market level: population, average real wage paid to employees in the industry, real per-capita income, county-level real medical benefits, and infant mortality rate. The term $h(\mathbf{z}_{mt},\boldsymbol{\theta}_{z})$ represents a quadratic function
of these five exogenous state variables. A drawback to this approach is that accounting profits observed in balanced sheet data can be substantially different than economic profits, especially in this setting as the two professions considered (dentists and chiropractors) are highly mobile. Indeed, one of the policy concerns with using entry subsidies is that the practitioners leave the needy areas after their contracted term of service is over. 

Given estimates of $\boldsymbol{\theta}$ parameters and market fixed effects $\omega_{m}$ in the regression equation (\ref{eq:dunne_regression}), the authors follow the empirical strategy in \cite{Pakes2007} to estimate fixed cost and entry cost parameters from the dynamic game. That paper leverages a discrete state space to generate a  matrix representation of the value function. To fit their data into that approach, and to reduce the dimension of the state space, Dunne et al assume that the only exogenous state variable in the dynamic game is the estimated index $h(\mathbf{z}_{mt},\widehat{\boldsymbol{\theta}}_{z})$ that is discretized it into ten categories. To control for persistent unobserved market-level heterogeneity, they also allow for a lower-dimensional vector of fixed effects formed by binning the estimated fixed effects from the regression equation (\ref{eq:dunne_regression}) (i.e., $\widehat{\omega}_{m}'s$). These simplifications are sufficient to allow the authors to form estimates of the continuation values at every point in the state space.

They find that profits decline quickly for dentists, but the same regression for chiropractors is not statistically significant. The implied net present values for these professions are reasonable, however. They estimate monopolist dentists in high-value markets have an average net present value of 1.3 million 1983 dollars, while chiropractors have less than half of that. For dental markets labeled as high need (and therefore subsidized), they estimate that entry costs are 11 percent lower. This leads to about one-half more firm per market on average, at the cost of about \$170,000 per additional entrant. A subsidy targeting the fixed costs of firms to keep them active has a much higher cost per retained firm, about half a million dollars, due primarily to infra-marginal firms that were not going to exit also receiving a subsidy. Targeting the subsidy to potential entrants is therefore far more cost effective.

\subsection{Innovation and market structure\label{innovation}}

Going back to \cite{shumpeter1942capitalism}, there has been interest in studying the causal relationships between innovation and competition, and, more specifically, the hypothesis that \emph{less} competition can have a positive impact on innovation. This interest was supercharged by the work in endogenous growth theory, such as \cite{romer1986increasing} and \cite{aghion1992model}, that placed the study of the determinants of economic growth at the forefront of economics. A line of work in this literature has been based on cross-industry regressions of innovation on competition, with \cite{aghion2005competition} being the most prominent example. In contrast, the recent work in industrial organization has tended to use the predictions of appropriately calibrated or estimated models of dynamic oligopoly in the \cite{Ericson1995} framework. This is due in part to the long held skepticism in IO of regressions of outcomes against market structure.\footnote{See \cite{berry2019increasing} for a discussion of the history of thought on this issue.} Furthermore, for many of the industries studied in the papers that we review in this section, such as hard drives or microprocessors, the effects of competition on innovation are likely to dwarf, in terms of welfare evaluation, the effects of competition on prices conditional on technology, given the vast decreases in costs that these industries have produced. 

\subsubsection{Microprocessor innovation: Intel vs AMD \label{section gg}}

\cite{Goettler2011} study competition between Intel and AMD in the PC microprocessors industry. The authors assess the question of whether there would have been more or less innovation without AMD. Indeed, given the rapid pace of technological change in the semiconductor industry, the welfare effect of reduced competition on innovation is the most important antitrust issue. \cite{Goettler2011} propose and estimate a dynamic game of investment in R\&D and dynamic price competition between Intel and AMD. Importantly, their model incorporates the durability of the product as a potentially important factor for innovation. In their model, there are two main forces driving innovation. First, because consumers value product quality (i.e., microprocessor speed) there is competition between firms to have a product at  the technological frontier. A second factor driving innovation is \textit{endogenous technological obsolescence}. Since microprocessors have little physical depreciation, firms have the incentive to innovate to generate a technological depreciation of the microprocessors (PCs) that consumers own and encourage consumers to upgrade. Note that duopolists are affected by these two forces to innovate, whereas a monopolist faces only the latter, but in a stronger way.

The demand side of the model is dynamic, with forward-looking consumers. The state variables in consumer $h$'s decision problem are: the quality of the PC (microprocessor) that the consumer currently owns, as measured by the logarithm of the microprocessor's speed in MHz, $q_{ht}^{*}$; the current quality of the product that each firm sells, $\mathbf{q}_{t} = (q_{jt}: j \in \{ \textit{Intel}, \textit{AMD} \})$; and the distribution of qualities of the products owned by all the consumers, $\boldsymbol{\Delta}_{t}$.\footnote{The model restricts each firm to selling only one product because the large computational burden of allowing multi-product firms in a model with dynamics in both demand and supply. \cite{Esteban2007} (for automobiles) and  \cite{gowrisankaran2012dynamics} (for digital cameras) estimate dynamic demand models of differentiated product with multi-product firms and forward-looking consumers but with supply side models that are substantially simpler than in Goettler and Gordon's study.} The distribution $\boldsymbol{\Delta}_{t}$ is part of a consumer's state variables because it affects her expectations about future prices. The vector of state variables in the firms' decision problems is $(q_{Intel,t}, q_{AMD,t}, \boldsymbol{\Delta}_{t})$. Given these state variables, firm $j$ makes two dynamic decisions: price, $p_{jt}$, and investment in R\&D $x_{jt}$ to enhance product quality. Note that, because computers are durable goods, firms face a Coasian pricing problem, so pricing has dynamic implications as it changes consumer holdings in the future.\footnote{\cite{Esteban2007} also study the effects of durability and secondary markets on dynamic price competition between automobile manufacturers.}

Every quarter $t$, a consumer decides whether to buy a new PC (microprocessor) or waiting and keeping her current PC with quality $q_{ht}^{*}$. The current utility of not buying is $u_{0,ht} = \gamma \text{ } q_{ht}^{*} + \varepsilon_{0,ht}$. The utility of buying brand $j \in \{\textit{Intel}, \textit{AMD} \}$ is $u_{j,ht} = \gamma \text{ } q_{jt} - \alpha \text{ } p_{jt} + \xi_{j} + \varepsilon_{j,ht}$, where $\xi_{j}$ is a brand fixed-effect, and the consumer taste shocks $(\varepsilon_{0,ht}, \varepsilon_{Intel,ht}, \varepsilon_{AMD,ht})$ are i.i.d. extreme value type 1. Consumers are forward-looking and maximize expected and discounted intertemporal utility.\footnote{This dynamic demand model is a simplified version of the model in \cite{gowrisankaran2012dynamics} which includes random coefficients, multi-product firms, and several product attributes.} Market shares for consumers currently owning a product with quality $q^{*}$ is:
\begin{equation}
    s_{jt}(q^{*}) 
    \text{ } = \text{ }
    \frac
    {
        \exp \{v_{j}^{con}(\mathbf{q}_{t}, 
        \boldsymbol{\Delta}_{t}, q^{*})\}
    }
    {
        \exp \{v_{0}^{con}(\mathbf{q}_{t}, 
        \boldsymbol{\Delta}_{t}, q^{*})\} +
        \exp \{v_{Intel}^{con}(\mathbf{q}_{t}, 
        \boldsymbol{\Delta}_{t}, q^{*})\} +
        \exp \{v_{AMD}^{con}(\mathbf{q}_{t}, 
        \boldsymbol{\Delta}_{t}, q^{*})\}
    }
\end{equation}
where $v_{j}^{con}$ is the conditional choice value function in a consumer's decision problem. Using the distribution of consumers' owned qualities, $\boldsymbol{\Delta}_{t}$, yields the market share of brand $j$:
\begin{equation}
    s_{jt} 
    \text{ } = \text{ }
    {\displaystyle \sum_{q^{*}}}
    s_{jt}(q^{*}) 
    \text{ }
    \Delta_{t}(q^{*})
\end{equation}
By definition, next period distribution of owned qualities, $\boldsymbol{\Delta}_{t+1}$, is a known closed-form function of $\boldsymbol{\Delta}_{t}$, $\mathbf{s}_{t}  \equiv (s_{Intel,t}, s_{AMD,t})$, and $\mathbf{q}_{t}$, that we can represent as $\boldsymbol{\Delta}_{t+1} = F_{\boldsymbol{\Delta}}(\boldsymbol{\Delta}_{t}, \mathbf{s}_{t},\mathbf{q}_{t})$.

In each period, microprocessor firms make an investment decision to try to reach a higher quality level. Change in quality, $q_{j,t+1} - q_{jt}$, can be zero (unsuccessful investment) or a positive constant $\delta$ (successful investment). The probability of success is  denoted $\chi_j$, and depends on the firm's investment $x_{jt}$, with the same functional form as \cite{Pakes1994}:
\begin{equation}
    \chi_j(x_{jt},q_{jt})
    \text{ } = \text{ }
    \frac{a_{jt}(q_{jt}) \text{ } x_{jt}}
    {1 \text{ } + \text{ }
    a_{jt}(q_{jt}) \text{ } x_{jt}}
\end{equation}
where the term $a_{jt}(q_{jt})$ represents firm $j$'s investment efficiency that has the following form:
\begin{equation}
    a_{jt}(q_{jt})
    \text{ } = \text{ }
    a_{0,j} \text{ }
    \max 
    \left[  
        1, a_1 \text{ }
        \left(
            \frac{\bar{q}_{t}-q_{jt}}
            {\delta}
        \right)^{1/2}
    \right]
\end{equation}
where $\bar{q}_{t} \equiv \max \{q_{Intel,t}, q_{AMD,t} \}$ is the frontier or highest quality in the industry at period $t$. This is an increasing function of the technology gap $\bar{q}_{t}-q_{jt}$, that captures the idea that generating successful innovations is more difficult for the leader that is at the technological frontier than for the follower that is catching up. This helps rationalize why AMD and Intel never drift to having too different quality levels. Parameters $a_{0,Intel}$ and $a_{0,AMD}$ allow for persistent differences in the investment efficiencies of the two firms, that can rationalize why AMD reached the the same microprocessor speed as Intel despite having a substantially smaller level of R\&D investment.

In addition, the non-frontier firm has marginal costs that are lower than the firm with the highest level of quality. The frontier firm has costs that are $\lambda_0$, while costs are reduced for the non-frontier firm by $\lambda_1(\bar{q}_{t}-q_{jt})$. That is, parameter $\lambda_1$ represents the dollar reduction in marginal cost per unit of quality difference with respect to the leader (as quality $q$ is the logarithm of microprocessor speed).

Note that the space of the state variables $\mathbf{q}_{t}$ and $\boldsymbol{\Delta}_{t}$ is unbounded, as they can increase forever at increments $\delta$. To deal with this issue, the authors impose the restriction that different structural functions are homogeneous of degree with respect to quality. This restriction makes it possible to recast the state space as one relative to the frontier $\bar{q}_{t}$. This modified state space is bounded.\footnote{This trick is used extensively to discuss balanced growth paths in the macroeconomics literature.}

For the estimation of the model, the authors estimate first the marginal cost parameters $\lambda_0$ and $\lambda_1$ using proprietary production costs data from In-Stat/MDR, a market research firm specializing in the microprocessor industry. The rest of the structural parameters -- both the demand parameters $\alpha$, $\gamma$, $\xi_{Intel}$, and $\xi_{AMD}$, and the supply side parameters $a_{0,Intel}$, $a_{0,AMD}$, and $a_{1}$ -- are estimated using the structure and predictions of the dynamic oligopoly model. \cite{Goettler2011} use a simulated method of moments estimator, similar to the approach used by \cite{gowrisankaran1997dynamic}. However, instead of assessing the gap between the data and the ergodic distribution predicted by the model, they look at the prediction from the model starting in the observed state in 1993 all the way out until 2004. They consider moments related to the firms' innovation rates, R\&D intensities, differential quality, frequency of quarters where Intel is the leader, gap to the quality frontier, average prices, and OLS coefficients in the regression of prices on qualities and in the regression of market shares on qualities. Parameter $\delta$ is fixed at $0.20$ (i.e., 20\%), and the discount factor is fixed at $0.90$ at the annual level.

The ratio between the estimates of $\gamma$ and $\alpha$ shows that consumers are willing to pay \$21 for enjoying a 20\% increase in quality during one quarter. According to the ratio between $\xi_{Intel}-\xi_{AMD}$ and $\alpha$, consumers are willing to pay \$194 for Intel over AMD. The model needs this strong brand effect to explain the fact that AMD's share never rises above 22 percent in the period during which AMD had a faster product. The innovation efficiencies $a_{0,j}$ are estimated to be $0.0010$ for Intel and $0.0019$ for AMD, as needed for AMD to occasionally be the technology leader while investing much less than Intel. 

The authors use the estimated model to implement two main sets of counterfactuals. The first set deals with the effects of competition on innovation. For instance, they solve and simulate the model under the counterfactual scenario of Intel monopoly and compare the results to the actual data. According to this experiment, the innovation rate (i.e., the growth rate in frontier quality $\bar{q}_{t}$) increases from $59.9\%$ to $62.4\%$; investment in R\&D more than doubles, increasing by 1.2 billion per quarter; price increases by \$102 (70\%); consumer surplus declines by \$121 million (4.2\%); industry profits increase by \$159 million; and social surplus increases by \$38 million (less than 1\%). Therefore, they find competition from AMD had a negative impact on the speed of innovation, but overall it has had a positive effect on consumer welfare because the competition effect on prices have than offset the lower quality. They also consider the counterfactual scenario of a symmetric duopoly where the two firms have the same demand brand fixed effects and innovation intensity parameters. The effects are basically the opposite to the first experiment: investment in R\&D, innovation rates, and average quality decline, but prices also decline and this effect more than offsets the quality decline such that consumer welfare increases by \$34 million (1.2\%), industry profits decline by \$8 million, and social surplus increases by \$26 million (less than 1\%). 

The finding that innovation by a monopoly exceeds that of a duopoly reflects two features of the model: a monopoly must innovate to induce consumers to upgrade; the monopoly is able to extract much of the potential surplus from these upgrades because of its pricing power. However, if there were a steady flow of new consumers into the market, such that most demand was not replacements of older computers, the monopoly would reduce innovation below that of the duopoly.

In a second set of counterfactuals, Goettler and Gordon study the claim that Intel used  anti-competitive foreclosure practices against AMD.\footnote{In 2009, Intel paid AMD \$1.25 billion to settle claims of anti-competitive practices to foreclose AMD from many consumers.} To study the effect of such practices on innovation, prices, and welfare, the authors perform a series of counterfactual simulations in which they vary the portion of the market to which Intel has exclusive access. Let $s_{Intel,t}^{mon}$ and $s_{Intel,t}^{duo}$  be Intel's market shares under monopoly and under free competition with AMD, respectively. The authors incorporate foreclosure using a simple model where the degree of foreclosure is measured by a parameter $\zeta \in [0,1]$ such that $s_{Intel,t} = \zeta \text{ } s_{Intel,t}^{mon} + (1-\zeta) \text{ } s_{Intel,t}^{duo}$. The authors solve and simulate the dynamic oligopoly model for a grid of values for parameter $\zeta$. Not surprisingly, margins monotonically rise steeply with $\zeta$. However, innovation exhibits an inverted U shape with a peak at $\zeta = 0.5$. Consumer surplus is actually higher when AMD is barred from a portion of the market, peaking at 40\% foreclosure. This finding highlights the importance of accounting for innovation in antitrust policy. The decrease in consumer surplus from higher prices can be more than offset by the compounding effects of higher innovation rates.

\subsubsection{Hard drive innovation: New products and cannibalization}

\cite{Igami2017} also studies the relationship between competition and innovation. He focuses on the propensity to innovate of new entrants relative to incumbents in the hard drive industry. Similarly to microprocessors, there has been dramatic fall in the price of hard drive storage. However, in contrast to microprocessors where Intel has had a dominant position for almost 50 years, the leading hard drive producers have changed several times over the last forty years. These shifts correspond to periods where the product format changed from 5.25 to 3.5 inch drives and from 3.5 to 2.5 inch. In addition, at some points in time, there are several dozen firms producing hard drives, but there has been gradual exit from this industry down to four firms. The active entry and exit of firms leads to a natural discussion on how the incentives to innovate differ between new entrants and incumbents, given that innovation tends to displace existing products. 

The key empirical evidence that motivates this paper is that the propensity to adopt a new product (e.g., producing the \textit{new} 3.5 inch format instead of the \textit{old} 5.25 inch) is substantially higher for new entrants than for incumbent firms. Igami focuses on the transition from 5.25 to 3.5 inch format, and studies three main factors that may contribute to the difference in the propensity to innovate of incumbents and new entrants: cannibalization, preemption, and differences in innovation costs. For an incumbent firm, the increase in sales and revenue from the introduction of a new product comes partly at the expense of cannibalizing its old products. This is not the case for a new entrant. Therefore, cannibalization may contribute to explain the higher propensity to product innovation by new entrants. The magnitude of this effect depends, among other things, on the degree of demand substitution between new and old products. Preemptive motives can encourage incumbent firms to early adoption of new products to 
deter entry and competition from potential entrants. This factor may partly offset the contribution of cannibalization. Last but not least, incumbents and new entrants can have different costs of adopting new products. This difference can go in either direction. Economies of scope between old and new products can imply lower adoption costs for incumbent firms. On the other hand, incumbent firms may exhibit \textit{organizational inertia} that makes it costly to abandon old practices and adjust the operation to the idiosyncrasies of the new product.

In Igami's model, there are four (endogenous) types of firms in the market, and the state of the market at period $t$ consists of the number of firms of each type: potential entrants $n_{t}^{pe}$, incumbents producing only the old product $n_{t}^{old}$, incumbents producing only the new product $n_{t}^{new}$, and incumbents producing both, $n_{t}^{both}$. Notice that this does not leave room for differences in market share between firms. The vector of state variables $\mathbf{x}_{t}$ is completed by demand shocks for the new and old products. $\xi_{t}^{new}$ and $\xi_{t}^{old}$. Every year $t$, potential entrants decide to enter with the old or new product, incumbents decide to exit or stay in the market, and old incumbents also decide whether to adopt the new product.\footnote{In principle, new incumbents might also choose to start producing the old product, and old incumbents might decide to stop producing the old product. However, these choices are never observed in this industry during the sample period.} There is one year time to build for these entry, exit, and adoption decisions to be effective. 

Demand has the structure of a static logit model between old and new products and an outside alternative. Following the standard structure in the \cite{Ericson1995} model, incumbent firms compete in prices \`a la Bertrand. To apply a full solution method for the estimation of the structural parameters, and to avoid the issue of the multiple equilibria in the counterfactual experiments, Igami imposes three restrictions that imply uniqueness of a MPBNE in his model: (i) the industry has a finite horizon $T$ that is certain and common knowledge; (ii) within each of the four types, firms are homogeneous up to i.i.d. private information shocks in entry, exit, and adoption costs; and (iii) every year $t$, firms take dynamic decisions according to a pre-established order that depends on firm type. In the benchmark version of the model, the order of moves is the following: first, old incumbents choose to exit, stay and innovate, or stay and not innovate; second, incumbents producing both products choose to exit or stay; third, new incumbents choose also between exit or stay; and finally, potential entrants decide whether to enter or not. Igami presents estimates of the model under other orders of moves. The estimates of the dynamic structural parameters (entry, exit, and adoption costs) are quite robust to the different orders of moves considered.\footnote{See, for instance, Table 6 in \cite{berry1992estimation}, for an example of the impact of these assumptions on ordering of moves on parameter estimates. In the context of repeated --- yearly --- interactions, it is plausible that ordering of move assumptions are less material. Table 4 in \cite{Igami2017} does a nice job of looking at the impact of alternative assumptions on the ordering of moves in a dynamic game context.}

Igami estimates the dynamic parameters using Rust's nested fixed point algorithm, as described in section \ref{sec:full solution methods} above. The state space that \cite{Igami2017} considers is large given that there may be dozens of firms within each endogenous type. There are over 38,000 states in his model. This makes the computation of the maximum likelihood estimator using the NFXP algorithm computationally intensive. To keep this cost tractable, Igami considers a parsimonious specification of the model with only three dynamic parameters: $\phi$, the fixed cost of operation; $\kappa^{inc}$, the sunk cost of adopting the new product for the old incumbents; and $\kappa^{ent}$, the sunk cost of entry with the new product for a new entrant.\footnote{The computation time of solving for the equilibrium of the dynamic game does not depend on the number of parameters. However, the number of iterations in the search of the parameter estimates does increase with the number of parameters.} Igami estimates demand parameters and marginal costs using standard static tools.

The parameter estimates show that the sunk cost of innovation is smaller for incumbents than for new entrants, $\kappa^{inc} < \kappa^{ent}$. That is, economies of scope seem more important than organizational inertia. The magnitude of the estimated sunk cost of innovation is between $0.6$ and $1.6$ billion dollars, which is comparable to the annual R\&D budget of specialized hard drive manufacturers like Seagate.

\cite{Igami2017} uses the estimated model to implement counterfactual experiments to evaluate the contribution of cannibalization and preemptive motives to the different innovation rates of incumbents and potential entrants. To isolate the effects of incumbents' incentive to avoid cannibalization, Igami divests each incumbent into a legacy firm and new product firm. That is, for every old incumbent firm, there is an independent firm that decides whether adopt or not the new product. This counterfactual incumbent type is more likely to enter the newer format as they do not internalize the cannibalization of the old product. The equilibrium under this scenario shows that the gap between entrants and incumbents in their innovation choices shrinks by 57 percent. To identify the effect of preemption, one needs to obtain an incumbent's behavior under the hypothetical scenario that the firm's own entry decision into the new format did not change what potential entrants would do. More specifically, Igami assumes that incumbents' strategies are the solution of a dynamic programming problem that assumes that potential entrants do not respond to the number of incumbent firms in the newer format ---they assume there are none of these in the market. This counterfactual shows that the long term number of incumbents that enter the newer format falls by 38 percent.\footnote{The counterfactual exercise of shutting down preemption motives is a difficult one. This issue has been confronted by a number of papers, \cite{chicu2013dynamic} and \cite{besanko2014economics} being perhaps the most notable. Like any deviation away from Nash Equilibrium, fixing a coherent system of beliefs is always tricky when shutting down one such mechanism. In this case, potential entrants may find it ex-ante unprofitable to enter in equilibrium given their incorrect beliefs.}

These counterfactuals show that both cannibalization and preemption play an important role in the decision of incumbent firm to adopt the new format. However, in the hard drive industry, the incentive to avoid cannibalization has dominated preemptive motives.

\subsubsection{Car innovation and quality ladders}

\cite{hashmi2016relationship} study the effect of market power on innovation in the automobile industry. They propose and estimate and model of innovation and quality competition that combines features of the static model of demand and price competition in \cite{berry1995automobile} (BLP hereinafter) and the dynamic quality ladder game in \cite{Pakes1994}. 

The automobile industry has many manufacturers and types of cars. Estimating a dynamic game with the enormous state space that results from this number of firms and products is impractical. The authors make substantial simplifying assumptions. The model starts with a stripped down version of BLP, in which consumers choose a manufacturer or brand (instead of a car model as in BLP). The utility of consumer $h$ if she chooses brand $j$ at period $t$ is $u_{h,jt} = \theta_p \text{ } p_{jt} + \xi_{jt} + e_{h,jt}$, where $p_{jt}$ and $\xi_{jt}$ are the brand's price and quality, respectively, and $e_{h,jt}$ is the usual extreme value type 1 shock.\footnote{The authors measure brand price as the weighted average of the prices of all the car models the manufacturer sells.}  

Firm quality $\xi_{jt}$ has a discrete and finite support and it evolves endogenously as the result of the firm's investment in R\&D.\footnote{In the BLP model, product quality is relative to the value of an outside alternative. This relative aspect makes more plausible the assumption of finite support, at least for the automobiles product category that does not present a trend in demand during the sample period.} The stochastic process for quality depends on two forces: depreciation and successful innovation. Depreciation makes quality decline in $\Delta_{\xi}$ units with an exogenous probability $\lambda^{d}$. Successful innovation can make quality increase in $\Delta_{\xi}$ units with and endogenous probability $\lambda_{jt}^{u}$ which depends on the firm's investment in R\&D, $x_{jt}$, and current quality according to the following equation:
\begin{equation}
    \lambda_{jt}^{u}
    = \exp \{
        - \exp \{\theta_{1}^{u} \text{ }
                \ln( x_{jt}+1 ) +
                \theta_{2}^{u} \text{ } \xi_{jt} +
                \theta_{3}^{u} \text{ } \xi_{jt}^{2} \}
    \}.
\end{equation}
Therefore, we have that $Pr(\xi_{j,t+1}-\xi_{jt} = \Delta_{\xi}) = (1-\lambda^{d})\lambda_{jt}^{u}$, and $Pr(\xi_{j,t+1}-\xi_{jt} = -\Delta_{\xi}) = \lambda^{d}(1-\lambda_{jt}^{u})$. 

The estimation of the demand parameter $\theta_{p}$ is based on a standard IV method. Quality $\xi_{jt}$ is obtained as a residual from the demand equation. Then, these qualities are discretized and the parameters $\lambda^{d}$, $\theta_{1}^{u}$, $\theta_{2}^{u}$, and $\theta_{4}^{u}$ in the stochastic process of quality are estimated by maximum likelihood. Note that the estimation of all these parameters does not use the predictions from the dynamic game.

Given the the vector of state variables $\mathbf{x}_{t} = (\xi_{jt}: j \in \mathcal{I})$ and a private information i.i.d. shock in the cost of R\&D invesntment, $\varepsilon_{jt}$, firms choose their investments in R\&D to maximize their expected present values. The cost of R\&D investment is given by the following cubic function:
\begin{equation}
    c(x_{jt},\varepsilon_{jt},\boldsymbol{\theta}_c)
    \text{ } = \text{ }
    \left(
        \theta_{c,1} + 
        \theta_{c,2} \text{ } x_{jt} +
        \theta_{c,3} \text{ } x_{jt}^{2} + 
        \theta_{c,4} \text{ } \varepsilon_{jt}
    \right)
    x_{jt}
\end{equation}
where $\boldsymbol{\theta}_c = (\theta_{c,1},\theta_{c,2},\theta_{c,3},\theta_{c,4})$ are parameters to be estimated  The parameters in the cost of R\&D are estimated from the predictions of the dynamic games using the estimation method in \cite{Bajari2007}. For the reduced form estimation of firms' strategy functions in the first step of BBL, the authors consider a specification that has the flavour of oblivious equilibrium or moment-based equilibrium, as the the explanatory variables are the firm's own quality and aggregate moments in the cross-sectional distribution of all the firms' qualities, i.e., mean, standard deviation, kurtosis, skewness, and inter-quartile difference. 

The final step of this exercise is to look at how changing the number of firms alters optimal investment decisions within the estimated model. For this experiment, and for computational reasons, the authors consider a strong simplification of their dynamic game, with at most five automobile manufacturers, and where quality is restricted to take 15 values. They solve for a MPBNE using the \cite{Pakes1994} algorithm. The authors find that adding another firm would lower the rate of innovation in this industry, and this effect is magnified with higher quality entrants. 

An important consideration in this modeling exercise is the degree to which cars are vertically versus horizontally differentiated. If cars were only vertically differentiated, then one might expect much more intense competition on the quality dimension. The simplified demand system that is being used in this exercise, while practical for shrinking down the state to a single firm dimension $\xi_j$, also shapes the conclusions of this exercise in a way that is difficult to assess. 

\cite{xu2008structural} study spillovers between firms'
R\&D investments in the Korean electric motor industry. The authors use the concept of oblivious equilibrium from \cite{weintraub2008markov} to simplify computation, which also allows for the use of standard estimation techniques like GMM. Indeed, in this type of work it is sometimes hard to distinguish competitive models from oligopolistic ones. \cite{Kryukov2010} analyzes the development of new drugs in the pharmaceutical industry. \cite{aw2011r} consider the synergies between firms' investment in R\&D and the decision to export.

On the quality ladder side, \cite{Borkovsky2012} present numerical experiments for this class of models. \cite{lin2015quality} develops and estimates a dynamic game of entry and exit and quality competition between nursing homes. Indeed, a large part of health care operates in an environment of regulated prices, either for most countries in world outside the US, or for a large fraction of health care expenditures in the United States. In this setting, competition acts more directly on quality rather than on prices.  

\subsubsection{Data on innovation}

A notable aspect of this literature is how differently it treats empirical work from the canonical empirical models presented in the previous sections. When looking into innovation into new products and technologies, it is very rare to have a panel of independent markets, or enough repeat innovations that are similar to be able to estimate much from the dynamic choices of agents.

There are some other papers that look at innovation through the lens of a dynamic model, but are able to utilize more cross-sectional data by looking at localized adoption of new technologies. \cite{Schmidt-Dengler2006} looks at the timing of MRI adoption, while \cite{caoui2019estimating} is concerned with adoption of digital projection in movie theatres in France. 

Overall, these studies contribute to filling in the theory literature on the relationship between competition and innovation. Given the flexibility of the \cite{Pakes1994} framework, these calibrations to specific industries both in terms of modeling details and parameter estimates help give us an idea of what predictions we should expect, essentially through the accumulation of computational case studies. However, much like much of the IO theory literature before it, many of the outcomes in this literature do depend on the details of the industry under study. This make extrapolation to new industries and innovations quite tricky, but also emphasizes the role of capturing industry level detail properly.

\subsection{Antitrust policy towards mergers \label{sec:antitrust}}

Merger policy occupies a central role in industrial organization, as antitrust is the most important area in which IO economists shape the debate on policy. \cite{whinston2007antitrust} provides an extensive discussion of antitrust policy on horizontal mergers. However, most of their survey takes a static viewpoint on the impact of mergers, essentially implying that all effects of a merger are realized immediately. In addition, in the Department of Justice and Federal Trade Commission's Horizontal Merger guidelines \cite{hmg2010}, only section 9 discusses the role of post-merger entry or exit.  

\subsubsection{Endogenous mergers}

One of the first applications of the \cite{Ericson1995} framework of dynamic oligopoly was the work of \cite{Gowrisankaran1999} which proposes a model of endogenous mergers and market structure. More recently, \cite{mermelstein2020internal} have attempted to address the problem of endogenous mergers from the perspective of a competition authority which is assessing different merger rules with and without commitment power. These papers address the thorny problem of how to deal with a sequence of mergers, considering that the free-rider problem from a merger shifts as the industry becomes more concentrated.\footnote{A free-rider problem in merger occurs when non-merging parties are liable to be the largest beneficiaries of the merger. See  \cite{farrell1990horizontal}.}  A full evaluation of a merger should take into account this type of effect. Modeling endogenous mergers is complicated as it involves a bargaining process between firms that should take into account, in one way or the other, the value added from different possible mergers. Thus, if firms are not identical, they need to evaluate all alternative merging parties, and indeed, the sequence of future mergers. Several of the figures outlining the protocol for merger choice in \cite{Gowrisankaran1999}, such as figure 1 and figure 2, can only be described as baroque, which underscores the complexity of modelling merger choice. 

Another attempt at embedding the negotiation process inside a dynamic game, in this case the hospital-insurance company problem first considered by \cite{ho2009insurer} is \cite{lee2013markov}. In this game, each period there is some probability that two parties get the chance to include a hospital in their network, and one party can make a take-it or leave it offer. The outside option in this bargaining game is the continuation value if agreement is not reached, much like the game described by \cite{shaked1984involuntary} following \cite{rubinstein1982perfect}. This paper also requires an enormous level of skill with computation to implement.

In practice, to compute these theory models, the authors need to make numerous choices as to the parameters that should be used, and these choices tend to be somewhat disconnected with the empirical reality in any given industry. For instance, both \cite{Gowrisankaran1999} and \cite{mermelstein2020internal} look at Cournot competition in a homogeneous product industry. A step in the direction of bringing data to bear on this issue is the \cite{IgamiConsolidation} model of mergers in hard drive manufacturing. One of the more difficult issues to resolve in this literature is the choice and timing of merging parties. \cite{IgamiConsolidation} make merging opportunities a random arrival process. It is difficult to see how empirical work can improve on this --- which is unfortunate since data does not seem to inform merger choices of firms much.\footnote{There is huge variation in the volume of merging activity over time that is not explained very well, e.g., so-called merger waves. See \cite{jovanovic2002q}, among others, for a discussion of the issues in this literature.}

\cite{IgamiConsolidation} propose and estimate a simpler dynamic game of endogenous mergers using the same industry and similar assumptions about market structure and profit function as in \cite{Igami2017}. In each period, a firm can make a merger offer to another firm, and it pays a sunk cost of making this offer. If an offer is made, the two parties, $i$ and $j$, negotiate an acquisition price $p_{ij}$, through different bargaining protocols such as take-it-or-leave-it or some form of Nash Bargaining.

Firms have productivity levels $\omega_{it}$, but this immediately poses the question of what happens to productivity when firms $i$ and $j$ merge. Does the merged firm have productivity $\omega_{it}$, or $\omega_{jt}$, a convex combination of the two productivities, the maximum of the two, 
or perhaps an even higher productivity due economies of scope? Following \cite{farrell1990horizontal}, \cite{IgamiConsolidation} assume that productivity after a merger becomes:
\begin{equation}
    \omega_{i,t+1}
    \text{ } = \text{ }
    \max
    \left\{
        \omega_{it} , \omega_{jt}
    \right\} +
    \Delta_{ij,t+1}
\end{equation}
where $\Delta_{ij,t+1}$ is a merger synergy term that is drawn from a Poisson distribution with parameter $\lambda$. Note that this parameter $\lambda$ and the cost of making a merger are key to guiding how quickly firms will want to merge. Furthermore, the change in productivity of firms following a merger can be used to identify the value of the $\lambda$ parameter. 

A merger authority needs to assess which mergers to let through, and the simplest possible merger policy is one that simply sets a minimum number of competitors for any industry. In the mobile phone market, there is a serious discussion of how markets perform with three or four competitors. Likewise, there is a discussion in the airline industry on whether the authority's merger decisions lead to too few competitors in the industry \parencite{olley2018}. In both cases, one can think of the competition authority as picking the minimum number of competitors. Note that \cite{mermelstein2020internal} consider more sophisticated issues in this policy discussion, such as whether the authority can credibly commit to its merger rules, or instead adapts sequentially its merger policy. 

\cite{IgamiConsolidation} find that, given their estimates for the hard drive industry, a threshold of $N=3$ firms is close to being socially optimal. 

\subsubsection{Evolving market structure and mergers}

An important component of antitrust scrutiny of a merger is the possibility of post-merger entry. Indeed, in the simplest steady-state model of market structure with identical firms, market structure is completely unaffected by merger activity. If the free entry condition dictated that four firms could be supported in a market before the merger, then there will be four firms in the market regardless of whether the merger goes through or not. This calculus can be altered by realistic dynamics in the entry and exit process, which lead to slow adjustments of market structure. Indeed, an important takeaway from the literature on firm dynamics is how slowly changes occur in many industries. For instance, \cite{collard2015reallocation} looks at the entry process of mini-mills into the production of steel. While this new technology does displace the older integrated producers in the steel bar product segment, and increasingly in the steel sheet segment, the entire process takes more than forty years, so the process of reallocation is quite slow. 

\cite{collard2014mergers} looks at the effects of a merger that would knock out a competitor in the ready-mix concrete market, using data on ready-mix concrete markets in isolated markets in the tradition of \cite{bresnahan1991entry}, and more specifically of the model of industry dynamics in 
\cite{Bresnahan1994}. Let $n_{mt}$ be the number of incumbent firms in a market $m$ at time $t$. This number evolves according to firms' entry and exit that determines a transition probability function  $P(n_{mt} | n_{m,t-1}, \mathbf{b}_{mt})$, where $\mathbf{b}_{mt}$ is a vector of exogenous market characteristics that also evolves over time. In his application, the model of \cite{abbring2010} is used to justify a single equilibrium in the entry and exit game, which imposes restrictions on the ordering of moves of firms, as well as the structure of the process for demand. This means that the entry and exit policy rules follow demand thresholds, where the gap between exit and entry thresholds is indicative of the importance of sunk costs. \cite{collard2014mergers} finds that a merger that initially transforms a duopoly market into a monopoly market would induce 9 to 10 years of monopoly relative to the benchmark of no merger. Indeed, the analysis of a merger in this market is closer to a situation where there is not post-merger entry all at. 

\cite{BenkardManu} propose a dynamic analysis of mergers in the airline industry. The airline industry works particularly well from an empirical perspective since the researcher can look at a cross-section of airline routes --- indeed this was the motivation behind how \cite{berry1992estimation} designed his analysis.\footnote{Clearly this approach ignores cross-market synergies in the airline routes. This is a topic of active research.} Moreover, there has been ongoing displacement of legacy airlines in the industry by newer entrants such as Southwest and JetBlue. Thus, dynamic entry considerations, for instance by the new carriers, can be important in assessing the impact of a merger in the airline industry. 

\cite{BenkardManu} estimate conditional choice probabilities for the likelihood of entering or exiting a route given by $P_{i}(a_{it}|\mathbf{x}_{t})$, where $\mathbf{x}_{t}$ includes a demand covariates (such as population at each endpoint), characteristics of an existing airline's network, and competitors routes. The authors also estimate the (Markovian) stochastic process of the exogenous variables in vector $\mathbf{x}_{t}$, given by $Q(\mathbf{z}_{t+1}|\mathbf{z}_{t})$. A merger will alter which firms offer service on different routes: they change the state of the market from $
\mathbf{x}^\text{no-merger}$ to $\mathbf{x}^\text{merger}$. Given the CCPs for firm's entry and exit choices and the process for exogenous states, the authors can forward simulate how the market evolves if a merger occurs or not and compare the expected outcomes on market structure. Key to this exercise is that a market structure simply changes the set of firms that participate in a route market but does not create new potential entrants in that route.\footnote{In these applications, an airline is considered a potential entrant in a route (say Chicago-New York route) if it operates flights in one of the two cities that define the route. After a merger, airlines' entry-exit decisions in different routes can generate changes in the set of potential entrants in other routes. \cite{BenkardManu} ignore these endogenous changes and impose the restriction that the set set of potential entrants in every route remains constant as in the first period when the merger occurs.} 

One of the main conclusions from this exercise is that the trends of the market matter. Indeed, much of the dynamic effects of mergers come not from changes in the number of competitors in the markets that experienced a merger, but instead from existing firms exiting markets in the absence of a merger. 

\subsubsection{Revealed merger efficiencies}
 
A final tack on looking at the impact of mergers is to attempt to uncover the cost efficiencies that they generate. Outside the world of antitrust litigation and their expert witnesses, these costs are thought to be very difficult to ascertain. Indeed, IO economists tend to be averse to using accounting cost data to assess synergies, and besides which, how would one know what these costs would have been in the counterfactual where the merger did not occur?
 
One approach to this problem is to use revealed preference to back out what perceived merger efficiency must be to rationalize the merger choices of firms. \cite{Jeziorski2014} applies this model to estimate cost efficiencies after the 1996 deregulation of US radio industry. Likewise, a working paper version of \cite{stahl2016effects} looks at revealed preference and mergers for the market for TV.\footnote{The working paper version of this paper was dynamic, but this was dropped in the published version.}

\subsection{Dynamic pricing}

The standard version of the \cite{Ericson1995} model assumes that price competition is static. This is a convenient assumption, both for computation and estimation, as it allows estimating parameters in demand and marginal costs using static methods. However, there are often good reasons to believe that firms' pricing decisions are dynamic and forward looking. Dynamics in demand, either because of durability or storability of products, introduces important forward looking considerations in pricing. We have already seen this in section 
\ref{section gg} in the model of \cite{Goettler2011}, where PCs/microprocessors are durable products and current prices affect 
consumers' replacement decisions, and therefore, future demand and profits. As mentioned in that section, \cite{Esteban2007} and \cite{gowrisankaran2012dynamics} are also good examples of applications with dynamic price competition because of product durability. \cite{hendel_nevo_aer_2013} estimate a dynamic game of firms' intertemporal price discrimination under product storability and consumer stockpiling using supermarket data on two-liter bottles of Coke and Pepsi. Other factors that introduce dynamics in pricing decisions are price adjustment costs, firm inventories and capacity, network effects, and learning. In this section, we review empirical applications of dynamic pricing games that incorporate these factors.

\subsubsection{Competition with price adjustment costs}

Dynamic models of price competition with price adjustment costs -- or more generally, \textit{sticky prices} -- have a long tradition in IO (\cite{rotemberg_jpe_1982}, \cite{gertner_1985}, \cite{rotemberg_saloner_aer_1987}). They are also among the first empirical applications of dynamic structural models in IO. \cite{slade1998optimal} proposes a model where the demand for a product in a store depends on a stock of goodwill that accumulates over time when the store charges low prices, and erodes when the price is high. The model incorporates also menu costs of changing prices. \cite{aguirregabiria1999dynamics} studies the relationship between inventories and prices in supermarkets. He proposes a model where retailers have lump-sum costs of placing orders, such as menu costs of changing prices, face substantial demand uncertainty, and experience stockouts. In addition, this paper was pathbreaking in that it was one of the first applications of the Hotz-Miller approach to firm decisions. \cite{kano2013menu} shows that strategic complementarity in price competition, together with menu costs, implies that firms may decide not to respond to firm-idiosyncratic shocks because they know that their competitors will keep their prices constant. \cite{mysliwski2020welfare} study price competition between manufacturers in the UK butter and margarine industry. They propose and estimate a dynamic game that incorporates both consumer brand switching costs and firms' price adjustment costs. \cite{ellison2018costs} study price competition between online sellers of computer components. They propose and estimate a dynamic game where managers have costs of acquiring information about other firms' prices, and of changing their own price.

\subsubsection{Limit pricing}

There has been a long concern in IO about the possibility that incumbents can deter entry. One such approach is limit pricing: setting a low enough price so that a potential entrant would find it unprofitable to enter the market. However, it is not clear that this type of strategy is subgame perfect, since an incumbent would not want to choose this low price in the subgame where the entrant comes in. 

A theoretically grounded motivation for limit pricing was provided by \cite{milgrom1982limit}. In their model, potential entrants have incomplete information about the cost of incumbents: they observe a signal for these costs but do not know them perfectly. Incumbents' prices contain information about their costs, and therefore potential entrants use also this information when they make their entry decisions. If a potential entrant believes that incumbents' costs are low enough, its best response is not to enter. This leads to the possibility that incumbents choose prices that are lower than  those they would choose in the absence of entry deterrence motives. While the \cite{milgrom1982limit} is certainly a reasonable way of rationalizing limit pricing behavior, it is not clear if this mechanism is relevant in actual markets. In particular, this mechanisms requires a reasonably large amount of imperfect information about costs, and that repeated interaction between firms should not uncover too much of this information over time. Empirical applications have provided evidence on the relevance of limit pricing in different industries.

An interesting piece of empirical evidence on limit pricing comes from the more reduced form work of \cite{goolsbee2008incumbents} on limit pricing in the airline industry. They use a novel source of exogenous variation in the probability of entry, based on Southwest entering the airports of two endpoints of a route, say starting to fly to Jacksonville and Tampa, but not currently serving the Tampa to Jacksonville route. They then observe incumbent airlines lowering their pricing in response, not to the actual entry of Southwest Airlines, but to the raised potential entry of Southwest. This is credible evidence that limit pricing motives are quantitatively relevant in the airline industry.

To fully bridge the model of \cite{milgrom1982limit} with the evidence from \cite{goolsbee2008incumbents}, one needs to build a structural model of limit pricing, and this is the purpose of \cite{sweeting2020model}'s paper. They study airline pricing in 109 routes with a dominant incumbent airline that faces potential entry from Southwest during 1993-2010. Air travel is a natural application as many (smaller) routes have a dominant incumbent with (changing) private information about its operating cost. For instance, incumbent hub-and-spoke carriers have private information about the profitability of connecting traffic. 

\cite{sweeting2020model} build a novel and analytically tractable model of dynamic limit pricing. There is an incumbent (with index $I$) and a potential entrant (with index $E$). The incumbent has a constant marginal cost $c_{I,t}$ that evolves according to a first order autoregressive process. Firms' products are differentiated, and consumer demand has a nested logit structure as in 
\cite{berry1994estimating}. Demand is common knowledge to the incumbent and the potential entrant. Every period $t$, the incumbent sets its price $p_{I,t}$. The potential entrant (i.e., Southwest) does not know the incumbent's cost $c_{I,t}$, but it uses the history of incumbent's prices and Bayesian updating to construct beliefs about this cost. Every period, the potential entrant decides whether to enter the market. In the absence of asymmetric information and entry deterrence motives, the model is a standard static model of Bertrand competition in a differentiated product industry. Note that, because the incumbent's marginal cost changes randomly over time, asymmetric information does not disappear over time for a potential entrant that never enters: it never perfectly learns the incumbent's cost from watching it's pricing.\footnote{This is in contrast to \cite{jovanovic1982selection}'s model of entry and asymmetric information, where a firm's type is constant over time.} \cite{sweeting2020model} show that this dynamic game has a unique fully-separating MPBNE for any finite number of periods. This equilibrium can be easily computed, enabling empirical implementation. 

The key parameters that need to be estimated relate to the distribution of entry costs and the variation in the incumbent's marginal cost $c_{I,t}$. In particular, the more variation in marginal costs for the incumbent, the greater the asymmetric information, and thus, the larger the incentives for limit pricing. Likewise, if marginal costs vary considerably over time, this information asymmetry does not shrink much over time, and \cite{sweeting2020model} find considerable variation in costs over time as inferred from pricing choices.

The estimation of this model shows that limit pricing substantially lowered prices, increasing consumer surplus by \$600 million and increasing total welfare by \$500 million on the 109 small routes studied. Subsidizing entry can have substantial welfare benefits, e.g., a subsidization program costing \$8,000 would have increased consumer welfare by \$9.7 million while lowering incumbent profit by \$4.7 million. The authors also present several extensions of the basic model, including “two-way learning” (in which the incumbent also learns over time about the entrant's cost), and endogenous evolution of marginal cost that depends on endogenous capacity and demand for connecting traffic. Perhaps most importantly, this paper shows that the limit pricing model of \cite{milgrom1982limit} may have real bite in certain markets. 

\subsubsection{Dynamic pricing with network effects}

The dynamic consequences of indirect network effects (\cite{katz1985network}) have been studied empirically, most extensively for video game consoles. Consumers purchase video game consoles not only because of the intrinsic value of this equipment but also because they grant access to libraries of video game titles. Moreover, software developers release products based on their expectation of console sales. This creates strong indirect network effects, where software developers value popular consoles, and consoles are more popular among consumers if more software has been developed for them. This sets up a dynamic game with positive spillovers in the actions of consumers and developers. For instance, a manufacturer that introduces a new video console has an incentive to start fixing a very low price -- it could be even below marginal cost -- to build a substantial group of clients that generate positive spillovers and future demand for its product --- echoing the dynamic incentives in \cite{Benkard2004}. The process of increasing a firm’s market share because of indirect network effects is denoted ``tipping.''

\cite{dube2010tipping} study dynamic price competition and tipping in the market of video game consoles. There are $N$ video console manufacturers indexed by $j$. Every period $t$, consumers who have not yet purchased a console choose between not purchasing (waiting at least one more period) and purchasing one of the $N$ consoles in the market. The value of purchasing a console $j$ is $\omega_{jt} + \delta_{j} - \alpha \text{ } p_{jt} + \xi_{jt} + \varepsilon_{h,jt}$, where $p_{jt}$ is price, $\xi_{jt}$ is a demand shock, $\delta_{j}$'s are brand fixed effects, $\varepsilon_{h,jt}$ is a logit consumer specific shock, and $\omega_{jt}$ is the expected and discounted value of using this product in the future, which is equal to $\mathbb{E}_{t}(\sum_{s=0}^{T} \gamma \text{ } n_{j,t+s})$ where $\gamma$ is a parameter and $n_{j,t+s}$ is the number of video game titles available for console $j$ at period $t+s$. Consumers are forward looking, and have rational expectations about future prices and titles of each console. 

In a model with indirect network effects, the number of titles $n_{jt}$ depends on the cumulative number of consumers who have purchased this console, represented by variable $y_{jt}$. The authors do not model explicitly (structurally) the behavior of software developers, and instead consider a simple reduced form function that relates $n_{jt}$ and $y_{jt}$.\footnote{As usual, this shortcut implies that some counterfactual experiments that modify structural parameters in consumers' demand or in the costs of console manufacturers can have an impact on the behavior of software developers, but this reduced form equation cannot account for this effect. This is the well-known \textit{Lucas critique} of reduced form models.} The state of the market at period $t$ is $\mathbf{x}_{t} = (y_{jt}, \xi_{jt}: j \in \mathcal{X})$, which is also the vector of state variables in consumers' and firms' dynamic decision problems. Every period $t$, firms choose the price of their video consoles to maximize expected and discounted intertemporal profits. They have an incentive to lower the price of their consoles, at least initially, in order to \textit{tip} market shares to their benefit.

The set of model parameters consists of the demand parameters, manufacturers' marginal costs, and the discount factors of consumers ($\beta_{c}$) and firms ($\beta_{f}$). The authors estimate demand parameters using the demand part of the model, the simulation-based estimation method in \cite{hotz1994simulation}, and fixing consumers' discount factor at $\beta_{c} = 0.90$. The estimate of the network effect parameter, $\gamma$, is positive and statistically significant. The economic significance of these network effects is evaluated using counterfactual experiments. The authors do not estimate marginal costs but instead use estimates from industry reports. Based on these parameters, the authors solve for an equilibrium of the dynamic game under different levels of network effects as measured by parameter $\gamma$, including the estimated value and $\gamma=0$ (no network effects). Firms' discount factor is fixed at $\beta_{f}=0.99$ under the argument that firms are more forward looking than consumers. These numerical experiments show that tipping is not a necessary outcome even if indirect network effects are present, but it appears when $\gamma$ becomes large enough. They also show that tipping can lead to a substantial increase in market concentration of 24 percentage
points or more. 

\cite{lee2013vertical} also estimates a structural model of network effects in the video console industry.  In contrast to \cite{dube2010tipping}, Lee's model also endogenizes the behavior of software developers. This paper focuses on the role of exclusive titles and vertical integration in the evolution of the market for consoles, but does not model the console's pricing decision. Exclusivity can either harm consumers by restricting the availability of software, or aid them, by making it easier for a new platform to enter the market. 

Lee's model is similar to the one by \cite{dube2010tipping}, but with three main differences: (i) the value to consumers of a particular portfolio of titles is explicitly included in consumer utility; (ii) the portfolio of titles for a platform/console is an endogenous variable that is the result of software developers' decisions;  (iii) software developers choose in which platforms/consoles to release their titles, including the option of releasing titles on multiple platforms, and (iv) platform pricing is exogenous in this model. Both consumers and console manufacturers are also forward looking.

Most of Lee's model is estimated using tools of dynamic demand. However, the software developers' porting costs are estimated using a moment inequality approach. The profits of the observed console porting choices must be greater than alternatives, say either the choice of only release on the PlayStation versus porting the title for release on both the PlayStation and the Xbox. The assumption that Lee makes is that the path console purchases and software releases is only affected by the changes in the lifetime hardware utility from these release decisions. Thus, Halo understand that if it chose to port to the PlayStation, this would alter the number of consumers who would adopt the console and future title availability. Lee's set estimate of porting costs is between \$150,000 to \$200,000, which is understandable given the large number of relatively unpopular games that get ported to multiple consoles. However, these small porting costs make the decision to release a popular game on a single console relatively unusual in the absence of exclusive contracts.  

\subsection{Regulation}

In this section, we review applications that study how regulations can have effects on firms' dynamic incentives, such as the environmental regulation studied by \cite{Ryan2012}. Note that a central aspect of regulation for IO economists, antitrust, has already been discussed in section \ref{sec:antitrust}. 

\subsubsection{Environmental regulation}

\cite{fowlie2016market} build on \cite{Ryan2012} to assess the efficacy and efficiency of various proposals to curb greenhouse gas emissions while guarding against \textit{emissions leakage}, which is the migration of pollution from regulated to unregulated jurisdictions. Using the US Portland cement industry as a backdrop, the authors examine several policy designs for allocating pollution permits in a cap-and-trade system where the economic environment is complicated by both market power and imports from unregulated jurisdictions. As \cite{buchanan1969external} points out, completely internalizing an externality through a Pigouvian tax is generally inefficient when firms have market power as they are already reducing their output. Additionally, when only a subset of sources are regulated (\textit{incomplete regulation}), the regulated firms are placed at a competitive disadvantage such that, as compliance costs increase, supply shifts from the regulated to the unregulated firms, potentially creating the ironic outcome of increasing overall pollution. Policymakers have sought tools to balance these countervailing forces. In this paper, the authors consider the long-run effects of allocating permits to domestic firms via four different mechanisms: a permit auction (equivalent to a carbon tax); a grandfathering scheme where a fraction of permits are given for free to incumbent firms based on, say, historic emissions levels; a dynamic updating scheme where permits are allocated each period in proportion to output or emissions in the last period; and finally, a border tax adjustment which penalizes imports according to their foreign carbon intensity.

The authors assume that the price of a pollution permit is equal to the social cost of carbon and consider a range of damages. They find that all four allocation mechanisms result in social losses for social damages below \$40 per ton of carbon dioxide. This is driven by the sum of losses in the product market and emissions leakage exceeding the benefits of the carbon abatement. The largest losses occur when firms have to bear the full cost of compliance, under the auctioning and grandfathering mechanisms. Policies that allocate permits on the basis of emissions or production do much better, since they (partially) address the welfare losses that are driven by reductions in domestic output. In all cases, welfare effects are magnified by firm exit, particularly when regulated firms do not have any compliance cost assistance. When damages are above \$40 per ton, dynamic permit allocations and the border tax adjustment scheme both result in social welfare gains. 

To highlight the effects of accounting for dynamics, the authors decompose their welfare measures into product market surplus, emissions reduction, and emissions leakage for both the static and dynamic case. In all cases, the static estimates look better than the dynamic estimates, as they miss the changes to market structure that the various permit mechanisms induce. This paper highlights the important role that dynamic games frameworks can play in assessing substantive problems in environmental regulation.

\subsubsection{Land use regulation}

\cite{Suzuki2013} considers the dynamic effects of land-use regulation (``zoning'') on business entry. Zoning restrictions may constrain building characteristics or uses within a certain geographic area. Examples include banning certain exterior materials (e.g. no visible siding from the street), requiring buildings to conform to stylistic templates (e.g. in historic districts all new buildings must look original to the neighborhood), limiting which types of businesses can operate (e.g. restricting commercial and industrial operations to be distant from residential areas or where alcohol-serving establishments can be located), and capping how tall buildings can be. 

Suzuki focuses on the effect of land-use regulations on mid-scale chains in the Texas lodging industry. This industry is a promising setting for examining these regulations: land-use regulation is a first-order cost component for hotels, competition is local, and the author argues that agents in this industry are aware that land-use regulations can serve as effective barriers to entry. Data on hotel revenue comes from quarterly taxation data collected on every hotel in Texas. Suzuki constructs firm-market revenue functions that depends on market characteristics, chain characteristics, and the degree of competition in the market. A key input to the study is the use of a land use regulation index to proxy for the stringency of regulation. He follows \cite{gyourko2008new}, who produces a range of residential land-use intensity indices (commercial indices were not available for his analysis) based on a survey of local governments. These indices include factors like the average number of months developers wait to receive building permits, whether there are density restrictions, and if developers have to pay for infrastructure upgrades related to their projects. The author focuses on six chains that account for 90 percent of the mid-scale chain hotels in Texas and defines the relevant market as a county. Similar to \cite{bresnahan1990entry}, he restricts his analysis to counties that have data on land use regulations, are not located in the four largest urban regions of Texas, have at least 50,000 residents, and have had at least four openings and closings during the sample period. Of 254 counties in Texas, this filters out all but 35. 

Suzuki builds a model of entry and exit in this industry for hotel chains. Players are chain hotel operators in the mid-scale segment. State variables include the number of hotels operated by each firm in the market and exogenous market-level characteristics such as population. Firms open and close hotels in local markets in each period to maximize expected discounted profits, with firms paying stochastic entry and exit costs. Since he observes (accounting) revenues directly, he also posits that firms pay fixed costs to operate. The author uses the method in \cite{Bajari2007} to recover the distributions of entry and exit costs. The revenue function is estimated using OLS, and the policy functions are estimated using a multinomial logit. In addition to the standard two-step procedure for estimating dynamic parameters, the author includes a third step where he regresses market-level cost estimates on the land-use indices to decompose which regulations drive costs.

Suzuki finds that the average hotel pays approximately \$250,000 each quarter per hotel. The cost of opening a new hotel is estimated to be \$2.4 million, with substantial heterogeneity by chain. These numbers are roughly in the range of what industry sources report as building costs, although this comparison is tempered again by the issue of accounting versus economic costs. Interestingly, the costs appear to be much lower for one of the chains; this may be driven by the imprecision of the policy functions for that chain, as it had relatively few entries and exits during the sample period.

The third-stage regression of operating and entry costs on land use stringency are unfortunately imprecise. This is mainly because all regulatory variation is cross-sectional and the set of markets in the estimation is very modest. To try to discern some deeper insights into what is happening, Suzuki runs several counterfactual experiments. Limiting the analysis to a subset of three counties and capping the number of active hotels per chain at three to make computation feasible, he solves the dynamic model and simulates outcomes under lenient (costs one standard deviation lower), observed, and stringent regimes (costs one standard deviation higher). He finds that the number of active hotels ranges by about one active hotel between the two extreme counterfactuals.

This paper illustrates both the promise and limitations of the dynamic games literature. The research question is very interesting, as land-use regulations are plausibly a first-order determinant of firm density, variety, and location. Indeed, a structural model of housing supply which is realistic, and as a function of this realism needs to be estimated from data, would be a huge innovation in the literature in urban economics. However, even with excellent data on revenues of all players in each market, which is data that is typically hard to come by outside of regulated settings, estimating a link between regulation and market outcomes is very difficult. Part of that is driven by the need to have clean market definition, which in this setting resulted in discarding almost all of the data. Some of the markets were thrown out for being too small and/or not having enough variation; we note that this is in some sense the exact opposite of the ideal market in
\cite{bresnahan1990entry}, where ideally one wants to observe the stable, unchanging long-run market configuration. This concern is amplified by the relatively high data requirements for estimating dynamic parameters; it is not sufficient to have high-quality data, it is also necessary to see a sufficient amount of variation across all actions that have associated parameters. That problem is exacerbated by the very nature of the strategic interactions that these models focus on: the fewer firms, the more market structure matters, but the lower the probability of observing sufficient variation to precisely recover underlying dynamic primitives.

\subsubsection{Product variety \label{sweet radio}}

Market structure consists not only of the number of firms and products, but also which goods they offer. \cite{Sweeting2013} examines the response of radio stations to the introduction of additional licensing fees for playing music. This paper is an example of how the primary welfare effects of a policy may be driven by a change in the types of products firms offer rather than through prices. The paper considers the effects of the Performance Rights Act of 2009 in the US, which stipulated that radio broadcasters should pay performance rights in addition to the composition rights that they already paid. The fees would convert to a flat rate for stations with revenues above a certain cap, while noncommercial and talk stations would be exempt. Ambiguity in the legislation led to the possibility of performance fees as high as 25 percent of advertising, an order of magnitude above composition fees. Sweeting develops a structural model to estimate the propensity of firms to switch from music formats to non-music formats as a result of these fees.

Radio stations are a good place to examine the effects of fees on product variety for several reasons. First, markets are local due to limited broadcasting range (echoing the isolated markets of \cite{bresnahan1990entry}), radio stations fall into generally easily-definable segments (e.g. classical, rock, country, top 40), spectrum constraints restrict the number of active radio stations in each market, and demographics vary widely across the sample, providing strong demand-side instruments. Heterogeneity in customer demand is important for two main reasons: match-quality between listener and station depends on tastes, which vary with observable demographics, and advertisers value different listeners by demographics. In a string of papers that use either regressions or a static model of entry, \parencite{waldfogel1999free,berry2001mergers,berry2016optimal} investigate issues of product variety in the radio industry such as the effect of mergers and free entry on product variety.  

The players in Sweeting's model are radio station operators in local markets. Each firm has a per-period profit function that consists of advertising revenues, fixed cost savings from operating several stations in the same format, repositioning costs that are incurred when a station changes format, and a vector of action-specific private information shocks. Advertising revenues are a function of listener demand, which is modeled using a discrete choice model. Utility is dependent on demographics, which are modeled to be slowly changing over time, introducing additional dynamics into the model. Finally, station quality is assumed to evolve according to an exogenous AR(1) process.

Sweeting estimates his model using a combination of methods, primarily variants of \cite{Aguirregabiria2007}, with a set of robustness checks following \cite{Bajari2007}. He estimates demand statically, recovering a rich set of preferences interacting demographics and station characteristics. For example, Black listeners have much higher marginal utility for urban formats than for country, while Hispanic listeners have particularly strong preferences for Spanish-language stations. Per-listener revenues are estimated as a function of demographics using a linear regression. The author reports four different sets of estimates, according to which estimator they came from: two variants of the pseudo-likelihood from \cite{Aguirregabiria2007}, one following \cite{Pakes2007}, and one using the forward-simulation approach of \cite{Bajari2007}. One of the key innovations that he has to make in the pseudo-likelihood method is the use of an approximation to the value function using basis functions. The first three approaches give roughly similar estimates for most of the coefficients in the model. Unsurprisingly, given how the estimators use statistical information from the model, the main difference is that the \cite{Pakes2007} estimator tends to be less efficient than the other two approaches. Sweeting reports estimates from the forward-simulation exercise using two different objective functions. The original BBL function is the squared error of inequalities that violate the optimality condition of qquation \ref{eqn:RyanInequality}, at all states. The second approach is inspired by \cite{pakes2015moment} (PPHI), where the optimality condition has to be true on average across states. The PPHI estimator trades off a loss of statistical information against the possibility of being more robust as it uses averaging, which may smooth out approximation errors in the estimated policy functions and simulation error from forward simulation. A second tradeoff is that the PPHI approach produces set-identified estimates, as there are only six inequalities. Comparing the estimates from the two forward-simulation estimates, Sweeting finds that the BBL point estimate lies outside the 95 percent confidence interval estimated by PPHI for six of the nine parameters, which may indicate that the BBL estimates are biased in his sample. He ends up using the PML estimates in the counterfactuals, as they generally were consistent with the estimates from PPHI. 

Sweeting simulates market evolution under two counterfactual fee schedules. He computes equilibrium outcomes for all markets when fees are set at zero, ten, and twenty percent of advertising revenues. He simulates out forty years and collects various summary statistics. The first finding is that a significant percentage of music-playing stations switch to non-music formats when fees are imposed, especially at the 20 percent level. He estimates that 578 stations would still be playing music after 40 years, starting from a base of 713. There is heterogeneity across formats, with the Urban format losing the least stations. Non-music formats gain in all three settings, partially reflecting demographic changes that increase the number of consumers that have preferences for Spanish-language stations. Much of the change takes place within five years, but the industry is still adapting at the forty-year mark. Indeed, the ability of dynamic models to assess the speed of adjustment to a new steady-state is an important improvement over the previous static literature on radio and product variety. The general takeaway is that which products are offered can be sensitive to policy choices. While no welfare numbers are reported, the loss of choice may have significant implications on consumer surplus and producer profits.

\subsubsection{Industrial policy}
\cite{kalouptsidi2018detection} examines the effects of government subsides in China on firm entry into shipbuilding, which is upstream to her previous work on the global shipping industry in \cite{Kalouptsidi2014} discussed in section \ref{Section: uncertainty} below. This is an interesting use of the dynamic games framework, since state-directed subsides are not publicized, as they may be in violation of international trade agreements, but they can be inferred using data on firms' actions and the principle of revealed preference. The author uses a dynamic framework to detect the subsidies and infer their size, and then computes a counterfactual world without the subsidies in order to calculate their incidence across domestic and foreign producers. There are dynamics on both the consumer and supply sides of the market for ships. Shipyards have backlogs that accumulate over time; there may be congestion (negative costs) or learning-by-doing (positive benefits) associated with these backlogs. On the demand side, ships are long-lived capital investments, and consumers consider expectations about future states of the world (including shipping demand and the evolution of shipping fleets) before making purchases, much like the dynamic demand model used in \cite{Goettler2011} for PC microprocessors, or \cite{gowrisankaran2012dynamics}'s study of the digital camera market.

The basic empirical strategy is to estimate cost structures in this industry before and after 2006, when China identified shipbuilding as a ``strategic industry'' in need of ``special oversight and support.'' Aggregate statistics show a large change after 2006, with a large amount of entry into the sector in China and a significant increase in Chinese market share. Variation in the cost structure of Chinese firms before and after 2006 is inferred to be the result of state-sponsored subsidies. The key identifying assumption, as in \cite{Ryan2012}, is that the subsidy policy was an unforeseen, permanent, and immediate change. 

The estimation method is a variant of the Hotz-Miller two-step method. A technical innovation in the paper is the use of sparse approximation techniques from the machine learning literature, namely, LASSO, to allow for a very large state space. There is a large number of state variables in firms' decision problem, such as the age distribution of the current fleet and the backlog of different shipyards. Doing a basis function approximation to the value function requires one to consider interactions between state variables, which could yield a basis with thousands of components. This makes a dimension reduction technique very attractive, and this particular paper uses LASSO to do so, even if the combination of value function iteration and LASSO is not well understood to our knowledge, with the closest papers in the economics literature being \cite{Arcidiacono2016}. 

The primary finding of the paper is that Chinese costs declined 13 to 20 percent, or 1.5 to 4.5 billion dollars, after 2006. She does not find similar declines for firms in other countries, which lends credence to the assumption that subsidies were behind the shift. What is important here is not only the direction of the change in subsidies --- these could be read off in part from the change in the Chinese market share over time --- but the magnitudes that are implied. 
With estimates of the cost structure in hand, she performs two primary counterfactuals. In the first, she removes all subsidies and simulates the resulting equilibrium. In the second, she removes investment subsidies but keeps cost subsidies; this helps us understand the relative importance of the two subsidies. Without any subsidies, she finds that the Chinese shipbuilding industry's market size would be half as large. The primary beneficiary would be Japan. Market prices would be higher as the subsidies shifted out the supply curve. The customers of this industry, oceangoing shippers, gained about 400 million dollars in surplus as a result of lower prices. These gains are relatively minor compared to the estimated four billion dollar cost of the subsidies. Finally, she estimates a significant allocative inefficiency as production shifted from low-cost Japanese shipyards to higher-cost Chinese firms. She does find evidence that there is significant learning-by-doing in this industry, which is often a stated rationale for subsidies.

\subsection{Retail \label{retail}}

In this section we discuss a variety of papers in the retail trade sector. In many ways, this part of the economy is suited for the the type of cross-market identification of the \cite{bresnahan1991entry} approach, as retail is a non-tradeable sector, so interactions between firms really are local. 

\subsubsection{Economies of density and cannibalization}

Walmart, the retail giant, started with a single store in Bentonville, Arkansas in 1962. It has since grown to over 3,000 stores in the United States. A startling fact is that Walmart always opened new stores near old ones; it never jumped to a distant location and then filled in the markets in between. \cite{Holmes2011} studies these patterns, posing and estimating a model of store location that accounts for two important countervailing economic forces: on one hand, placing a new store near an old one can lead to cannibalization of sales from the old store. On the other hand, Walmart experiences economies of density due to the use of regional distribution centers where large stocks of items are kept. Keeping stores close to distribution centers cuts down on trucking costs and speeds up restocking times. Using data on store-level sales, Holmes is able to estimate a significant negative cannibalization effect, while he uses a profit-maximizing revealed preference argument to bound the benefits of economies of density. The basic strategy is to perturb the sequence of store openings; under the assumption that the observed policy (which stores to open where and in which order) is optimal, alternative sequences should generate lower profits. Holmes finds that economies of density are significantly positive: locating a store one mile closer to a distribution center reduces annual costs by approximately \$3,500. Given the scale of Walmart's operations across the country, these economies of density play a key role in Walmart's successful business model.

Holmes focuses on the decision about where and when to open two types of stores: regular stores that sell general merchandise and supercenters that also sell groceries. He takes other choices, like how many stores to open and analogous decisions about distribution centers, as given. There are $L$ possible locations for these stores, consisting of all the census blocks in the US (approximately, 11 million locations). Let $a_{\ell t}^{s} \in \{0,1\}$ be the indicator of the event ``Walmart has a store of type $s \in \{regular, supercenter \}$ in location $\ell$ at year $t$,'' and let $\boldsymbol{a}_{t} \equiv \{a_{\ell t}^{s}: \ell=1,2,...,L; s=regular, supercenter\}$ be the vector describing the map of Walmart's stores in the US at period $t$. The heart of the analysis is the present discounted profits associated with a sequence, $(\boldsymbol{a}_{1}, \boldsymbol{a}_{2}, ..., \boldsymbol{a}_{t}, ...)$ of store openings:\footnote{In this paper, Walmart's store openings are assumed irreversible. Closing a store is not possible. It has not been until recently that Walmart started closing stores. Note that the irreversibility of the entry decisions implies that we cannot use the finite dependence properties described in section \ref{sec:finite dependence} to derive relatively simple optimality conditions for the estimation of the model.}
\begin{equation}
    \max_{(\boldsymbol{a}_{1}, \boldsymbol{a}_{2}, ..., \boldsymbol{a}_{t}, ...)} 
    \sum_{t=1}^{\infty} (\rho_t \beta)^{t-1} 
    \left[ 
        \sum_{s} 
        \sum_{\ell=1}^{L}
            a_{\ell t}^{s} \text{ }    
            \left( 
                R_{\ell t}^s - VC_{\ell t}^s 
                - f_{\ell t}^s -\tau \text{ } d_{\ell t}^s 
            \right) 
    \right].
\end{equation}
Profits of a store type $s$ in location $\ell$ consist of revenue, $R_{\ell t}^s$, variable costs, $VC_{\ell t}^s$, exogenous fixed costs, $f_{\ell t}^s$, and economies of density, $\tau \text{ } d_{jt}^s$, where $d$ is the distance to the nearest distribution center. The discount factor has two terms in it: the usual intertemporal discount rate, $\beta$, and an additional term, $\rho_t$, that captures the fact that per-store revenues are growing over time. Revenues $R_{\ell t}^s$ are obtained from the estimation of a nested logit consumer demand model using revenue data from Walmart stores in 2005. This demand system captures cannibalization between Walmart's stores. Variable costs $VC_{\ell t}^s$ are obtained using data labor costs, land value, and price-cost margins at the local level, and calibrating some parameters. The exogenous fixed cost $f_{\ell t}^s$ depends on population density in location $\ell$, $m_{\ell t}$, according to function, $f_{\ell t}^{s} = \omega_{0} + \omega_{1} \text{ } \ln(m_{\ell t}) + \omega_{2} \text{ } [\ln(m_{\ell t})]^{2}$. This feature of the model contributes to explain Walmart's propensity to open stores in locations with low population density. The dynamic structural model is used to estimate fixed cost parameters $(\omega_{0},\omega_{1},\omega_{2})$, and economies of density parameter $\tau$.

Using the inequality average approach from \cite{pakes2015moment} and \cite{Bajari2007}, Holmes considers deviations from Walmart's observed behavior that consist of pairwise resequencing in the opening dates of two stores. For instance, if store number 1 actually opened in 1962 and store number 2 opened in 1964, a pairwise resequencing would be to open store number 2 in 1962, store number 1 in 1964, leaving everything else the same. This is a clever strategy, because Holmes assumes that outside of those swaps, the entire future sequence remains constant. This implies that those future streams of profits difference out, leading to a clean, simple estimator. He considers three broad types of swaps: density-decreasing swaps where he switches the order of an early store located close to a distribution center with a later store that is located farther away; density-increasing swaps which move in the other direction; and population-density swaps which hold density constant but changes the sequence of stores that face different population densities. The target of the first two types of deviations is the economies of density parameter $\tau$, and the target of the third type of deviation is the vector of $\omega$ parameters in the fixed cost.  

Holmes finds a tight bound around \$3,500 per mile as the cost savings of locating closer to a distribution center. A back-of-the-envelope calculation suggests that this is about four times as much as would be implied by trucking costs alone; the remainder could be interpreted as the benefits of increased flexibility to respond to demand shocks.

This moment inequality approach to deal with the complexity of dynamic choice models is pathbreaking, which makes it  surprising that this approach has not really been picked up in subsequent work. One issue is that in models with competing agents, such as \cite{jia2008happens}'s analysis of the entry decision of Walmart in competition with Kmart (but a static analysis), one cannot simply look at a deviation without thinking through how rivals will react to these deviations, both in the current period and in the future. This makes computing deviations substantially more difficult. Moreover, there are relatively few instances where the researcher is interested in payoff parameters per se, without needing to work through their implications on firm behavior. 
   
\subsubsection{Chains}
\cite{hollenbeck2017economic} investigates the role that demand-side factors may play in firms organizing their production into chains---defined in the paper as ``any business that operates multiple outlets offering similar goods or services under the same banner'' --- in the context of the Texas hotel industry, the same industry considered by \cite{Suzuki2013}. On the supply side, firms may form chains to exploit economies of scale and scope. On the demand side, consumers may view chain affiliation as a form of quality signaling in a market for experience goods; rather than take their chances with a single-location motel in west Texas they may decide to go to the nearest Motel 6. Of course, this benefit is not free, as firms have to pay affiliation fees to join chains, so in equilibrium not all firms will join a chain; furthermore, the decision to affiliate may also be a function of market competition and other market-specific factors.

The decision to build a hotel of a certain quality and whether to associate with a chain are both dynamic decisions: significant irreversible costs are incurred today for the promise of higher returns in the long run while accounting for the strategic responses of rivals. The Texas hotel industry provides a nice environment for studying this question: one can reasonably partition the (enormous) state of Texas into a significant number of discrete markets that do not directly compete with each other, \`a la \cite{bresnahan1989empirical}; the state of Texas collects a hotel occupancy tax, which means that high-quality revenue data is available for every establishment in the state; and AAA publishes information about the quality (e.g. number of stars) and characteristics of each hotel, including chain affiliation.

In a first step, Hollenbeck uses this information to build a revenue model as a function of market characteristics, market structure, and chain affiliation. He finds that chain affiliation is associated with a 27 percent premium in revenue per available room. This estimate is slightly lower if estimated from a subset of hotels that switched affiliation during the sample period. In a second step, he recovers costs associated with running a hotel as an independent versus as having a chain affiliation. This paper is one of the very first to use the methods of \cite{Arcidiacono2011} to estimate dynamic parameters in the presence of possible correlated unobservables. Hollenbeck finds that higher-quality hotels have higher costs, but, significantly, chain hotels do not have a cost advantage over independents. Entry costs are estimated to be higher for chain firms, but the difference is on the order of the chain affiliation fee. Accounting for the unobserved heterogeneity is critical to these findings, as failing to do so significantly biases the effect of being a chain on costs. He documents a declining chain premium over time, which is consistent with the idea that the increase in online information about the quality of hotels is substituting for the signalling effect of chain affiliation, further developed in \cite{hollenbeck2018online}.

This paper highlights a number of appealing features of the setting. First, it is close to the ideal data set discussed previously: there are a small number of firms competing in large number of distinct markets; there is high quality data on their product offerings and revenues, partially driven by the fact that a tax authority collects and reports the data; the technology in the industry is relatively simple and slow moving; and finally, the set of dynamic parameters of interest are both relatively small and, perhaps most importantly, transparently and directly connected to moments of the data.

\subsubsection{Unobserved heterogeneity and entry in retail}

\cite{Igami2016} also examine the role of unobserved heterogeneity in dynamic models using the empirical setting of hamburger chains in Canada. Firms appear to prefer to locate in places with many other competitors. The authors argue that this is due to unobserved geographic heterogeneity and not positive spillover effects from being close to competitors, such as higher consumer traffic. Entry and exit in this setting happens at the level of openings and closings of chain stores, more than entire firms entering or leaving the industry. As in \cite{Holmes2011}, there is a cannibalization concern that opening an additional outlet harms sales at extant stores. On the other hand, there may be preemption motives due to the threat of competitor entry. Hamburger chains are a good setting to study these incentives, as there are both many chains and stores that compete in local markets. This generates lots of variation for empirical analysis. Additionally, chains compete primarily on the entry/exit margin instead of prices or product variety, as those are often uniform across chains in a given region for marketing reasons.

The empirical approach in this paper mixes together three separate approaches. First, they use a nonparametric finite mixture method from \cite{Kasahara2009} to recover a minimum number of market types, which vary in their profitability in a way that is not captured by observable variables. Second, they use the estimation technique of \cite{Arcidiacono2011} to obtain firm entry and exit strategies conditional on those market types. Finally, they follow \cite{Bajari2007} and use forward simulation to recover firm profits and the cost of entry. Linking together the number of types from the first step with the latter two estimation steps is a significant methodological innovation. The primary takeaway from this paper is to highlight the empirical bias of ignoring unobserved heterogeneity and proposing methodology to handle it.

\subsubsection{Effect of Walmart on rival grocers}

A complementary paper to the previous two works is \cite{Arcidiacono2016}, adapting continuous time methods to develop a model of retail competition, specifically investigating the effect of Walmart's entry on the retail grocery market. In contrast to Holmes' work, this paper considers the entry decisions of Walmart and seven competing supermarket chains along with a fringe of dozens of single-store competitors. Markets are characterized by population levels and growth rates and are allowed to have unobserved heterogeneity. This specification generates a very high-dimension state space---up to 157 million states across 205 markets. They adapt continuous time methods to deal with the the problem of computing equilibria in a such a large state space. In contrast to the general merchandising retail sector, they estimate that Walmart's primary effect was on other grocery chains rather than on independent grocers---in fact, they estimate that independents actually benefit from Walmart's entry via a change in product market competition. As with \cite{Igami2016}, unobserved heterogeneity is a key input to obtaining unbiased estimates of Walmart's effect on competitors; without it, independent grocery stores would have been uniformly worse off after Walmart's entry.

\subsubsection{Exit in declining industries}

\cite{Takahashi2015} looks at strategic exit when demand is declining. Strategic delay may lead to suboptimal outcomes, as firms have incentives to free ride on the capacity reductions (or at the extreme, exit) of their competitors, as studied at a theoretical level by \cite{ghemawat1985exit} or \cite{fudenberg1986theory}. In the presence of uncertainty, there is also a real option value of waiting for more information before making an irreversible decision to exit. Takahashi studies the US movie theatre industry in the 1950s, which was facing a combination of a long-run decline in demand due to the increased adoption of home televisions and a large initial stock of theaters. The author compares the profits that firms earn in the observed equilibrium against two counterfactuals. In the first, firms are non-strategic and exit when operating profits are equal to fixed costs ("coordination benchmark"). The difference in profits between this outcome and the observed data is interpreted as the cost of strategic behavior. In the second counterfactual, firms exit in a coordinated fashion to maximize total industry profits ("regulator benchmark"). The difference in profits between this counterfactual and the coordination benchmark is interpreted as the cost of oligopolistic competition. He finds that the delay in exit from strategic interactions is 2.7 years on average. Less than four percent of that delay is due to strategic behavior, while 96 percent is due to oligopolistic competition. This implies a loss of a little less than five percent of optimal profits in the median market.

Compared to much of the dynamic games literature, this paper has some unique features. The first is that the model is a modified version of \cite{fudenberg1986theory}, who provide a theory of exit in duopoly with incomplete information. The model is in continuous time. Theaters are endowed with a time-invariant fixed cost of operation but do not know the costs of their competitors (although they know the common distribution that generates fixed costs for all players). This generates a strategic motive to delay exit, as it is possible that some of their competitors will exit instead and residual profits increase. This is balanced against the cost of delay, as revenues are declining over time. In the unique equilibrium of this game, firms exit in the order of their fixed costs from high to low. Another advantage of this approach is that the computational cost of finding the unique equilibrium is low, so Takahashi can utilize a full-solution estimation approach. Another difference from much of the literature is that the focus is on exit alone; a benefit of this approach is that the set of potential exiters in each period is observed, as opposed to the ad hoc modeling assumptions about the pool of potential entrants that is typically required in other settings. Indeed, in general it is easier to model counterfactuals with exit rather than entry for this reason. 

\subsubsection{Repositioning}
\cite{ellickson2012repositioning} consider repositioning decisions of retail firms. As they highlight, the firm's choice of where to position itself in the market (e.g., which segments to compete in, how it brands itself, or which pricing strategy it uses) is a dynamic decision with costs that may exceed the costs associated with entry, investment, or exit in a given market. For example, if McDonald's decided to become a chain of upscale French bistros, it would face significant repositioning costs in moving away from its current branding and business practices as a family-friendly fast food restaurant. Those costs are likely larger, and therefore more strategically important, than decisions relating to entry or exit in marginal markets. Furthermore, these costs are also likely much higher for existing incumbents than new entrants, who do not have to overcome existing brand capital when deciding how to position themselves. \cite{ellickson2012repositioning} focus on supermarket pricing strategy, specifically the choice between relatively static ``everyday low prices'' (EDLP) and ``promotional pricing'' where prices vary periodically due to discounts. The authors leverage the entrance of Walmart Supercenters, which follow the EDLP strategy, as a shock event to local market structure. The authors estimate the set of payoffs associated with each pricing format, conditional on competiting firms' decisions, and use the gains/losses associated with switching pricing strategies to infer repositioning costs. They find that repositioning costs are both large and asymmetric among formats. Moving to promotional pricing from EDLP is associated with a \$2.3 million cost, while the reverse is estimated at about six times as much. Conditioning on competitive conditions is also important, as the presence or absence of Walmart is a significant driver of profitability across and within pricing strategies.

The specification of revenue and cost functions includes dummy variables for whether the chain focuses on EDLP or promotional pricing. However, the model assumes that the decision of pricing format is market-specific and there are not spillover effects across markets. This seems a rough way of capturing economies of scale or scope in the choice of pricing format between stores of the same chain. This restriction is imposed to avoid having to solve for the equilibrium of a dynamic game over all the markets, would be computationally very challenging. The authors also mention the need to model unobserved heterogeneity more systematically; the literature is finally making strides in that direction almost a decade after this paper was written.

\subsubsection{Advertising}

\cite{dube2005empirical} is an interesting twist on the standard empirical approach taken by most papers. They assess the question of whether ``pulsing'' in advertising, that is, periods of promotions followed by periods of regular pricing, can be sustained in a Markov-perfect equilibrium. Their goal is not to estimate structural parameters and perform a counterfactual, per se, but rather to, one, illustrate that such a strategy is feasibly profitable, and two, illustrate some conditions on the demand-advertising relationship that are necessary for that profitability. This echos the goal of \cite{Benkard2004} or \cite{sweeting2020model} in showing the pricing consequences of a dynamic model with learning by doing and limit pricing. The study these questions, the authors look at the market of frozen entrees, which is an advertising intensive industry and studied in \cite{sutton1991sunk}, for example. 

The paper proceeds in two distinct steps. First, using a long panel of consumer purchases, they estimate demand for frozen entrees as a function of price and advertising. The length and richness of the panel allows them to estimate market-specific fixed effects, which helps dealing with issues of unobserved heterogeneity, and also estimate the nonlinear relationship between advertising and demand. Critically, they find that there is a threshold of advertising that is necessary for a demand response to advertising. Below that threshold the impact is negligible; above the threshold, they find persistent effects.

Given some combination of convexity in advertising costs or concavity in advertising's marginal return, the existence of the minimum threshold suggests that "pulsing" can be an optimal strategy: firms engage in relatively high advertising first to build up brand equity, and once declining marginal returns set in, they then find it optimal to not engage in any advertising at all until that brand equity depreciates to a sufficiently low level. To test this conjecture, they take the estimated demand model and calibrated costs using accounting data, and plug them in a dynamic game of pricing and advertising. As the authors state, their objective is not to obtain an in-sample fit of any particular time series of prices, but rather to see if it is possible to generate a pattern of prices. The answer to that question is positive, and they show that the manufacturers in their equilibrium alternate between periods of intensive advertising and periods without any. They also show that this pattern crucially depends on the non-convexity of the advertising return function; without the threshold, firms do not exhibit pulsing behavior.

While this paper has a minimum of estimated dynamics, all of which come through the demand side via the advertising state variable, and supply-side parameters are minimal and calibrated to accounting cost data, it is useful to point out a major strength of their approach: the paper is transparently clear about how non-convexities in demand lead to differences in dynamic outcomes. While the frontier in the dynamic games literature has moved forward, with increasing demands for better identification, higher modeling complexity, and broader scope of counterfactuals, this paper serves as a refreshing counterpoint that simple models with good data can also be useful at elucidating the broader relationship between primitives and equilibrium outcomes.

\subsection{Uncertainty and firms’ investment decisions \label{Section: uncertainty}}

The role of uncertainty on investment, and particularly the option value of waiting, is a fundamental application of dynamic models in economics (\cite{dixit1994investment}). This is a critical channel of public policies aimed at business cycles, since variation in uncertainty not only changes firms' investment choices but the entire relationship between investment and the underlying state space. Much of the recent work in industrial organization has looked into the role of uncertainty in the firm's investment choices, both in oligopoly and competitive contexts.

\subsubsection{Firm investment under uncertainty \label{sec:bloom_framework}}

\cite{bloom2009impact} provides a simple and helpful empirical framework to study the impact of uncertainty on firms' investment. This framework has been quite influential in recent empirical work in IO and macroeconomics. Firms have a Cobb-Douglas sales generating production function, $q_{it} = f(\omega_{it}, k_{it}, \ell_{it}) = \exp\{\omega_{it}\} \text{ } k_{it}^{\alpha_{k}} \text{ } \ell_{it}^{\alpha_{\ell}}$, where $k_{it}$ is capital stock, $\ell_{it}$ is labor, and $\omega_{it}$ is the logarithm of the firm's total factor productivity (TFP) which is a composite of firm level efficiency and demand shocks. The process for the evolution of $\omega_{it}$ is given by an heteroscesatic random walk:
\begin{equation}
    \omega_{it}  = \omega_{i,t-1} +  \ln  \left( 1+ \sigma_{i,t-1} u_{it} \right),    
\end{equation}
where $\sigma_{i,t-1} \text{ } u_{it}$ is a random shock, with $u_{it}$ i.i.d. standard normal, and $\sigma_{i,t-1} > 0$ represents the variance of the shock to productivity. Uncertainty is measured by $\sigma_{t-1}$, and it varies over time according to a Markov chain with two points of support, $\sigma_L$ or $\sigma_H$, and a transition matrix $\mathbf{F}_{\sigma}$. In this framework, the term \textit{uncertainty shock} means that $\sigma_{t-1}$ has shifted from $\sigma_L$ or $\sigma_H$.

In the absence of intertemporal concerns, a (mean preserving) change in uncertainty would not be particularly interesting, as it would simply affect the variance of investment but not its mean or its intertemporal allocation. Irreversibility, sunk costs, adjustment costs, or time-to-build introduce the intertemporal linkages that can generate rich and interesting effects of uncertainty. In Bloom's model, these dynamic or intertemporal concerns are introduced through adjustment costs in capital and labor inputs. For instance, Bloom considers the following specification for capital adjustment costs:\footnote{See also \cite{cooper_haltiwanger_2006} for a similar specification of capital adjustment costs function and for a structural estimation of the parameters in this function using firm panel data.}
\begin{equation}
    \begin{array}
    [c]{rll}
        AC(\omega_{it}, k_{it}, \ell_{it}, a_{it})
        & = & 
        1\{a_{it} > 0\} \text{ } a_{it} 
        - (1-\theta_{r}) 1\{a_{it} < 0\}  \text{ } a_{it}
        \\
        &  &
        \\
        & + & \theta_{q} \text{ } 1\{a_{it} \neq 0\} \text{ }
        f(\omega_{it}, k_{it}, \ell_{it}) 
        + \theta_{k} \text{ } k_{it} \text{ }
        \left( 
            {\displaystyle \frac{a_{it}}{k_{it}}} 
        \right)^2,
    \end{array}
\end{equation}
where $a_{it}$ is capital investment, and $\theta_{r}$, $\theta_{q}$, and $\theta_{k}$ are parameters: $\theta_{r}$ represents the cost of reselling capital; $\theta_{q}$ captures adjustment costs that do not depend on the investment amount but are proportional to the amount of sales (e.g., because stoppage of the production process to install or disinstall capital equipment); and $\theta_{k}$ captures a quadratic adjustment cost.

\cite{bloom2009impact} estimates the adjustment costs parameters using GMM, by matching moments on the variability of investment to identify the quadratic cost $\theta_{k}$, the frequency of zero investment to get parameter $\theta_{q}$, and the asymmetry around zero in the distribution of investment to get $\theta_{r}$. In the estimated model, uncertainty shocks -- changes from $\sigma_L$ to $\sigma_H$ -- alter the firm's investment policy function. In particular, they increase the size of the inaction region, the states where firms decide neither to invest nor to sell capital. This has an effect on aggregate productivity because it slows the reallocation of factors to more productive firms.

\cite{asker2014dynamic} study how differences in the magnitude of productivity shocks across industries and countries change investment decisions and the alignment of productivity and investment. Echoing the findings in \cite{bloom2009impact}, greater uncertainty on productivity, as measured by $\sigma_c$ (where $c$ indicates a country or country/industry), leads to lower alignment between productivity and capital. Through the lens of \cite{hsieh2009misallocation}'s model, this may appear as more misallocation in capital, but in fact, in this very model there is no inefficiency whatsoever. Furthermore, differences in measured $\sigma_c$ explain most of the variation in measured misallocation between countries. 

\subsubsection{Uncertainty and oil drilling in Texas}

\cite{Kellogg2014} studies the impact of uncertainty on drilling decisions of oil producers in Texas. Drilling is an interesting investment decision: it is a one-time decision (irrevesible), and it is mainly a binary decision as the intensive margin of drilling is mainly determined by geological factors. As such, it fits well the ``option value of waiting'' effect of uncertainty (\cite{dixit1994investment}) better than the continuous investment decisions considered in section \ref{sec:bloom_framework} above, where uncertainty can either raise or lower investment levels such as described by \cite{caballero1996investment}. Moreover, while \cite{bloom2009impact} and \cite{asker2014dynamic} are concerned with uncertainty in productivity shocks, the main source of uncertainty in the oil industry revolves around the price of oil. In this context, there are well developed financial tools to measure uncertainty, either by looking at the gap between futures prices and option prices for oil, or by looking at daily changes in oil prices and backing out implied volatility through a GARCH model. 

Using very detailed data on drilling activity in Texas, \cite{Kellogg2014} finds that firms do pull back on drilling activity when uncertainty goes up. Moreover, this response is more closely linked to measures of uncertainty that come from futures and option prices which are forward looking, rather than measures of volatility based on past changes in the price of oil. This is an important finding, as the normative implications of uncertainty have been well studied theoretically, but the positive effects of uncertainty -- i.e., how does uncertainty empirically shape firm's decision making -- are much less studied due to issues of how to measure uncertainty with either realization of shocks or more direct elicitation of expectations. 

\subsubsection{Uncertainty in shipping}

Other papers that look into the role of uncertainty in firms' investment decisions are \cite{Kalouptsidi2014} in the bulk shipping industry and \cite{jeon2020learning} in the container shipping industry. These industries involve shipping of commodities across the globe, and thus, respond stronger to changes in global economic activity, such as the great recession of 2008. Moreover, ships are long-lived assets that take several years to build, so there is a natural delay in the response of the industry to demand shocks. The main difference between these two shipping sectors is that bulk shipping is a relatively unconcentrated industry where shipments are made on the spot, while container shipping is more concentrated and most of the routes are fixed. This leads to very different empirical approaches. 

Firms in the bulk shipping industry ship bulk commodities, such as coal and wheat, that occupy the entire hold of a ship. Commodity prices fluctuate wildly over the business cycle, and these are passed through into large changes in shipping rates. The typical ship lasts 20 to 30 years, after which they are scrapped for recycled steel. Ships take a year or more to build, and shipyards have limited capacity, so they maintain order books. This means that the time from ordering a ship until delivery depends critically on the backlog at the shipyard, which is itself endogenous. 

In Kalouptsidi's model, a firm is a shipowner (indexed by $i$) that owns only one ship, and a ship is characterized by its age.\footnote{The model does not allow for firms with multiple ships.} Let $k_{it} \in \{0, 1, ..., K\}$ be the age at period $t$ of the ship owned by firm $i$. The state of the industry at period $t$ has three components, $\mathbf{x}_{t} = (\mathbf{n}_{t}, \mathbf{b}_{t}, d_{t})$: $\mathbf{n}_{t} = (n_{0t}, n_{1t}, ..., n_{Kt})$ is the vector with the age distribution of all the ships active in the industry, where $n_{kt}$ is the number of ships with age $k$ at period $t$; $\mathbf{b}_{t} = (b_{1t}, n_{2t}, ..., n_{Tt})$ is the vector with the backlog of orders at shipyards, where $b_{st}$ is the number of ships to be delivered at period $t+s$; and $d_{t}$ is the aggregate demand of shipping services, that follows an exogenous Markov process. The transition of the state variables in $\mathbf{n}_{t}$ and $\mathbf{b}_{t}$ is quite straightforward. For instance, the number of ships with age $k>0$ at $t+1$ is equal to the number of ships with age $k-1$ at period $t$ minus the number of ships that are scrapped. Likewise, backlog state variables evolve according to a simple rule. 

At a given period (quarter), a firm can be an incumbent or a potential entrant, depending on whether it owns a ship or not. Every period, incumbents decide whether to scrap their ships or continue operating, and potential entrants decide to enter (i.e., order a ship) or not. The scrap value of a ship is $\phi_{it}$, and it is private information of the firm and i.i.d over time and firms. The Bellman equation describing an incumbent's decision problem is:
\begin{equation}
    V(k_{it},\mathbf{x}_{t})
    =
    \pi(k_{it},\mathbf{n}_{t},d_{t}) 
    + \beta \text{ } \mathbb{E}_{\phi}
    \left(
        \max 
        \left\{ 
            \phi_{it} \text{ } , \text{ }
            \mathbb{E} \left[
                V(k_{it}+1,\mathbf{x}_{t+1}) 
                \text{ } | \text{ } \mathbf{x}_{t}
            \right]
        \right\}
        \text{ }
    \right)
\end{equation}
A potential entrant chooses to order a new ship if the expected value after entry is greater than entry costs, which are represented by function $\kappa(\mathbf{n}_{t})$. Note that entry costs depend on the state of the market. Importantly, there is time to build, that also evolves endogenously as a function of the backlog vector. More specifically, the order of a ship at period $t$ is delivered after $T(\mathbf{b}_{t})$ periods. Accordingly, a potential entrant decides to enter at period $t$ if the following condition holds:
\begin{equation}
    \beta^{T(\mathbf{b}_{t})} \text{ }
    \mathbb{E} 
    \left[
        V(0,\mathbf{x}_{t+T(\mathbf{b}_{t})}) 
        \text{ } | \text{ } 
        \mathbf{x}_{t}
    \right]
    \text{ } > \text{}
    \kappa(\mathbf{n}_{t}),
\end{equation}
where the expression in the left hand side is the expected discounted value of an incumbent with a new ship $T(\mathbf{b}_{t})$ periods in the future.

Kalouptsidi solves this dynamic game by looking at a quasi competitive version of the model, more specifically, by invoking the arguments from \cite{weintraub2008markov} on oblivious equilibrium. For the estimation of the model, the author exploits in an ingenious way information from the resale market for ships. For bulk shipping, there is good information on transaction prices in the resale market of ships. Let $p_{k,t}$ be the transaction price of a ship of age $k$ at period $t$. Under the assumption of perfect competition, no transaction costs, and no asymmetric information in the resale market, we have that $p_{k,t}= V(k,\mathbf{x}_{t})$, such that transaction prices provide direct information on the realized values of function $V(.)$. Kalouptsidi uses data on transaction prices to estimate the whole value function based on the regression equation: 
\begin{equation}
    p_{k,t} \text{ } = \text{ }
    V(k,\mathbf{x}_{t}) + \varepsilon_{k,t}
\end{equation}
for every $k=1,2,...,K$ and sample period $t$. This is a nonparametric regression model where the dimension of the vector of explanatory variables $\mathbf{x}_{t}$ is extremely large (94 variables), and the sample includes only a few hundred observations (the number of ages times the number of quarters in the sample). Therefore, this nonparametric regression is subject to a huge curse of dimensionality problem. To deal with this issue, Kalouptsidi combines aggregation restrictions on the vector of state variables $\mathbf{x}_{t}$ and machine learning techniques such as 
clustering and LASSO.\footnote{Though Kalouptsidi's approach to estimate the value function using transaction prices is quite ingenious, it is tricky to apply to many other industries. Most resale markets of capital are characterized by substantial transaction costs and asymmetric information. This is well known for cars and trucks, with perhaps the exception of aircraft (\cite{gavazza2011leasing,gavazza2011role}). See for instance the evidence on aerospace plants in \cite{ramey2001displaced}. These frictions in capital resale markets imply that there is not a straightforward relationship between transaction prices and firms' values. Nevertheless, Kalouptsidi's approach could be applied using other measures of firm valuation, such as stock market values.}

Given the estimated values $\widehat{V}(k,\mathbf{x}_{t})$ and data on entry and exit decisions, Kalouptsidi estimates the distribution of scrappage values $\phi_{it}$, the entry cost function $\kappa(\mathbf{n}_{t})$, and the time to build function $\kappa(\mathbf{b}_{t})$. Then, she looks into the impact of time to build and its endogeneity through backlog. In comparison to a fixed time to build, backlogs make time to build longer during booms, and shorter during downturns. Indeed, simulating the evolution of the industry, endogenous backlogs lower the volatility of investment by 45 percent compared to constant time to build. In addition, time to build slows the entry response to demand shocks, and leads to a fleet that is 15 percent bigger. In much of the literature on entry in industrial organization, a partial equilibrium stance is taken on entry costs: they are a fixed parameter or distribution. \cite{Kalouptsidi2014} is a nice example of what changes once a more general equilibrium view on the supply of entrants is considered.

\cite{jeon2020learning} studies the role of demand uncertainty in the cyclical investment fluctuations in the container shipping industry. This industry is more concentrated than bulk shipping and some companies own many vessels, but it is also subject to large swings in demand. A distinguishing feature of Jeon's paper within this literature is that firms' uncertainty is not limited to future unpredictable demand shocks but they have also uncertainty about demand parameters. Firms learn over time about these parameters using a form of adaptive learning. 

The state variables related to firm $i$ are the aggregate capacity of all its ships, $k_{it}$, and the backlog of the firm's orders of new ships, $b_{it}$, which is also measured in capacity units. The state of the industry is given by $\mathbf{x}_{t} = (k_{it}, b_{it}, z_{t}^{A}, z_{t}^{B}: i \in \mathcal{I})$, where $z_{t}^{A}$ and $z_{t}^{B}$ represent the state of demand in the route from Europe to Asia (the most travelled shipping route) and elsewhere, respectively. Jeon assumes that each of these demand variables follows an exogenous AR(1) process. For route $s \in \{A,B\}$:
\begin{equation}
    z_{t}^{s}
    \text{ } = \text{ } 
    \rho_{0}^{s} \text{ } + \text{ } 
    \rho_{1}^{s} \text{ } z_{t-1}^{s} 
    \text{ } + \text{ } 
    \sigma^{s} \text{ } \omega_{t}^{s},
\end{equation}
where $\omega_{t}^{s}$ is i.i.d. standard normal. As usual, firms have uncertainty about future realizations of $\omega$ shocks. Jeon considers that firms also have uncertainty about the parameters $(\rho_{0}^{s},\rho_{1}^{s},\sigma^{s}:s \in \{A,B\})$ that govern the stochastic process of these variables. Following the macroeconomics literature on agents' learning (\cite{evans2012learning}), Jeon assumes that firms in this industry use a form of adaptive learning to update their beliefs about these parameters. A parameter $\lambda$, which controls the weight that new data receives in the updating rule of beliefs, plays a key role in this learning mechanism. A higher value of $\lambda$ increases the responsiveness of firm beliefs to a downturn in demand. 

\cite{jeon2020learning} assumes that the industry outcomes come from a \textit{moment based Markov equilibrium} (MME), as defined by \cite{Weintraub2017}, that we have discussed in section \ref{large state space section}. This equilibrium concept aids in reducing the size of the state space by focusing on moments of the distribution of firms. In this model, without rational expectations, the parameters in firms' beliefs and learning process should be estimated together with the rest of the structural parameters in investment costs and scrap values from the predictions of the dynamic game. That is, observed firm behavior reveals not only firms' "preferences" bu also their beliefs. Jeon estimates all these parameters using a full-solution method of simulated moments. The estimates show that the weight in the learning process of a 10-year-old observation relative to a current observation is 45\%. 

The author presents counterfactual experiments to evaluate the impact of uncertainty about demand parameters on the level, volatility, and pattern of firms' investment. Removing uncertainty about demand parameters reduces aggregate investment by 17\% and its volatility by 22\%, and reallocates investment across the demand cycle: it reduces the positive response  of investment during boom years. Interestingly, there is also a very substantial impact on welfare, increasing producer surplus by 85\%, but having only a small negative impact on consumer surplus. This paper shows the potentially important impact on industry outcomes of sources of firm uncertainty which have not been included in the most standard dynamic models of competition in IO.

\subsection{Network competition in the airline industry}

An airline's network is the set of city-pairs that the airline connects via non-stop flights. The choice of network structure is one of the most important strategic decisions of an airline. Indeed, one of the assumptions that is the most difficult to accept in \cite{berry1992estimation}, is that entry decisions are solely about origin and destinations, rather than the entire route network. Two network structures that have received particular attention in studies of the airline industry are hub-and-spoke networks and point-to-point networks. In a hub-and-spoke network, an airline concentrates most of its operations in one airport, called the hub. All other cities in the network (i.e., the spokes) are connected to the hub by non-stop flights such that travelers between two spoke cities must take a connecting flight to the hub. In contrast, in a point-to-point network, all cities are connected with each other through nonstop flights. Like the work of \cite{Holmes2011} on Walmart's distribution network, it is quite challenging to consider the entire network formation process, since the underlying set of networks is the power set of all origin destination pairs. Moreover, while the field of network economics is quite large, there is a paucity of work on strategic network formation.  

Soon after deregulation of the US airline industry in 1978, most airline companies adopted hub-and-spoke networks to organize their routes. Different hypotheses have been suggested to explain airlines' adoption of hub-and-spoke networks. According to demand-side explanations, some travelers value the services associated with the scale of operation of an airline in the hub airport, e.g., more convenient check-in and landing facilities and higher flight frequencies. Cost-side explanations argue that an airline can exploit economies of scale and scope by concentrating most of its operation in a hub airport. Larger planes are cheaper to fly on a per-seat basis, and airlines can exploit these economies of scale by seating in a single plane, flying it to the hub city with passengers with different final destinations. The are also economies of scope as part of the fixed costs of operating a route, such as maintenance and labor costs that may be common across different routes in the same airport. Another hypothesis that has been suggested to explain hub-and-spoke networks is that it can be an effective strategy to deter the entry of competitors (\cite{hendricks1997entry}). In a hub-and-spoke network, the profit function of an airline is supermodular with respect to its entry decisions for different city-pairs. This complementarity implies that a hub-and-spoke airline may be willing to operate non-stop flights for a city-pair even when profits are negative because operating between that city-pair can generate positive profits connected with other routes. Potential entrants are aware of this, and therefore, it may deter entry.\footnote{This argument for entry deterrence does not suffer from several limitations that hinder other more standard arguments of predatory conduct. In particular, it does not require a sacrifice on the part of the incumbent (i.e., a reduction in current profits) that will be compensated for in the future only if competitors do not enter the market. Furthermore, it is not subject to well-known criticisms of some arguments and models of spatial entry deterrence (see \cite{judd1985credible}).}

Despite the attractive features of the \cite{hendricks1997entry} entry deterrence argument, there were no previous studies that empirically explore this entry deterrence motive in airlines' use of hub-and-spoke networks. Part of the reason for this lack of empirical evidence is the absence of structural models of dynamic network competition that incorporate this hypothesis and that were flexible and realistic enough to be estimated with actual data. This limitation in the literature motivated \cite{aguirregabiria2012airline} to develop an estimable dynamic game of airline network competition that incorporates the demand, cost, and strategic factors described above. 

In their model, every quarter airline companies decide the city-pairs where they operate non-stop flights and the fares for each route-product they serve. The structure of the model follows \cite{Ericson1995}: direct strategic interactions between firms occur only through the effect of prices on demand; price competition is static; and firms' entry decisions in city-pairs is dynamic or forward looking and it affects other firms' profits only indirectly through its effect on equilibrium prices. While static entry models such as \cite{berry1992estimation} and \cite{ciliberto2009market} provide measures of the effects of hubs on fixed operating costs, endogenizing the existence of hubs and, more generally, the structure of airlines' networks is important for multiple reasons. Treating hub size as a variable that is endogenously determined in the equilibrium of the model is important for some predictions and counterfactual experiments using these structural models, such as the medium and long run effects of a merger. 

The model is estimated using data from the Airline Origin and Destination Survey (DB1B) of the US Bureau of Transportation Statistics with information on quantities, prices, and route entry and exit decisions for every airline company in the routes between the 55 largest US cities, for a total of 1,485 city-pairs. 

Given the huge dimension of the state space in this network game, the authors need to develop several methodological contributions for the estimation and for the solution of an equilibrium in this model. They propose a method to reduce the dimension of the state space in dynamic games that extends to dynamic games the inclusive values approach in \cite{hendel2006measuring}, \cite{nevo2008approach}, or \cite{gowrisankaran2012dynamics}. The main contribution of their approach to model inclusive values is that they endogenize the transition probabilities of the inclusive values such that one can use the estimated model to make counterfactual experiments that take into account how these transition probabilities depend on the strategies of all the players, and therefore how they change in the counterfactual scenario. They also propose and implement a relatively simple homotopy method to deal with multiple equilibria when making counterfactual experiments with the estimated model.

Their empirical results show that an airline's number of connections in an airport has a statistically significant effect on consumer demand, unit costs, fixed operating costs, and costs of starting a new route ("entry costs"). However, the economically most substantial impact is on entry costs. Counterfactual experiments show that eliminating this effect on entry costs would reduce very substantially airlines' propensity to use hub-and-spoke networks. For some of the larger carriers, \cite{hendricks1997entry}'s strategic entry deterrence motive is the second most important factor to explain this network choice.
    
\subsection{Dynamic matching \label{dynamic matching}}

A newer strand of literature has looked into market equilibrium in industries characterized by search and matching frictions, primarily focused on the market for taxi service in New York City. This literature is somewhat apart from the literature on dynamic games that uses an \cite{Ericson1995} framework. Instead, these papers focus on dynamic competitive equilibrium in the tradition of \cite{hopenhayn1992entry}. However, intuitively, there should be a close correspondence between competitive models and an oligopoly model with many firms, as has been exploited by \cite{weintraub2008markov} and \cite{Weintraub2017}. Moreover, these dynamic competitive models allow for substantially simpler computation, as well as more theoretical clarity given that these models yield second welfare theorems. 

\cite{Buchholz} looks into the equilibrium of the New York City taxicab market. In this environment, cabs are driving through the city looking for riders. The friction that prevents matching between the two sides of the market is space: empty cabs and riders are in different places. In addition, even if cabs and riders are in the same neighborhood, they may not see each other. Some of this could be simply be about cabs and riders being on different street corners, which is a key friction modelled by \cite{frechette2019frictions} in a similar study of the New York City taxi market. As well, drivers and riders might simply have difficulty finding each other in front of Penn Station. 

A second inefficiency in this market is that fares are functions -- pre-established by the regulator -- of distance, time, and a flag fall fee. This precludes, in particular, dealing with differences in demand at the origin and destination of a ride. For instance, many people are looking for a ride from Queens to Manhattan on Friday mornings, but not in the other direction. Thus, the social planner may want to charge riders based on a richer set of characteristics, such as locational pricing, in this case a higher Queens price than Manhattan price for the same trip.

Locations in the city are indexed by $\ell \in \{1,2, ..., L \}$. \cite{Buchholz} models the state of location $\ell$ at time $t$ in terms of two variables: (i) the probability that a rider shows up in this location, denoted $\lambda_{\ell t}$, and where she wants to travel to, $d_{\ell t}$, which is allowed to vary by time of day in a predictable manner; and (ii) the number of vacant cabs in that location at time $t$ denoted as $v_{\ell t}$. The latter state is the main endogenous object in this market, and also needs to incorporate cabs in transit. For example, a rider may be going to the airport, and this means that a vacant cab will show up in 45 minutes at LaGuardia airport. Given the number of riders and vacant cabs in a location, a matching function $m(\lambda_{\ell t},v_{\ell t})$ determines the number of riders finding a  vacant cab. 

Every period, vacant taxi drivers choose in which location to search for riders.  They can decide to search in their current location or drive to another location in the city where there may be more riders. To make this decision, they compare the value of searching for riders at every location in the city. Solving this model via the approach of \cite{Ericson1995} is clearly intractable given the thousands of cabs in New York searching over dozens of neighbourhoods at different times of day. Instead, \cite{Buchholz} uses a dynamic competitive model following \cite{hopenhayn1992entry}, which assumes that cabs are atomistic; i.e., small enough so that they do not believe that their actions alter the equilibrium of the market. Moreover, as there are no aggregate shocks in this model, one can solve it as if all agents have perfect foresight over the evolution of the market over the day. This makes computation far easier as it implies that one just needs to solve for the number of vacant cabs in each location at every time of the day. It also avoids the issues around multiple equilibria in this environment. 

\cite{Buchholz} uses his model to look at the effect of using more sophisticated location based pricing, so charging prices based on origin and destination and time of day, rather than simply metering by distance and time. He finds that the total number of trips could be increased by 28 percent, and that welfare would increase around 8 percent. This suggests that more complex pricing mechanisms could be useful in the New York City taxi market.\footnote{In a related paper, \cite{frechette2019frictions} compare decentralized matching protocols --- cabs picking up hails on the street -- versus a centralized dispatch protocol --- Uber assigning cabs. This paper relies heavily on the topography of New York City's street grid to model how cabs travel to assess the efficiency gains from a better dispatch algorithm.}

\cite{Brancaccio2020} develop a dynamic spatial equilibrium model of the interaction between world trade and oceanic transportation services. In this model, forward-looking ship owners and exporters participate in a decentralized matching process where exporters decide where to export and ship owners choose which ports to move their vessels to. The authors estimate this model using detailed data on vessel movements, shipping prices, and trade flows. An interesting fact in this industry is that prices differ substantially by the direction of travel. For instance, it is far cheaper to ship cargo from China to Australia than the other way around, at least for bulk shipments like coal. This means that the characterization of equilibrium in this type of market needs to incorporate the directional flows of traffic across the globe. A contribution of this paper is to endogenize the shipping costs paid by exporters, as they depend on the shipper's decisions of what routes to take. The model provides a nice tool to study the effects on shipping costs and the patterns of exports on interesting worldwide economic events, such as the opening of the Arctic to shipping, or changes in fuel prices.

While there is a sharp discontinuity between the theory models of dynamic oligopoly of \cite{Ericson1995} versus the dynamic competitive ones of \cite{hopenhayn1992entry}, for much of the empirical work on industry dynamics, this boundary has started to blur. The computational approaches used for dynamic competitive equilibrium with aggregate shocks are quite similar to the MME and oblivious equilibrium concepts of  \cite{Weintraub2017} and \cite{weintraub2008markov}. Likewise, beyond correctly specifying the state space used by agents, there is not much practical difference between CCP approaches applied to competitive versus oligopolistic markets. Thus, we expect some convergence between the empirical literature on dynamic versus oligopolistic competition. 

\subsection{Natural resources\label{sec natural resources}}

\cite{Huang2014} investigate dynamics in a common pool setting, where an exhaustible common resource is used by many independent agents. Exploitation of a common pool resource can give rise to several externalities that interfere with efficient behavior: consumption by one agent reduces the stock of the resource for all other agents, which can induce over-extraction of the good. Dynamically, this externality can also distort the optimal time pattern of resource use, shifting extraction either too early or too late relative to the social planner. Finally, there is a static externality that is caused by overcrowding during extraction. This congestion externality increases the costs of extraction, which may lead to lower surplus, allocative inefficiencies (e.g., the wrong firms extracting the resource), and, potentially, may countervail the stock externality.

Huang and Smith examine these economic forces in the context of the North Carolina shrimp industry. There are several characteristics to this industry that make it amenable to this analysis. First, the dynamics of the resource are well-understood. The shrimp life cycle fits neatly into a year, and, importantly for modeling considerations, the species is able to reproduce at a sufficiently high rate, such that it is reasonable to assume that the stock renews completely each calendar year. This implies that one can model the essential dynamics in a repeated finite-horizon model that runs from January to December. This is relatively unusual in this literature, as most settings are concerns with long-lived firms that are modeled as having an infinite horizon. This also means that the model can be solved through backward induction, which also guarantees a unique solution conditional on each stage game having a single equilibrium. The biological basis for stock dynamics also informs the functional forms used in the structural analysis. There are also a number of nice weather-based exogenous supply shifters; increased wind speeds and wave heights make the harvesting process more difficult and therefore are excellent instruments for shifting supply. North Carolina is also a very small part of the globally-integrated shrimp industry, which means prices for input and outputs can be taken as given. The data is also unusually detailed, as the state collects information on every commercial shrimping boat trip.

The model consists of $N$ individual shrimp boats indexed by $i$; a state space which includes the present stock of shrimp, input and output prices, and current weather conditions; and transitions from state to state that depend on the present state vector and actions of shrimp boats in the present period. Shrimp boats decide whether to go fishing once a day. The one-period payoff function is:
\begin{equation}
    \pi_{it}(a_i) = 
    \begin{cases}
        \alpha \text{ } p_t \text{ } \mathbb{E}(h_{it}) - 
        \mathbf{z}_{it} \beta + 
        \varepsilon_{it}(1) 
        & \text{ if } a_{it} = 1 
        \\
        \varepsilon_{it}(0) 
        & \text{ if } a_{it} = 1.
    \end{cases}
\end{equation}
Expected revenue is the product of shrimp price, $p_{t}$, and expected harvest, $\mathbb{E}(h_{it})$. The term $\mathbf{z}_{it} \text{ } \beta$ captures the cost of a trip, and $\mathbf{z}_{it}$ is a vector of exogenous variables such as the legnth of the vessel, wind speed, wave height, fuel price, an indicator for weekend days, and the fish stock. Variables $\varepsilon_{it}(0)$ and $\varepsilon_{it}(1)$ are action-specific idiosyncratic shocks which are unobservable to the researcher and are i.i.f. type I extreme value, such that they generate the familiar logit choice probability when integrated. The authors assume that harvest depends on whether conditions ($w_{t}$), the stock of shrimp ($s_{t}$), the total number of vessels on the water that day ($n_{t} \equiv  \sum_{i=1}^N a_{it}$), the vessel's time invariant productivity ($\eta_{i}$), and a productivity shock ($u_{it}$), according to the following exponential function: 
\begin{equation}
    h_{it} \text{ } = \text{ }
    s_{t} \text{ } \times \text{ }
    \exp
    \{
        \gamma \text{ } n_{t} +
        w_{t} + \eta_{i} + u_{it}
    \}.
\end{equation}
The term $\gamma \text{ } n_{t}$ captures the agglomeration (if $\gamma > 0$) or congestion (if $\gamma < 0$) externality, depending on the sign of $\gamma$. 

Huang and Smith allow for more complex transitions between states than other dynamic settings. This is facilitated in part by exogeneity assumptions and the availability of high-frequency data. Price, wind speed, and wave heights are all modeled as a vector AR(1) process, with wind speed and wave heights correlated. The price of fuel is modeled as a function of the week of the year, and shrimp stocks are modeled as a latent stochastic difference equation that comes from a biological model. Since the actions today influence the state vector through the stock of shrimp, agents choose the best action today given the strategies of their competitors and the choice-specific continuation values.

The authors estimate the harvest production function as part of a first step, outside the dynamic model. The estimate of the externality parameter $\gamma$ implies that one unit increase in the total number of vessels implies a 0.127\% reduction in each vessel's harvest. Given that the average number of vessels per day is around 60, this parameter value implies a substantial congestion externality. The approach for the estimation of the dynamic model is a mix of \cite{Aguirregabiria2007} and \cite{Bajari2007}. The authors first estimate the transitions of exogenous state variables using time series methods. They then estimate the conditional choice probability using a logit that is saturated with state variables, their powers, and their interactions. In the stage game, profits depend on the number of other vessels on the water. Conditional on the equilibrium played in the data, one can integrate out the expected actions of all other boats on that day using the conditional choice probability function estimated in the prior step. The only remaining step is to compute continuation values to put into the likelihood function. With policy functions in hand, one can use the forward simulation method from \cite{Bajari2007} to approximate the continuation value. Once the continuation value is known, one can then maximize a pseudo log-likelihood.

The primary counterfactual evaluates the efficiency gains from using a centralized vessel allocation policy, where a social planner, who internalizes the externalities in this setting, decides how many (and which) shrimp boats will go fishing in a day. To perform this counterfactual, the authors discretize the ending shrimp stock and work backwards from the terminal date, solving the value function by filling in the optimal social policy at each point in the state space as they go. Once the value function is filled out for all points in the discretized state space, the social planner can pick the path that delivers the highest surplus. The authors find that the observed equilibrium shifts too much of the harvest early in the year, due to the extraction externality, and this then translates to too little of the harvest happening later in the year, as stocks would have been higher. There are too many vessels early on, and there is also an allocative inefficiency as some of the boats are lower productivity vessels that should not have been dispatched. Finally, they also examine the dynamics of the industry when congestion is eliminated; they find that congestion actually has a positive effect in equilibrium as it helps offset the extraction externality. This is a particularly compelling counterfactual for the use of the dynamic model, as otherwise one would incorrectly conclude in a static world that the congestion externality was welfare reducing through its imposition of higher fishing costs.

This paper has the flavor of both  single-agent dynamics and the multi-agent tools described above. There are many agents in this model, and their behavior only influences a particular agent through an aggregate quantity, which is the total number of vessels on the water in a given day. In that sense, this paper presages some of the work by \cite{Buchholz} and others. It is also an example of a paper where the policy question of interest is directly estimable from the data---the authors do not compute any counterfactual equilibria with strategic agents (the social planner is a single-agent problem).\footnote{\cite{ryan2012heterogeneity} is another example of where the counterfactuals are contained within the support of the observed data.} Rather, the authors are able to simulate in-sample counterfactuals that remain within the support of the observed state variables. Any counterfactuals that change the agent's profit incentives outside those bounds would necessitate solving the entire equilibrium; the finite-horizon assumption makes this computationally feasible (if expensive), but one would have to address issues of multiple equilibria in the stage game. The paper also highlights the incorporation of very rich data, with nontrivial dynamics, through their use as exogenous state variables. This contrasts with our previous discussion that focused on the need to typically simplify the endogenous dynamics as much as possible in order to facilitate estimation and simulation.

\section{Concluding remarks}

Over the last three decades, the work on dynamic oligopoly has moved from being a primarily methods-oriented line of research towards fulfilling its promise as a central tool in the empirical IO literature, paralleling transitions from theory to empirical implementation in demand estimation or vertical relationships. The models, methods, and applications we have outlined in this chapter are critical to understanding questions at the heart of industrial organization. In many ways, vast progress has been made. To an observer in the mid-1990s, the idea of a research agenda that delivered realistic empirical dynamic oligopoly models that could account for heterogeneous firms, complex state spaces evolving in response to both exogenous forces and endogenous strategic decisions, non-trivial dynamics on both supply and demand, and complicated payoff structures may have seemed completely out of reach. Indeed, in 2006, Tim Bresnahan colorfully compared the chances of this endeavor to winning a land war in Asia.
Thirty years later, these are seen as difficult, but solvable, problems. The literature has also started to deliver on its promise of quantifying the importance of dynamics in a wide range of settings. Yet, there are many remaining challenges in this literature, which we put into five different categories.

First, computation is still enormously difficult. Indeed, the state spaces and computational problems considered by the earliest papers in this literature, such as \cite{Pakes1994} or \cite{Gowrisankaran1999}, are embarrassingly close to some of the problems considered in the most recent papers in this survey. One might have thought that increasing computational power coming from the semiconductor industry would eliminate this as an issue, at least one that economists have to deal with, but this has not happened. One reason is that increases in computing power are simply exhausted by making the models slightly more complex. This leads both large delays in getting work published, as well as severe restrictions on the size of the state space any given application which affect the plausibility of this type of analysis.

When the CCP-inspired estimation papers, such as \cite{Bajari2007}, \cite{Aguirregabiria2007}, or \cite{Pesendorfer2008} were coming out, there was a hope that we had cracked the problem of computation, at least as far as the issue of estimation was concerned. Indeed, there are many applications that estimate parameters in models that have never been computed. However, there is a certain hollowness to estimation of parameters without being able to draw out their implications through a computed model, such as by running counterfactuals. Most of the parameters estimated in dynamic models do not have stand-alone policy implications, and even those that do are better understood by putting their implications into an equilibrium context.

Second, two decades of empirical work on dynamic oligopoly has revealed that both the right data can be particularly hard to find and that there can be an enormous disconnect between what the ideal empirical model asks for and what the data can actually deliver. At a bare minimum, one needs detailed data on all the participants in an industry, while a longitudinal panel spanning years or decades is even better. To use CCP-based methods, one needs enough observations, by enough independent agents, to estimate reduced-form policy functions describing agent behavior at all possible states. Ideally, one has observations on a large number of firms; it is even better if they are spread across independent markets. This makes using the CCP-based approach difficult for modeling globally-integrated markets, such as those for semiconductors or hard drives. This is in contrast to the large datasets that are commonly used for CCP-based papers in labor economics.\footnote{See, for instance, \cite{traiberman2019occupations,ransom2021labor,llull2018immigration,hincapie2020entrepreneurship} for recent papers in labor using CCP's, and in particular the large datasets employed in these analyses.}  Moreover, the relevant characteristics of firms need to be summarized into a parsimonious number of states, which can often require some heroic modeling assumptions. As a result of these data and specification challenges, many of the successful papers in this literature examine industries where institutional details of the industry generate data that is similar to that considered in the original structural studies of entry in \cite{bresnahan1990entry} and \cite{bresnahan1991entry} and where the essentially dynamics are interesting and necessary without being too complex.

Third, many recent applications of dynamic games applying two-step methods to estimate models with very large state spaces use very restrictive parametric specifications of reduced form CCP functions in the first step of the method. Recent work in the econometrics literature using machine learning techniques to improve small-sample performance in high-dimensional settings may be useful in this context. For example, \cite{nekipelov2018moment} show their method can be applied to the first-stage estimation in \cite{Bajari2007}, flexibly estimating policy functions while also accounting for the fact that different equilibria may be played across different markets. Further efforts to apply machine learning-based model selection techniques to identify the ``best'' specification of the reduced form CCP functions can be helpful in this context. It is also important to consider that, if the goal is the precise estimation of structural parameters in the second step, the ``best'' estimation of CCPs in the first step is not the specification that provides the lowest standard errors of reduced form parameters, and not even the one that provides the lowest mean square error in the first step. Often in two-step semiparametric procedures the first step nonparametric estimator is under-smoothed to deal with bias in the second stage parametric estimator (e.g. \cite{abadie2011bias}). This is an exciting area for future research.

Fourth, the agenda of computational-based theory outlined in \cite{Pakes1994} has not lead to a particularly well-organized body of work as to the theoretical predictions of these models. Indeed, the researcher first computing the solution to a dynamic game may have very little intuition of why the results end up the way they do: John Asker has qualified this type of work of unpacking computational results on dynamic games as ``forensic''. 

Fifth, multiplicity of equilibria remains an important challenge in empirical applications of dynamic games, especially in the implementation of counterfactual experiments. Two-step methods partially circumvent this problem by conditioning on the equilibrium played in the data, but one must either assume the same equilibrium is played in all markets or lose precision by estimating policy functions independently. In any case, this solution only applies to the estimation and not the computation of counterfactuals. \cite{besanko2010learning} introduce a homotopy path-following method for tracing out some (but not necessarily all) of the equilibria in a dynamic game. In a related paper, \cite{besanko2014economics}, use this homotopy method to trace out equilibria in a model of predation. They show that policy interventions not only change the behavior of firms within an equilibrium, but may also change the set of equilibria. Interpreting the difference between the two is crucial, but, at least for now, the tools necessary to show this remain limited. For instance, these papers – in the context of much more stylized models than those in empirical applications – reveal a correspondence between structural parameters and equilibrium outcomes that is chaotic, discontinuous, and non-intuitive. Infinitesimal changes to parameters induce jumps from single to many equilibria, with different comparative statics implications. Peering into such a Rorschach inkblot, one gets the impression that there are no robust predictions for some important counterfactual experiments.

All of that said, we conclude on a note of optimism. As this chapter has outlined, there is now a large body of empirical work looking at dynamic games that will inform future policy debates in economics. There has been an expansion in the types of industries that are considered, moving us away from the Bresnahan-Reiss program of looking at geographically isolated markets with a small number of relatively similar competitors. Instead, recent work looks at industries with firms with complex characteristics, global integrated markets, and markets with large numbers of firms in them. Furthermore, there is now a mature set of tools to both compute solutions to dynamic oligopoly problems with large state spaces and many firms, and an even more developed set of estimation techniques for these settings that can  incorporate differences in beliefs or cross-market heterogeneity. Just as one could not have completely foreseen all of the methodological advances in the field thirty years ago when the Markov-perfect Nash equilibrium foundations were being constructed by Maskin, Pakes, and Tirole, we are hopeful that the next wave of research in this area will successfully address the outstanding problems in the dynamic games literature. In particular, the field of machine learning is quickly evolving to handle problems with very large state spaces which could further extent the purview of these methods to realistic analysis of ever more complex and interesting markets. 

\newpage 

\printbibliography

@book{sutton1991sunk,
	author = {Sutton, John},
	publisher = {MIT press},
	title = {Sunk costs and market structure: Price competition, advertising, and the evolution of concentration},
	year = {1991}}

@online{hmg2010,
	author = {{U.S. Department of Justice} and {Federal Trade Commission}},
	title = {Horizontal Merger Guidelines},
	year = {2010}}

@article{hincapie2020entrepreneurship,
	author = {Hincapi{\'e}, Andr{\'e}s},
	journal = {International Economic Review},
	number = {2},
	pages = {617--681},
	publisher = {Wiley Online Library},
	title = {Entrepreneurship over the life cycle: Where are the young entrepreneurs?},
	volume = {61},
	year = {2020}}

@article{llull2018immigration,
	author = {Llull, Joan},
	journal = {The Review of Economic Studies},
	number = {3},
	pages = {1852--1896},
	publisher = {Oxford University Press},
	title = {Immigration, wages, and education: A labour market equilibrium structural model},
	volume = {85},
	year = {2018}}

@article{ransom2021labor,
	author = {Ransom, Tyler},
	journal = {Journal of Human Resources},
	pages = {0219--10013R2},
	publisher = {University of Wisconsin Press},
	title = {Labor market frictions and moving costs of the employed and unemployed},
	year = {2021}}

@article{traiberman2019occupations,
	author = {Traiberman, Sharon},
	journal = {American Economic Review},
	number = {12},
	pages = {4260--4301},
	title = {Occupations and import competition: Evidence from Denmark},
	volume = {109},
	year = {2019}}

@misc{1912.06680,
	author = {OpenAI and : and Christopher Berner and Greg Brockman and Brooke Chan and Vicki Cheung and Przemys{\l}aw D{\k e}biak and Christy Dennison and David Farhi and Quirin Fischer and Shariq Hashme and Chris Hesse and Rafal J{\'o}zefowicz and Scott Gray and Catherine Olsson and Jakub Pachocki and Michael Petrov and Henrique P. d. O. Pinto and Jonathan Raiman and Tim Salimans and Jeremy Schlatter and Jonas Schneider and Szymon Sidor and Ilya Sutskever and Jie Tang and Filip Wolski and Susan Zhang},
	eprint = {arXiv:1912.06680},
	title = {Dota 2 with Large Scale Deep Reinforcement Learning},
	year = {2019}}

@incollection{berndt1974estimation,
	author = {Berndt, Ernst R and Hall, Bronwyn H and Hall, Robert E and Hausman, Jerry A},
	booktitle = {Annals of Economic and Social Measurement},
	editor = {Sanford V. Berg},
	number = {4},
	pages = {653--665},
	publisher = {NBER},
	title = {Estimation and inference in nonlinear structural models},
	volume = {3},
	year = {1974}}

@article{chesher2017generalized,
	author = {Chesher, Andrew and Rosen, Adam M},
	journal = {Econometrica},
	number = {3},
	pages = {959--989},
	publisher = {Wiley Online Library},
	title = {Generalized instrumental variable models},
	volume = {85},
	year = {2017}}

@article{ackerberg2014asymptotic,
	author = {Ackerberg, Daniel and Chen, Xiaohong and Hahn, Jinyong and Liao, Zhipeng},
	journal = {Review of Economic Studies},
	number = {3},
	pages = {919--943},
	publisher = {Oxford University Press},
	title = {Asymptotic efficiency of semiparametric two-step GMM},
	volume = {81},
	year = {2014}}

@article{bai2021two,
	author = {Bai, Yuehao and Santos, Andres and Shaikh, Azeem M},
	journal = {Journal of Business \& Economic Statistics},
	pages = {1--11},
	publisher = {Taylor \& Francis},
	title = {A Two-Step Method for Testing Many Moment Inequalities},
	year = {2021}}

@techreport{belloni2019subvector,
	author = {Belloni, Alexandre and Bugni, Federico A and Chernozhukov, Victor},
	institution = {cemmap working paper},
	title = {Subvector inference in partially identified models with many moment inequalities},
	year = {2019}}

@article{bugni2010bootstrap,
	author = {Bugni, Federico A},
	journal = {Econometrica},
	number = {2},
	pages = {735--753},
	publisher = {Wiley Online Library},
	title = {Bootstrap inference in partially identified models defined by moment inequalities: Coverage of the identified set},
	volume = {78},
	year = {2010}}

@article{canay2010inference,
	author = {Canay, Ivan A},
	journal = {Journal of Econometrics},
	number = {2},
	pages = {408--425},
	publisher = {Elsevier},
	title = {EL inference for partially identified models: Large deviations optimality and bootstrap validity},
	volume = {156},
	year = {2010}}

@article{chernozhukov2016locally,
	author = {Chernozhukov, Victor and Escanciano, Juan Carlos and Ichimura, Hidehiko and Newey, Whitney K and Robins, James M},
	journal = {arXiv preprint arXiv:1608.00033},
	title = {Locally robust semiparametric estimation},
	year = {2016}}

@article{romano2014practical,
	author = {Romano, Joseph P and Shaikh, Azeem M and Wolf, Michael},
	journal = {Econometrica},
	number = {5},
	pages = {1979--2002},
	publisher = {Wiley Online Library},
	title = {A practical two-step method for testing moment inequalities},
	volume = {82},
	year = {2014}}

@article{chernozhukov2019inference,
	author = {Chernozhukov, Victor and Chetverikov, Denis and Kato, Kengo},
	journal = {The Review of Economic Studies},
	number = {5},
	pages = {1867--1900},
	publisher = {Oxford University Press},
	title = {Inference on causal and structural parameters using many moment inequalities},
	volume = {86},
	year = {2019}}

@article{chernozhukov2013intersection,
	author = {Chernozhukov, Victor and Lee, Sokbae and Rosen, Adam M},
	journal = {Econometrica},
	number = {2},
	pages = {667--737},
	publisher = {Wiley Online Library},
	title = {Intersection bounds: estimation and inference},
	volume = {81},
	year = {2013}}

@article{andrews2013inference,
	author = {Andrews, Donald WK and Shi, Xiaoxia},
	journal = {Econometrica},
	number = {2},
	pages = {609--666},
	publisher = {Wiley Online Library},
	title = {Inference based on conditional moment inequalities},
	volume = {81},
	year = {2013}}

@article{andrews2010inference,
	author = {Andrews, Donald WK and Soares, Gustavo},
	journal = {Econometrica},
	number = {1},
	pages = {119--157},
	publisher = {Wiley Online Library},
	title = {Inference for parameters defined by moment inequalities using generalized moment selection},
	volume = {78},
	year = {2010}}

@article{arulkumaran2017deep,
	author = {Arulkumaran, Kai and Deisenroth, Marc Peter and Brundage, Miles and Bharath, Anil Anthony},
	journal = {IEEE Signal Processing Magazine},
	number = {6},
	pages = {26--38},
	publisher = {IEEE},
	title = {Deep reinforcement learning: A brief survey},
	volume = {34},
	year = {2017}}

@article{chow1989complexity,
	author = {Chow, Chef-Seng and Tsitsiklis, John N},
	journal = {Journal of complexity},
	number = {4},
	pages = {466--488},
	publisher = {Elsevier},
	title = {The complexity of dynamic programming},
	volume = {5},
	year = {1989}}

@article{shao2019survey,
	author = {Shao, Kun and Tang, Zhentao and Zhu, Yuanheng and Li, Nannan and Zhao, Dongbin},
	journal = {arXiv preprint arXiv:1912.10944},
	title = {A survey of deep reinforcement learning in video games},
	year = {2019}}

@article{sweeting2015selective,
	author = {Sweeting, Andrew and Bhattacharya, Vivek},
	journal = {international Journal of industrial Organization},
	pages = {189--207},
	publisher = {Elsevier},
	title = {Selective entry and auction design},
	volume = {43},
	year = {2015}}

@misc{chernozhukov2018double,
	author = {Chernozhukov, Victor and Chetverikov, Denis and Demirer, Mert and Duflo, Esther and Hansen, Christian and Newey, Whitney and Robins, James},
	publisher = {Oxford University Press Oxford, UK},
	title = {Double/debiased machine learning for treatment and structural parameters},
	year = {2018}}

@article{nekipelov2018moment,
	author = {Nekipelov, Denis and Novosad, Paul and Ryan, Stephen P},
	title = {Moment Forests},
	year = {2021}}

@article{abbring2010,
	author = {Abbring , Jaap H. and Campbell, Jeffrey R.},
	journal = {Econometrica},
	number = {5},
	pages = {1491--1527},
	title = {Last-in first-out oligopoly dynamics},
	volume = {78},
	year = {2010}}

@article{hollenbeck2017economic,
	author = {Hollenbeck, Brett},
	journal = {The RAND Journal of Economics},
	number = {4},
	pages = {1103--1135},
	title = {The economic advantages of chain organization},
	volume = {48},
	year = {2017}}

@article{Aguirregabiria2014,
	author = {Aguirregabiria, Victor and Suzuki, Junichi},
	journal = {Quantitative Marketing and Economics},
	number = {3},
	pages = {267-304},
	title = {Identification and counterfactuals in dynamic models of market entry and exit},
	volume = {12},
	year = {2014}}

@article{Aguirregabiria2013,
	author = {Aguirregabiria, Victor and Nevo, Aviv},
	journal = {Advances in economics and econometrics},
	pages = {53--122},
	publisher = {Cambridge University Press Cambridge},
	title = {Recent developments in empirical IO: Dynamic demand and dynamic games},
	volume = {3},
	year = {2013}}

@article{Aguirregabiria2019incomplete,
	author = {Aguirregabiria, Victor and Mira, Pedro},
	journal = {Quantitative Economics},
	number = {4},
	pages = {1659--1701},
	title = {Identification of games of incomplete information with multiple equilibria and unobserved heterogeneity},
	volume = {10},
	year = {2019}}

@article{Aguirregabiria2007,
	author = {Aguirregabiria, Victor and Mira, Pedro},
	journal = {Econometrica},
	number = {1},
	pages = {1--53},
	title = {Sequential estimation of dynamic discrete games},
	volume = {75},
	year = {2007}}

@article{aguirregabiria_marcoux_2021_qe,
	author = {Aguirregabiria, Victor and Marcoux, Mathieu},
	journal = {Quantitative Economics},
	title = {Imposing equilibrium restrictions in the estimation of dynamic discrete games},
	volume = {Forthcoming},
	year = {2021}}

@incollection{aguirregabiria2013Euler,
	author = {Aguirregabiria, Victor and Magesan, Arvind},
	booktitle = {Structural Econometric Models (Advances in Econometrics)},
	pages = {3-44},
	publisher = {Emerald Group Publishing Limited},
	title = {Euler equations for the estimation of dynamic discrete choice structural models},
	volume = {31},
	year = {2013}}

@article{aguirregabiria2012airline,
	author = {Aguirregabiria, Victor and Ho, Chun-Yu},
	journal = {Journal of Econometrics},
	number = {1},
	pages = {156-173},
	title = {A dynamic oligopoly game of the US airline industry: Estimation and policy experiments},
	volume = {168},
	year = {2012}}

@article{aguirregabiria1999dynamics,
	author = {Aguirregabiria, Victor},
	journal = {The Review of Economic Studies},
	number = {2},
	pages = {275--308},
	title = {The dynamics of markups and inventories in retailing firms},
	volume = {66},
	year = {1999}}

@article{aguirregabiria_alonso_2014,
	author = {Aguirregabiria, Victor and Alonso-Borrego, Cesar},
	journal = {Economic Inquiry},
	number = {2},
	pages = {930--957},
	title = {Labor contracts and flexibility: evidence from a labor market reform in Spain},
	volume = {52},
	year = {2014}}

@article{aguirregabiria_mira_2002,
	author = {Aguirregabiria, Victor and Mira, Pedro},
	journal = {Econometrica},
	number = {4},
	pages = {1519--1543},
	title = {Swapping the nested fixed point algorithm: A class of estimators for discrete Markov decision models},
	volume = {70},
	year = {2002}}

@article{aguirregabiria_2005_econletters,
	author = {Aguirregabiria, Victor},
	journal = {Economics Letters},
	number = {3},
	pages = {393--398},
	publisher = {Elsevier},
	title = {Nonparametric identification of behavioral responses to counterfactual policy interventions in dynamic discrete decision processes},
	volume = {87},
	year = {2005}}

@article{asker2020computational,
	author = {Asker, John and Fershtman, Chaim and Jeon, Jihye and Pakes, Ariel},
	journal = {The RAND Journal of Economics},
	number = {3},
	title = {A Computational Framework for Analyzing Dynamic Auctions: The Market Impact of Information Sharing},
	volume = {51},
	year = {2020}}

@article{backus2020productivity,
	author = {Backus, Matthew},
	journal = {Econometrica},
	number = {6},
	pages = {2415--2444},
	title = {Why is productivity correlated with competition?},
	volume = {88},
	year = {2020}}

@article{barwick2015costs,
	author = {Barwick, Panle Jia and Pathak, Parag A},
	journal = {The RAND Journal of Economics},
	number = {1},
	pages = {103--145},
	title = {The costs of free entry: an empirical study of real estate agents in Greater Boston},
	volume = {46},
	year = {2015}}

@article{benkard2015oblivious,
	author = {Benkard, C Lanier and Jeziorski, Przemyslaw and Weintraub, Gabriel Y},
	journal = {The RAND Journal of Economics},
	number = {4},
	pages = {671--708},
	title = {Oblivious equilibrium for concentrated industries},
	volume = {46},
	year = {2015}}

@article{berry2001mergers,
	author = {Berry, Steven T and Waldfogel, Joel},
	journal = {The Quarterly Journal of Economics},
	number = {3},
	pages = {1009--1025},
	publisher = {MIT Press},
	title = {Do mergers increase product variety? Evidence from radio broadcasting},
	volume = {116},
	year = {2001}}

@book{bertsekas1996neuro,
	author = {Bertsekas, Dimitri P and Tsitsiklis, John N},
	publisher = {Athena Scientific},
	title = {Neuro-dynamic programming},
	year = {1996}}

@article{camerer_ho_2004,
	author = {Camerer, Colin F and Ho, Teck-Hua and Chong, Juin-Kuan},
	journal = {The Quarterly Journal of Economics},
	number = {3},
	pages = {861--898},
	publisher = {MIT Press},
	title = {A cognitive hierarchy model of games},
	volume = {119},
	year = {2004}}

@article{bishop2008dynamic,
	author = {Bishop, Kelly C},
	journal = {Unpublished, Washington University in St. Louis},
	title = {A dynamic model of location choice and hedonic valuation},
	volume = {5},
	year = {2008}}

@article{chernozhukov_hong_2007,
	author = {Chernozhukov, Victor and Hong, Han and Tamer, Elie},
	journal = {Econometrica},
	number = {5},
	pages = {1243--1284},
	title = {Estimation and confidence regions for parameter sets in econometric models 1},
	volume = {75},
	year = {2007}}

@article{copeland_monnet_2009,
	author = {Copeland, Adam and Monnet, Cyril},
	journal = {The Review of Economic Studies},
	number = {1},
	pages = {93--113},
	publisher = {Wiley-Blackwell},
	title = {The welfare effects of incentive schemes},
	volume = {76},
	year = {2009}}

@article{depinto_nelson_2009,
	author = {De Pinto, Alessandro and Nelson, Gerald C},
	journal = {Environmental and Resource Economics},
	number = {2},
	pages = {209--229},
	publisher = {Springer},
	title = {Land use change with spatially explicit data: a dynamic approach},
	volume = {43},
	year = {2009}}

@article{doraszelski2003r,
	author = {Doraszelski, Ulrich},
	journal = {The RAND Journal of Economics},
	number = {1},
	pages = {20--42},
	publisher = {JSTOR},
	title = {An R\&D race with knowledge accumulation},
	volume = {34},
	year = {2003}}

@article{ellickson_misra_2008,
	author = {Ellickson, Paul B and Misra, Sanjog},
	journal = {Marketing science},
	number = {5},
	pages = {811--828},
	publisher = {INFORMS},
	title = {Supermarket pricing strategies},
	volume = {27},
	year = {2008}}

@article{ellickson2012repositioning,
	author = {Ellickson, Paul B and Misra, Sanjog and Nair, Harikesh S},
	journal = {Journal of Marketing Research},
	number = {6},
	pages = {750--772},
	publisher = {SAGE Publications Sage CA: Los Angeles, CA},
	title = {Repositioning dynamics and pricing strategy},
	volume = {49},
	year = {2012}}

@article{egesdal_lai_2015,
	author = {Egesdal, Michael and Lai, Zhenyu and Su, Che-Lin},
	journal = {Quantitative Economics},
	number = {3},
	pages = {567--597},
	title = {Estimating dynamic discrete-choice games of incomplete information},
	volume = {6},
	year = {2015}}

@article{farias2012approximate,
	author = {Farias, Vivek and Saure, Denis and Weintraub, Gabriel Y},
	journal = {The RAND Journal of Economics},
	number = {2},
	pages = {253--282},
	title = {An approximate dynamic programming approach to solving dynamic oligopoly models},
	volume = {43},
	year = {2012}}

@article{gayle_xie_2018,
	author = {Gayle, Philip G and Xie, Xin},
	journal = {Economic Inquiry},
	number = {3},
	pages = {1898--1924},
	title = {Entry deterrence and strategic alliances},
	volume = {56},
	year = {2018}}

@article{ghemawat1985exit,
	author = {Ghemawat, Pankaj and Nalebuff, Barry},
	journal = {The RAND Journal of Economics},
	number = {2},
	pages = {184--194},
	publisher = {JSTOR},
	title = {Exit},
	volume = {16},
	year = {1985}}

@article{green1984noncooperative,
	author = {Green, Edward J and Porter, Robert H},
	journal = {Econometrica},
	number = {1},
	pages = {87--100},
	title = {Noncooperative collusion under imperfect price information},
	volume = {52},
	year = {1984}}

@article{han_hong_2011,
	author = {Han, Lu and Hong, Seung-Hyun},
	journal = {Journal of Business \& Economic Statistics},
	number = {4},
	pages = {564--578},
	publisher = {Taylor \& Francis},
	title = {Testing cost inefficiency under free entry in the real estate brokerage industry},
	volume = {29},
	year = {2011}}

@article{hansen_singleton_1982,
	author = {Hansen, Lars Peter and Singleton, Kenneth J},
	journal = {Econometrica},
	number = {5},
	pages = {1269--1286},
	publisher = {JSTOR},
	title = {Generalized instrumental variables estimation of nonlinear rational expectations models},
	volume = {50},
	year = {1982}}

@incollection{heckman_1981_incidental,
	author = {Heckman, James},
	booktitle = {in C. Manski and D. McFadden (eds.), Structural Analysis of Discrete Data with Econometric Applications},
	publisher = {MIT Press},
	title = {The incidental parameters problem and the problem of initial conditions in estimating a discrete time - discrete data stochastic process},
	year = {1981}}

@article{hendel2006measuring,
	author = {Hendel, Igal and Nevo, Aviv},
	journal = {Econometrica},
	number = {6},
	pages = {1637--1673},
	title = {Measuring the implications of sales and consumer inventory behavior},
	volume = {74},
	year = {2006}}

@article{huang_2015_ms,
	author = {Huang, Yan and Singh, Param Vir and Ghose, Anindya},
	journal = {Management Science},
	number = {12},
	pages = {2825--2844},
	publisher = {INFORMS},
	title = {A structural model of employee behavioral dynamics in enterprise social media},
	volume = {61},
	year = {2015}}

@article{krusell1998income,
	author = {Krusell, Per and Smith, Jr, Anthony A},
	journal = {Journal of Political Economy},
	number = {5},
	pages = {867--896},
	publisher = {The University of Chicago Press},
	title = {Income and wealth heterogeneity in the macroeconomy},
	volume = {106},
	year = {1998}}

@article{kydland1982time,
	author = {Kydland, Finn E and Prescott, Edward C},
	journal = {Econometrica},
	number = {6},
	pages = {1345--1370},
	publisher = {JSTOR},
	title = {Time to build and aggregate fluctuations},
	volume = {50},
	year = {1982}}

@article{lim_yurukoglu_2018,
	author = {Lim, Claire SH and Yurukoglu, Ali},
	journal = {Journal of Political Economy},
	number = {1},
	pages = {263--312},
	publisher = {University of Chicago Press Chicago, IL},
	title = {Dynamic natural monopoly regulation: Time inconsistency, moral hazard, and political environments},
	volume = {126},
	year = {2018}}

@article{liu_zhou_2017,
	author = {Liu, Xiaodong and Zhou, Jiannan},
	journal = {Economics Letters},
	pages = {86--89},
	title = {A social interaction model with ordered choices},
	volume = {161},
	year = {2017}}

@article{lin_xu_2017,
	author = {Lin, Zhongjian and Xu, Haiqing},
	journal = {The Econometrics Journal},
	number = {3},
	pages = {S86--S102},
	publisher = {Oxford University Press Oxford, UK},
	title = {Estimation of social-influence-dependent peer pressure in a large network game},
	volume = {20},
	year = {2017}}

@article{murphy2018dynamic,
	author = {Murphy, Alvin},
	journal = {American economic journal: economic policy},
	number = {4},
	pages = {243--67},
	title = {A dynamic model of housing supply},
	volume = {10},
	year = {2018}}

@article{nagel_1995,
	author = {Nagel, Rosemarie},
	journal = {American Economic Review},
	number = {5},
	pages = {1313--1326},
	publisher = {JSTOR},
	title = {Unraveling in guessing games: An experimental study},
	volume = {85},
	year = {1995}}

@article{nevo2008approach,
	author = {Nevo, Aviv and Rossi, Federico},
	journal = {Economics Letters},
	number = {1},
	pages = {49--52},
	title = {An approach for extending dynamic models to settings with multi-product firms},
	volume = {100},
	year = {2008}}

@article{newey_1994_asymptotic,
	author = {Newey, Whitney K},
	journal = {Econometrica},
	number = {6},
	pages = {1349--1382},
	publisher = {JSTOR},
	title = {The asymptotic variance of semiparametric estimators},
	volume = {62},
	year = {1994}}

@article{Pakes2007,
	author = {Ariel Pakes and Michael Ostrovsky and Steven Berry},
	journal = {The RAND Journal of Economics},
	number = {2},
	pages = {373--399},
	title = {Simple estimators for the parameters of discrete dynamic games (with entry/exit examples)},
	volume = {38},
	year = {2007}}

@article{Pakes2001,
	author = {Pakes, Ariel and McGuire, Paul},
	journal = {Econometrica},
	number = {5},
	pages = {1261-1281},
	title = {Stochastic algorithms, symmetric markov perfect equilibrium, and the `curse' of dimensionality},
	volume = {69},
	year = {2001}}

@article{Pakes1994,
	author = {Pakes, Ariel and McGuire, Paul},
	journal = {The RAND Journal of Economics},
	number = {4},
	pages = {555-589},
	title = {Computing Markov-perfect nash equilibria: numerical implications of a dynamic differentiated product model},
	volume = {25},
	year = {1994}}

@book{powell2007approximate,
	author = {Powell, Warren B},
	publisher = {John Wiley \& Sons},
	title = {Approximate Dynamic Programming: Solving the curses of dimensionality},
	volume = {703},
	year = {2007}}

@article{scott2014dynamic,
	author = {Scott, Paul},
	publisher = {TSE Working Paper},
	title = {Dynamic discrete choice estimation of agricultural land use},
	year = {2014}}

@article{srisuma_linton_2012,
	author = {Srisuma, Sorawoot and Linton, Oliver},
	journal = {Journal of Econometrics},
	number = {2},
	pages = {320--341},
	title = {Semiparametric estimation of Markov decision processes with continuous state space},
	volume = {166},
	year = {2012}}

@article{hollenbeck2018online,
	author = {Hollenbeck, Brett},
	journal = {Journal of Marketing Research},
	number = {5},
	pages = {636--654},
	publisher = {SAGE Publications Sage CA: Los Angeles, CA},
	title = {Online reputation mechanisms and the decreasing value of chain affiliation},
	volume = {55},
	year = {2018}}

@article{stahl_wilson_1995,
	author = {Stahl, Dale O and Wilson, Paul W},
	journal = {Games and Economic Behavior},
	number = {1},
	pages = {218--254},
	title = {On players' models of other players: Theory and experimental evidence},
	volume = {10},
	year = {1995}}

@article{tomlin2014exchange,
	author = {Tomlin, Ben},
	journal = {International Journal of Industrial Organization},
	pages = {12--28},
	title = {Exchange rate fluctuations, plant turnover and productivity},
	volume = {35},
	year = {2014}}

@techreport{vreugdenhil2020booms,
	author = {Vreugdenhil, Nicholas},
	institution = {Arizona State University},
	title = {Booms, Busts, and Mismatch in Capital Markets: Evidence from the Offshore Oil and Gas Industry},
	year = {2020}}

@article{jia2008happens,
	author = {Jia, Panle},
	journal = {Econometrica},
	number = {6},
	pages = {1263--1316},
	title = {What happens when Wal-Mart comes to town: An empirical analysis of the discount retailing industry},
	volume = {76},
	year = {2008}}

@article{waldfogel1999free,
	author = {Waldfogel, Joel and Berry, Steven T},
	journal = {The RAND Journal of Economics},
	number = {3},
	pages = {397--420},
	title = {Free Entry and Social Inefficiency in Radio Broadcasting},
	volume = {30},
	year = {1999}}

@article{berry2016optimal,
	author = {Berry, Steven and Eizenberg, Alon and Waldfogel, Joel},
	journal = {The RAND Journal of Economics},
	number = {3},
	pages = {463--497},
	title = {Optimal product variety in radio markets},
	volume = {47},
	year = {2016}}

@inbook{bresnahan1989empirical,
	author = {Bresnahan, Timothy F},
	booktitle = {Handbook of Industrial Organization},
	editor = {Richard Schmalensee and Robert D. Willig},
	pages = {1011--1057},
	publisher = {Elsevier},
	title = {Empirical studies of industries with market power},
	volume = {2},
	year = {1989}}

@techreport{dee2020,
	author = {{Jan Victor} Dee},
	institution = {Concordia University},
	title = {A Dynamic Structural Model for Pay-Per-Bid Auctions: Explaining the Excess Revenue Puzzle in Online Auctions},
	year = {2020}}

@article{pakes1986patents,
	author = {Pakes, Ariel},
	journal = {Econometrica},
	number = {4},
	pages = {755--784},
	title = {Patents as Options: Some Estimates of the Value of Holding European Patent Stocks},
	volume = {54},
	year = {1986}}

@article{jovanovic1982selection,
	author = {Jovanovic, Boyan},
	journal = {Econometrica},
	number = {3},
	pages = {649--670},
	title = {Selection and the Evolution of Industry},
	volume = {50},
	year = {1982}}

@article{goolsbee2008incumbents,
	author = {Goolsbee, Austan and Syverson, Chad},
	journal = {The Quarterly Journal of Economics},
	number = {4},
	pages = {1611--1633},
	title = {How do incumbents respond to the threat of entry? Evidence from the major airlines},
	volume = {123},
	year = {2008}}

@article{besanko2010learning,
	author = {Besanko, David and Doraszelski, Ulrich and Kryukov, Yaroslav and Satterthwaite, Mark},
	journal = {Econometrica},
	number = {2},
	pages = {453--508},
	title = {Learning-by-doing, organizational forgetting, and industry dynamics},
	volume = {78},
	year = {2010}}

@article{stahl2016effects,
	author = {Stahl, Jessica Calfee},
	journal = {American Economic Review},
	number = {8},
	pages = {2185--2218},
	title = {Effects of deregulation and consolidation of the broadcast television industry},
	volume = {106},
	year = {2016}}

@incollection{olley2018,
	author = {Olley, G. S. and Town, R.},
	booktitle = {The Antitrust Revolution: Economics, Competition and Policy, 7th Edition},
	editor = {Kwoka, J and White, L.},
	publisher = {Oxford University Press},
	title = {End of an Era: The American Airlines-US Airways Merger},
	year = {2018}}

@article{syverson2004market,
	author = {Syverson, Chad},
	journal = {Journal of Political Economy},
	number = {6},
	pages = {1181--1222},
	title = {Market structure and productivity: A concrete example},
	volume = {112},
	year = {2004}}

@article{foster2008reallocation,
	author = {Foster, Lucia and Haltiwanger, John and Syverson, Chad},
	journal = {American Economic Review},
	number = {1},
	pages = {394--425},
	title = {Reallocation, firm turnover, and efficiency: Selection on productivity or profitability?},
	volume = {98},
	year = {2008}}

@article{hotz1993conditional,
	author = {Hotz, V Joseph and Miller, Robert A},
	journal = {The Review of Economic Studies},
	number = {3},
	pages = {497--529},
	title = {Conditional choice probabilities and the estimation of dynamic models},
	volume = {60},
	year = {1993}}

@article{hotz1994simulation,
	author = {Hotz, V Joseph and Miller, Robert A and Sanders, Seth and Smith, Jeffrey},
	journal = {The Review of Economic Studies},
	number = {2},
	pages = {265--289},
	title = {A simulation estimator for dynamic models of discrete choice},
	volume = {61},
	year = {1994}}

@article{jeziorski2016dynamic,
	author = {Jeziorski, Przemyslaw and Krasnokutskaya, Elena},
	journal = {The RAND Journal of Economics},
	number = {4},
	pages = {751--791},
	title = {Dynamic auction environment with subcontracting},
	volume = {47},
	year = {2016}}

@article{hopenhayn1992entry,
	author = {Hopenhayn, Hugo A},
	journal = {Econometrica},
	number = {5},
	pages = {1127--1150},
	title = {Entry, exit, and firm dynamics in long run equilibrium},
	volume = {60},
	year = {1992}}

@techreport{hopenhayn2016bidding,
	author = {Hopenhayn, Hugo and Saeedi, Maryam},
	institution = {National Bureau of Economic Research},
	title = {Bidding dynamics in auctions},
	year = {2016}}

@article{groeger2014study,
	author = {Groeger, Joachim R},
	journal = {International Economic Review},
	number = {4},
	pages = {1129--1154},
	title = {A study of participation in dynamic auctions},
	volume = {55},
	year = {2014}}

@book{shumpeter1942capitalism,
	address = {New York},
	author = {Shumpeter, Joseph},
	publisher = {{Harper and Brothers}},
	title = {Capitalism, socialism and democracy},
	year = {1942}}

@article{doraszelski2010computable,
	author = {Doraszelski, Ulrich and Satterthwaite, Mark},
	journal = {The RAND Journal of Economics},
	number = {2},
	pages = {215--243},
	title = {Computable Markov-perfect industry dynamics},
	volume = {41},
	year = {2010}}

@article{milgrom1982limit,
	author = {Milgrom, Paul and Roberts, John},
	journal = {Econometrica},
	number = {2},
	pages = {443--459},
	title = {Limit pricing and entry under incomplete information: An equilibrium analysis},
	volume = {50},
	year = {1982}}

@article{bresnahan1990entry,
	author = {Bresnahan, Timothy F and Reiss, Peter C},
	journal = {The Review of Economic Studies},
	number = {4},
	pages = {531--553},
	title = {Entry in Monopoly Market},
	volume = {57},
	year = {1990}}

@article{sweeting2020model,
	author = {Sweeting, Andrew and Roberts, James W and Gedge, Chris},
	journal = {Journal of Political Economy},
	number = {3},
	pages = {1148--1193},
	title = {A model of dynamic limit pricing with an application to the airline industry},
	volume = {128},
	year = {2020}}

@article{maskin1988theory1,
	author = {Maskin, Eric and Tirole, Jean},
	journal = {Econometrica},
	number = {3},
	pages = {549--569},
	publisher = {JSTOR},
	title = {A theory of dynamic oligopoly, I: Overview and quantity competition with large fixed costs},
	volume = {56},
	year = {1988}}

@article{maskin1988theory2,
	author = {Maskin, Eric and Tirole, Jean},
	journal = {Econometrica},
	number = {3},
	pages = {571--599},
	publisher = {JSTOR},
	title = {A theory of dynamic oligopoly, II: Price competition, kinked demand curves, and Edgeworth cycles},
	volume = {56},
	year = {1988}}

@book{laffont1993theory,
	author = {Laffont, Jean-Jacques and Tirole, Jean},
	publisher = {MIT press},
	title = {A theory of incentives in procurement and regulation},
	year = {1993}}

@phdthesis{gowrisankaran1995dynamic,
	author = {Gowrisankaran, Gautam},
	school = {Dissertation, Yale University},
	title = {A Dynamic Analysis of Mergers},
	year = {1995}}

@article{hsieh2009misallocation,
	author = {Hsieh, Chang-Tai and Klenow, Peter J},
	journal = {Quarterly Journal of Economics},
	number = {4},
	pages = {1403--1448},
	title = {Misallocation and manufacturing TFP in China and India},
	volume = {124},
	year = {2009}}

@article{caballero1996investment,
	author = {Caballero, Ricardo and Pindyck, Robert},
	journal = {International Economic Review},
	number = {3},
	pages = {641--662},
	title = {Investment, uncertainty and industry evolution},
	volume = {37},
	year = {1996}}

@article{ramey2001displaced,
	author = {Ramey, Valerie A and Shapiro, Matthew D},
	journal = {Journal of Political Economy},
	number = {5},
	pages = {958--992},
	title = {Displaced capital: A study of aerospace plant closings},
	volume = {109},
	year = {2001}}

@article{gavazza2011role,
	author = {Gavazza, Alessandro},
	journal = {American Economic Review},
	number = {4},
	pages = {1106--43},
	title = {The role of trading frictions in real asset markets},
	volume = {101},
	year = {2011}}

@article{gavazza2011leasing,
	author = {Gavazza, Alessandro},
	journal = {Journal of Political Economy},
	number = {2},
	pages = {325--377},
	title = {Leasing and secondary markets: Theory and evidence from commercial aircraft},
	volume = {119},
	year = {2011}}

@article{jeon2020learning,
	author = {Jeon, Jihye},
	journal = {The RAND Journal of Economics},
	pages = {Forthcoming},
	title = {Learning and investment under demand uncertainty in container shipping},
	year = {2020}}

@book{dixit1994investment,
	author = {Dixit, Avinash K and Pindyck, Robert},
	publisher = {Princeton University Press},
	title = {Investment under uncertainty},
	year = {1994}}

@article{ho2009insurer,
	author = {Ho, Katherine},
	journal = {American Economic Review},
	number = {1},
	pages = {393--430},
	title = {Insurer-provider networks in the medical care market},
	volume = {99},
	year = {2009}}

@techreport{berry2020instrumental,
	author = {Berry, Steven T and Compiani, Giovanni},
	institution = {National Bureau of Economic Research},
	title = {An instrumental variable approach to dynamic models},
	year = {2020}}

@unpublished{lee2013markov,
	author = {Lee, Robin S and Fong, Kyna},
	note = {Manuscript, Harvard University},
	title = {Markov perfect network formation: An applied framework for bilateral oligopoly and bargaining in buyer-seller networks},
	year = {2013}}

@article{rubinstein1982perfect,
	author = {Rubinstein, Ariel},
	journal = {Econometrica},
	number = {1},
	pages = {97--109},
	title = {Perfect equilibrium in a bargaining model},
	volume = {50},
	year = {1982}}

@article{collard2015reallocation,
	author = {Collard-Wexler, Allan and De Loecker, Jan},
	journal = {American Economic Review},
	number = {1},
	pages = {131--71},
	title = {Reallocation and technology: Evidence from the US steel industry},
	volume = {105},
	year = {2015}}

@article{shaked1984involuntary,
	author = {Shaked, Avner and Sutton, John},
	journal = {Econometrica},
	number = {6},
	pages = {1351--1364},
	title = {Involuntary unemployment as a perfect equilibrium in a bargaining model},
	volume = {52},
	year = {1984}}

@article{berry1995automobile,
	author = {Berry, Steven and Levinsohn, James and Pakes, Ariel},
	journal = {Econometrica},
	number = {4},
	pages = {841--890},
	title = {Automobile prices in market equilibrium},
	volume = {63},
	year = {1995}}

@unpublished{caoui2019estimating,
	author = {Caoui, El Hadi},
	note = {Manuscript, University of Toronto},
	title = {Estimating the Costs of Standardization: Evidence from the Movie Industry},
	year = {2019}}

@article{jovanovic2002q,
	author = {Jovanovic, Boyan and Rousseau, Peter L},
	journal = {American Economic Review},
	number = {2},
	pages = {198--204},
	title = {The Q-theory of mergers},
	volume = {92},
	year = {2002}}

@article{Takahashi2015,
	author = {Takahashi, Yuya},
	journal = {American Economic Review},
	number = {7},
	pages = {2204-2241},
	title = {Estimating a war of attrition: the case of the US movie theater industry},
	volume = {105},
	year = {2015}}

@article{Sweeting2013,
	author = {Sweeting, Andrew},
	journal = {Econometrica},
	number = {5},
	pages = {1763--1803},
	title = {Dynamic product positioning in differentiated product markets: the effect of fees for musical performance rights on the commercial radio industry},
	volume = {81},
	year = {2013}}

@article{Suzuki2013,
	author = {Suzuki, Junichi},
	journal = {International Economic Review},
	number = {2},
	pages = {495-523},
	title = {Land use regulation as a barrier to entry: evidence from the texas lodging industry},
	volume = {54},
	year = {2013}}

@article{Su2012,
	author = {Su, Che-Lin and Judd, Kenneth L},
	journal = {Econometrica},
	number = {5},
	pages = {2213--2230},
	title = {Constrained optimization approaches to estimation of structural models},
	volume = {80},
	year = {2012}}

@unpublished{Schmidt-Dengler2006,
	author = {Schmidt-Dengler, Philipp},
	note = {Manuscript, London School of Economics},
	title = {The timing of new technology adoption: the case of MRI},
	year = {2006}}

@article{Ryan2012,
	author = {Ryan, Stephen P.},
	journal = {Econometrica},
	number = {3},
	pages = {1019--1061},
	title = {The costs of environmental regulation in a concentrated industry},
	volume = {80},
	year = {2012}}

@article{Pesendorfer2010,
	author = {Pesendorfer, Martin and Schmidt-Dengler, Philipp},
	journal = {Econometrica},
	number = {2},
	pages = {833--842},
	title = {Sequential estimation of dynamic discrete games: a comment},
	volume = {78},
	year = {2010}}

@article{Pesendorfer2008,
	author = {Pesendorfer, Martin and Schmidt-Dengler, Philipp},
	journal = {The Review of Economic Studies},
	number = {3},
	pages = {901-928},
	title = {Asymptotic least squares estimators for dynamic games},
	volume = {75},
	year = {2008}}

@article{berry1992estimation,
	author = {Berry, Steven T},
	journal = {Econometrica},
	number = {4},
	pages = {889--917},
	title = {Estimation of a Model of Entry in the Airline Industry},
	volume = {60},
	year = {1992}}

@article{berry1994estimating,
	author = {Berry, Steven T},
	journal = {The RAND Journal of Economics},
	number = {2},
	pages = {242--262},
	title = {Estimating discrete-choice models of product differentiation},
	volume = {25},
	year = {1994}}

@article{farrell1990horizontal,
	author = {Farrell, Joseph and Shapiro, Carl},
	journal = {American Economic Review},
	number = {1},
	pages = {107--126},
	publisher = {JSTOR},
	title = {Horizontal mergers: an equilibrium analysis},
	volume = {80},
	year = {1990}}

@unpublished{Kryukov2010,
	author = {Kryukov, Yaroslav},
	note = {Manuscript. Tepper School of Business. Carnegie Mellon University.},
	title = {Dynamic R\&D and the effectiveness of policy intervention in the pharmaceutical industry},
	year = {2010}}

@article{Kellogg2014,
	author = {Kellogg, Ryan},
	journal = {American Economic Review},
	number = {6},
	pages = {1698--1734},
	title = {The effect of uncertainty on investment: evidence from texas oil drilling},
	volume = {104},
	year = {2014}}

@article{Kasahara2012,
	author = {Kasahara, Hiroyuki and Shimotsu, Katsumi},
	journal = {Econometrica},
	number = {5},
	pages = {2303--2319},
	title = {Sequential estimation of structural models with a fixed point constraint},
	volume = {80},
	year = {2012}}

@article{Kasahara2009,
	author = {Kasahara, Hiroyuki and Shimotsu, Katsumi},
	journal = {Econometrica},
	number = {1},
	pages = {135--175},
	title = {Nonparametric identification of finite mixture models of dynamic discrete choices},
	volume = {77},
	year = {2009}}

@article{Kalouptsidi2014,
	author = {Kalouptsidi, Myrto},
	journal = {American Economic Review},
	number = {2},
	pages = {564--608},
	title = {Time to build and fluctuations in bulk shipping},
	volume = {104},
	year = {2014}}

@article{Jofre-Bonet2003,
	author = {Jofre-Bonet, Mireia and Pesendorfer, Martin},
	journal = {Econometrica},
	number = {5},
	pages = {1443--1489},
	title = {Estimation of a dynamic auction game},
	volume = {71},
	year = {2003}}

@article{rust1987optimal,
	author = {Rust, John},
	journal = {Econometrica},
	number = {5},
	pages = {999--1033},
	publisher = {JSTOR},
	title = {Optimal replacement of GMC bus engines: An empirical model of Harold Zurcher},
	vol = {55},
	year = {1987}}

@article{Jeziorski2014,
	author = {Jeziorski, Przemys{\l}aw},
	journal = {The RAND Journal of Economics},
	number = {4},
	pages = {816--846},
	title = {Estimation of cost efficiencies from mergers: application to US radio},
	volume = {45},
	year = {2014}}

@article{IgamiConsolidation,
	author = {Igami, Mitsuru and Uetake, Kosuke},
	journal = {The Review of Economic Studies},
	number = {6},
	title = {Mergers, innovation, and entry-exit dynamics: consolidation of the hard disk drive industry, 1996-2016},
	volume = {87},
	year = {2020}}

@article{Igami2016,
	author = {Igami, Mitsuru and Yang, Nathan},
	journal = {Quantitative Economics},
	number = {2},
	pages = {483--521},
	title = {Unobserved heterogeneity in dynamic games: Cannibalization and preemptive entry of hamburger chains in Canada},
	volume = {7},
	year = {2016}}

@article{hashmi2016relationship,
	author = {Hashmi, Aamir and Biesebroeck, Johannes Van},
	journal = {The Review of Economics and Statistics},
	number = {1},
	pages = {192--208},
	title = {The relationship between market structure and innovation in industry equilibrium: a case study of the global automobile industry},
	volume = {98},
	year = {2016}}

@article{Igami2017,
	author = {Igami, Mitsuru},
	journal = {Journal of Political Economy},
	number = {3},
	pages = {798-847},
	title = {Estimating the innovator's dilemma: structural analysis of creative destruction in the hard disk drive industry, 1981--1998},
	volume = {125},
	year = {2017}}

@article{Weintraub2017,
	author = {Ifrach, Bar and Weintraub, Gabriel Y},
	journal = {The Review of Economic Studies},
	number = {3},
	pages = {1106--1150},
	title = {A framework for dynamic oligopoly in concentrated industries},
	volume = {84},
	year = {2017}}

@article{Huang2014,
	author = {Huang, Ling and Smith, Martin D.},
	journal = {American Economic Review},
	number = {12},
	pages = {4071--4103},
	title = {The dynamic efficiency costs of common-pool resource exploitation},
	volume = {104},
	year = {2014}}

@article{Holmes2011,
	author = {Holmes, Thomas J.},
	journal = {Econometrica},
	number = {1},
	pages = {253--302},
	title = {The diffusion of wal-mart and economies of density},
	volume = {79},
	year = {2011}}

@article{katz1985network,
	author = {Katz, Michael L and Shapiro, Carl},
	journal = {American Economic Review},
	number = {3},
	pages = {424--440},
	title = {Network externalities, competition, and compatibility},
	volume = {75},
	year = {1985}}

@article{slade1998optimal,
	author = {Slade, Margaret E},
	journal = {The Review of Economic Studies},
	number = {1},
	pages = {87--107},
	title = {Optimal pricing with costly adjustment: evidence from retail-grocery prices},
	volume = {65},
	year = {1998}}

@article{kano2013menu,
	author = {Kano, Kazuko},
	journal = {International Journal of Industrial Organization},
	number = {1},
	pages = {102--118},
	title = {Menu costs and dynamic duopoly},
	volume = {31},
	year = {2013}}

@techreport{ellison2018costs,
	author = {Ellison, Sara Fisher and Snyder, Christopher and Zhang, Hongkai},
	institution = {National Bureau of Economic Research},
	title = {Costs of managerial attention and activity as a source of sticky prices: Structural estimates from an online market},
	year = {2018}}

@techreport{mysliwski2020welfare,
	author = {Mysliwski, Mateusz and Sanches, Fabio M and Junior, Daniel Silva and Srisuma, Sorawoot},
	institution = {CeMMAP Working Paper, CWP35/20},
	title = {The Welfare Effects of Promotional Fees},
	year = {2020}}

@article{Gowrisankaran1999,
	author = {Gowrisankaran, Gautam},
	journal = {The RAND Journal of Economics},
	number = {1},
	pages = {56-83},
	title = {A dynamic model of endogenous horizontal mergers},
	volume = {30},
	year = {1999}}

@article{Goettler2011,
	author = {Goettler, Ronald L. and Gordon, Brett R.},
	journal = {Journal of Political Economy},
	number = {6},
	pages = {1141-1200},
	title = {Does AMD spur Intel to innovate more?},
	volume = {119},
	year = {2011}}

@article{Fershtman2012,
	author = {Fershtman, Chaim and Pakes, Ariel},
	journal = {The Quarterly Journal of Economics},
	number = {4},
	pages = {1611-1661},
	title = {Dynamic games with asymmetric information: a framework for empirical work},
	volume = {127},
	year = {2012}}

@article{Esteban2007,
	author = {Esteban, Susanna and Shum, Matthew},
	journal = {The RAND Journal of Economics},
	number = {2},
	pages = {332--354},
	title = {Durable-goods oligopoly with secondary markets: the case of automobiles},
	volume = {38},
	year = {2007}}

@article{Ericson1995,
	author = {Ericson, Richard and Pakes, Ariel},
	journal = {The Review of Economic Studies},
	number = {1},
	pages = {53-82},
	title = {Markov-perfect industry dynamics: a framework for empirical work},
	volume = {62},
	year = {1995}}

@article{aghion1992model,
	author = {Aghion, P and Howitt, P},
	journal = {Econometrica},
	number = {2},
	title = {A Model of Growth through Creative Destruction},
	volume = {60},
	year = {1992}}

@article{aghion2005competition,
	author = {Aghion, Philippe and Bloom, Nick and Blundell, Richard and Griffith, Rachel and Howitt, Peter},
	journal = {The Quarterly Journal of Economics},
	number = {2},
	pages = {701--728},
	title = {Competition and innovation: An inverted-U relationship},
	volume = {120},
	year = {2005}}

@article{bloom2009impact,
	author = {Bloom, Nicholas},
	journal = {Econometrica},
	number = {3},
	pages = {623--685},
	title = {The impact of uncertainty shocks},
	volume = {77},
	year = {2009}}

@article{frechette2019frictions,
	author = {Frechette, Guillaume R and Lizzeri, Alessandro and Salz, Tobias},
	journal = {American Economic Review},
	number = {8},
	pages = {2954--92},
	title = {Frictions in a competitive, regulated market: Evidence from taxis},
	volume = {109},
	year = {2019}}

@article{asker2014dynamic,
	author = {Asker, John and Collard-Wexler, Allan and De Loecker, Jan},
	journal = {Journal of Political Economy},
	number = {5},
	pages = {1013--1063},
	title = {Dynamic inputs and resource (mis) allocation},
	volume = {122},
	year = {2014}}

@article{aw2011r,
	author = {Aw, Bee Yan and Roberts, Mark J and Xu, Daniel Yi},
	journal = {American Economic Review},
	number = {4},
	pages = {1312--44},
	title = {R\&D investment, exporting, and productivity dynamics},
	volume = {101},
	year = {2011}}

@article{romer1986increasing,
	author = {Romer, Paul M},
	journal = {Journal of Political Economy},
	number = {5},
	pages = {1002--1037},
	title = {Increasing returns and long-run growth},
	volume = {94},
	year = {1986}}

@article{Dunne2013,
	author = {Dunne, Timothy and Klimek, Shawn D. and Roberts, Mark J. and Xu, Daniel Yi},
	journal = {The RAND Journal of Economics},
	number = {3},
	pages = {462--487},
	title = {Entry, exit, and the determinants of market structure},
	volume = {4},
	year = {2013}}

@inbook{Doraszelski2007,
	author = {Doraszelski, Ulrich and Pakes, Ariel},
	booktitle = {Handbook of Industrial Organization},
	chapter = {20},
	pages = {1887-1966},
	title = {A framework for applied dynamic analysis in IO},
	volume = {3},
	year = {2007}}

@article{Doraszelski2018,
	author = {Doraszelski, Ulrich and Lewis, Gregory and Pakes, Ariel},
	journal = {American Economic Review},
	number = {3},
	pages = {565--615},
	title = {Just starting out: learning and equilibrium in a new market},
	volume = {108},
	year = {2018}}

@article{Doraszelski2012,
	author = {Doraszelski, Ulrich and Judd, Kenneth L.},
	journal = {Quantitative Economics},
	number = {1},
	pages = {53-93},
	title = {Avoiding the curse of dimensionality in dynamic stochastic games},
	volume = {3},
	year = {2012}}

@article{gyourko2008new,
	author = {Gyourko, Joseph and Saiz, Albert and Summers, Anita},
	journal = {Urban Studies},
	number = {3},
	pages = {693--729},
	publisher = {Sage Publications Sage UK: London, England},
	title = {A new measure of the local regulatory environment for housing markets: The Wharton Residential Land Use Regulatory Index},
	volume = {45},
	year = {2008}}

@article{Doraszelski2010,
	author = {Doraszelski, Ulrich and Escobar, Juan F.},
	journal = {Theoretical Economics},
	number = {3},
	pages = {369-402},
	title = {A theory of regular Markov perfect equilibria in dynamic stochastic games: Genericity, stability, and purification},
	volume = {5},
	year = {2010}}

@article{collard2014mergers,
	author = {Collard-Wexler, Allan},
	journal = {American Economic Journal: Microeconomics},
	number = {4},
	pages = {407--47},
	title = {Mergers and sunk costs: An application to the ready-mix concrete industry},
	volume = {6},
	year = {2014}}

@article{Collard-Wexler2013,
	author = {Collard-Wexler, Allan},
	journal = {Econometrica},
	number = {3},
	pages = {1003--1037},
	title = {Demand fluctuations in the ready-mix concrete industry},
	volume = {81},
	year = {2013}}

@unpublished{Buchholz,
	author = {Buchholz, Nicholas},
	note = {Manuscript. Princeton University.},
	title = {Spatial equilibrium, search frictions and dynamic efficiency in the taxi industry},
	year = {2018}}

@article{bresnahan1991entry,
	author = {Bresnahan, Timothy F. and Reiss, Peter C.},
	journal = {Journal of Political Economy},
	number = {5},
	pages = {977--1009},
	title = {Entry and competition in concentrated markets},
	volume = {99},
	year = {1991}}

@article{Bresnahan1994,
	author = {Bresnahan, Timothy F. and Reiss, Peter C.},
	journal = {Annales D'{\'E}conomie et de Statistique},
	pages = {183-217},
	title = {Measuring the importance of sunk costs},
	volume = {31},
	year = {1994}}

@article{Brancaccio2020,
	author = {Brancaccio, Giulia and Kalouptsidi, Myrto and Papageorgiou, Theodore},
	journal = {Econometrica},
	pages = {Forthcoming},
	title = {Geography, search frictions and endogenous trade costs},
	year = {2020}}

@article{dube2005empirical,
	author = {Dub{\'e}, Jean-Pierre and Hitsch, G{\"u}nter J and Manchanda, Puneet},
	journal = {Quantitative Marketing and Economics},
	number = {2},
	pages = {107--144},
	publisher = {Springer},
	title = {An empirical model of advertising dynamics},
	volume = {3},
	year = {2005}}

@article{dube2010tipping,
	author = {Dub{\'e}, Jean-Pierre H and Hitsch, G{\"u}nter J and Chintagunta, Pradeep K},
	journal = {Marketing Science},
	number = {2},
	pages = {216--249},
	title = {Tipping and concentration in markets with indirect network effects},
	volume = {29},
	year = {2010}}

@article{lin2015quality,
	author = {Lin, Haizhen},
	journal = {International Economic Review},
	number = {4},
	pages = {1261--1290},
	title = {Quality choice and market structure: A dynamic analysis of nursing home oligopolies},
	volume = {56},
	year = {2015}}

@article{Borkovsky2012,
	author = {Borkovsky, Ron N. and Doraszelski, Ulrich and Kryukov, Yaroslav},
	journal = {Quantitative Marketing and Economics},
	number = {2},
	pages = {197--229},
	title = {A dynamic quality ladder model with entry and exit: Exploring the equilibrium correspondence using the homotopy method},
	volume = {10},
	year = {2012}}

@article{gowrisankaran1997dynamic,
	author = {Gowrisankaran, Gautam and Town, Robert J},
	journal = {Journal of Economics \& Management Strategy},
	number = {1},
	pages = {45--74},
	title = {Dynamic equilibrium in the hospital industry},
	volume = {6},
	year = {1997}}

@article{gowrisankaran2012dynamics,
	author = {Gowrisankaran, Gautam and Rysman, Marc},
	journal = {Journal of Political Economy},
	number = {6},
	pages = {1173--1219},
	title = {Dynamics of consumer demand for new durable goods},
	volume = {120},
	year = {2012}}

@article{lee2013vertical,
	author = {Lee, Robin S},
	journal = {American Economic Review},
	number = {7},
	pages = {2960--3000},
	title = {Vertical integration and exclusivity in platform and two-sided markets},
	volume = {103},
	year = {2013}}

@unpublished{BenkardManu,
	author = {Benkard, C. Lanier and Bodoh-Creed, Aaron and Lazarev, John},
	note = {Manuscript, Yale University.},
	title = {Simulating the Dynamic Effects of Horizontal Mergers: U.S. Airlines},
	year = {2010}}

@article{Benkard2004,
	author = {Benkard, C. Lanier},
	journal = {The Review of Economic Studies},
	number = {3},
	pages = {581--611},
	title = {A dynamic analysis of the market for wide-bodied commercial aircraft},
	volume = {71},
	year = {2004}}

@article{Benkard2000,
	author = {Benkard, C. Lanier},
	journal = {American Economic Review},
	number = {4},
	title = {Learning and forgetting: the dynamics of aircraft production},
	volume = {90},
	year = {2000}}

@article{Bajari2007,
	author = {Bajari, Patrick and Benkard, C. Lanier and Levin, Jonathan},
	journal = {Econometrica},
	number = {5},
	pages = {1331--1370},
	title = {Estimating dynamic models of imperfect competition},
	volume = {75},
	year = {2007}}

@article{Arcidiacono2011,
	author = {Arcidiacono, Peter and Miller, Robert A.},
	journal = {Econometrica},
	number = {6},
	pages = {1823--1867},
	title = {Conditional choice probability estimation of dynamic discrete choice models with unobserved heterogeneity},
	volume = {79},
	year = {2011}}

@article{Arcidiacono2016,
	author = {Arcidiacono, Peter and Bayer, Patrick and Blevins, Jason R and Ellickson, Paul B},
	journal = {The Review of Economic Studies},
	number = {3},
	pages = {889--931},
	publisher = {Wiley-Blackwell},
	title = {Estimation of dynamic discrete choice models in continuous time with an application to retail competition},
	volume = {83},
	year = {2016}}

@article{weintraub2008markov,
	author = {Weintraub, Gabriel Y and Benkard, C Lanier and Van Roy, Benjamin},
	journal = {Econometrica},
	number = {6},
	pages = {1375--1411},
	title = {Markov perfect industry dynamics with many firms},
	volume = {76},
	year = {2008}}

@article{whinston2007antitrust,
	author = {Whinston, Michael D},
	journal = {Handbook of industrial organization},
	pages = {2369--2440},
	title = {Antitrust policy toward horizontal mergers},
	volume = {3},
	year = {2007}}

@article{mermelstein2020internal,
	author = {Mermelstein, Ben and Nocke, Volker and Satterthwaite, Mark A and Whinston, Michael D},
	journal = {Journal of Political Economy},
	number = {1},
	pages = {301--341},
	title = {Internal versus external growth in industries with scale economies: A computational model of optimal merger policy},
	volume = {128},
	year = {2020}}

@unpublished{chicu2013dynamic,
	author = {Chicu, Mark},
	note = {Manuscript, Harvard University},
	title = {Dynamic investment and deterrence in the US cement industry},
	year = {2013}}

@article{xu2008structural,
	author = {Xu, Daniel Yi and Yanyou Chen},
	journal = {Journal of Industrial Economics},
	title = {A structural empirical model of R\&D, firm heterogeneity, and industry evolution},
	year = {2020}}

@article{guerre2000optimal,
	author = {Guerre, Emmanuel and Perrigne, Isabelle and Vuong, Quang},
	journal = {Econometrica},
	number = {3},
	pages = {525--574},
	title = {Optimal nonparametric estimation of first-price auctions},
	volume = {68},
	year = {2000}}

@article{judd1985credible,
	author = {Judd, Kenneth L},
	journal = {The RAND Journal of Economics},
	number = {2},
	pages = {153--166},
	publisher = {JSTOR},
	title = {Credible spatial preemption},
	volume = {16},
	year = {1985}}

@article{ciliberto2009market,
	author = {Ciliberto, Federico and Tamer, Elie},
	journal = {Econometrica},
	number = {6},
	pages = {1791--1828},
	title = {Market structure and multiple equilibria in airline markets},
	volume = {77},
	year = {2009}}

@article{hendricks1997entry,
	author = {Hendricks, Ken and Piccione, Michele and Tan, Guofu},
	journal = {The RAND Journal of Economics},
	number = {2},
	pages = {291--303},
	publisher = {JSTOR},
	title = {Entry and exit in hub-spoke networks},
	volume = {28},
	year = {1997}}

@article{pakes2015moment,
	author = {Pakes, Ariel and Porter, Jack and Ho, Kate and Ishii, Joy},
	journal = {Econometrica},
	number = {1},
	pages = {315--334},
	title = {Moment inequalities and their application},
	volume = {83},
	year = {2015}}

@article{kalouptsidi2018detection,
	author = {Kalouptsidi, Myrto},
	journal = {The Review of Economic Studies},
	number = {2},
	pages = {1111--1158},
	publisher = {Oxford University Press},
	title = {Detection and impact of industrial subsidies: The case of Chinese shipbuilding},
	volume = {85},
	year = {2018}}

@article{besanko2014economics,
	author = {Besanko, David and Doraszelski, Ulrich and Kryukov, Yaroslav},
	journal = {American Economic Review},
	number = {3},
	pages = {868--97},
	title = {The economics of predation: What drives pricing when there is learning-by-doing?},
	volume = {104},
	year = {2014}}

@article{bain1951relation,
	author = {Bain, Joe S},
	journal = {The Quarterly Journal of Economics},
	number = {3},
	pages = {293--324},
	publisher = {MIT Press},
	title = {Relation of profit rate to industry concentration: American manufacturing, 1936--1940},
	volume = {65},
	year = {1951}}

@article{fudenberg1986theory,
	author = {Fudenberg, Drew and Tirole, Jean},
	journal = {Econometrica},
	number = {4},
	pages = {943--960},
	publisher = {JSTOR},
	title = {A theory of exit in duopoly},
	volume = {54},
	year = {1986}}

@book{bain1956,
	author = {Joseph Bain},
	location = {Cambridge, MA.},
	publisher = {Harvard University Press},
	title = {Barriers to New Competition},
	year = {1956}}

@article{ryan2012heterogeneity,
	author = {Ryan, Stephen P and Tucker, Catherine},
	journal = {Quantitative Marketing and Economics},
	number = {1},
	pages = {63--109},
	publisher = {Springer},
	title = {Heterogeneity and the dynamics of technology adoption},
	volume = {10},
	year = {2012}}

@article{buchanan1969external,
	author = {Buchanan, James M},
	journal = {American Economic Review},
	number = {1},
	pages = {174--177},
	publisher = {JSTOR},
	title = {External diseconomies, corrective taxes, and market structure},
	volume = {59},
	year = {1969}}

@article{elyakime1994first,
	author = {Elyakime, Bernard and Laffont, Jean Jacques and Loisel, Patrice and Vuong, Quang},
	journal = {Annales d'Economie et de Statistique},
	pages = {115--141},
	publisher = {JSTOR},
	title = {First-price sealed-bid auctions with secret reservation prices},
	volume = {34},
	year = {1994}}

@article{berry2019increasing,
	author = {Berry, Steven and Gaynor, Martin and Scott Morton, Fiona},
	journal = {Journal of Economic Perspectives},
	number = {3},
	pages = {44--68},
	title = {Do increasing markups matter? lessons from empirical industrial organization},
	volume = {33},
	year = {2019}}

@article{attanasio2000consumer,
	author = {Attanasio, Orazio P},
	journal = {The Review of Economic Studies},
	number = {4},
	pages = {667--696},
	publisher = {JSTOR},
	title = {Consumer durables and inertial behaviour: Estimation and aggregation of (S, s) rules for automobile purchases},
	volume = {67},
	year = {2000}}

@article{otsu2016pooling,
	author = {Otsu, Taisuke and Pesendorfer, Martin and Takahashi, Yuya},
	journal = {Quantitative Economics},
	number = {2},
	pages = {523--559},
	title = {Pooling data across markets in dynamic Markov games},
	volume = {7},
	year = {2016}}

@article{fowlie2016market,
	author = {Fowlie, Meredith and Reguant, Mar and Ryan, Stephen P.},
	journal = {Journal of Political Economy},
	number = {1},
	pages = {249--302},
	publisher = {University of Chicago Press Chicago, IL},
	title = {Market-based emissions regulation and industry dynamics},
	volume = {124},
	year = {2016}}

@article{mcfadden1989method,
	author = {McFadden, Daniel},
	journal = {Econometrica},
	number = {5},
	pages = {995--1026},
	publisher = {JSTOR},
	title = {A method of simulated moments for estimation of discrete response models without numerical integration},
	volume = {57},
	year = {1989}}

@article{pakes1989simulation,
	author = {Pakes, Ariel and Pollard, David},
	journal = {Econometrica},
	number = {5},
	pages = {1027--1057},
	publisher = {JSTOR},
	title = {Simulation and the asymptotics of optimization estimators},
	volume = {57},
	year = {1989}}

@article{berry_compiani_2021,
	author = {Berry, Steven and Compiani, Giovanni},
	journal = {Annual Review of Economics},
	pages = {309--334},
	publisher = {Annual Reviews},
	title = {Empirical Models of Industry Dynamics with Endogenous Market Structure},
	volume = {13},
	year = {2021}}

@article{arcidiacono_ellickson_2011,
	author = {Arcidiacono, Peter and Ellickson, Paul},
	journal = {Annual Review of Economics},
	number = {1},
	pages = {363--394},
	publisher = {Annual Reviews},
	title = {Practical methods for estimation of dynamic discrete choice models},
	volume = {3},
	year = {2011}}

@article{harsanyi1973games,
	author = {Harsanyi, John},
	journal = {International Journal of Game Theory},
	number = {1},
	pages = {1--23},
	publisher = {Springer},
	title = {Games with randomly disturbed payoffs: A new rationale for mixed-strategy equilibrium points},
	volume = {2},
	year = {1973}}

@inbook{rust_1994_worldcongress,
	author = {Rust, John},
	booktitle = {Advances in Econometrics: Sixth World Congress},
	collection = {Econometric Society Monographs},
	pages = {119--170},
	place = {Cambridge},
	publisher = {Cambridge University Press},
	series = {Econometric Society Monographs},
	title = {Estimation of dynamic structural models, problems and prospects: discrete decision processes},
	volume = {2},
	year = {1994}}

@article{blevins_2014,
	author = {Blevins, Jason},
	journal = {Quantitative Economics},
	number = {3},
	pages = {531--554},
	title = {Nonparametric identification of dynamic decision processes with discrete and continuous choices},
	volume = {5},
	year = {2014}}

@article{rust_1994_hbook,
	author = {Rust, John},
	journal = {Handbook of Econometrics},
	pages = {3081--3143},
	publisher = {Elsevier},
	title = {Structural estimation of Markov decision processes},
	volume = {4},
	year = {1994}}

@article{magnac_thesmar_2002,
	author = {Magnac, Thierry and Thesmar, David},
	journal = {Econometrica},
	number = {2},
	pages = {801--816},
	publisher = {JSTOR},
	title = {Identifying dynamic discrete decision processes},
	volume = {70},
	year = {2002}}

@article{abbring_daljord_2020,
	author = {Abbring, Jaap and Daljord, {\O}ystein},
	journal = {Quantitative Economics},
	number = {2},
	pages = {471--501},
	title = {Identifying the discount factor in dynamic discrete choice models},
	volume = {11},
	year = {2020}}

@article{hu_shum_2012,
	author = {Hu, Yingyao and Shum, Matthew},
	journal = {Journal of Econometrics},
	number = {1},
	pages = {32--44},
	title = {Nonparametric identification of dynamic models with unobserved state variables},
	volume = {171},
	year = {2012}}

@incollection{hu_shum_2013,
	author = {Hu, Yingyao and Shum, Matthew},
	booktitle = {Structural Econometric Models},
	publisher = {Emerald Group Publishing Limited},
	title = {Identifying Dynamic Games with Serially Correlated Unobservables},
	year = {2013}}

@article{depaula_tang_2012,
	author = {De Paula, Aureo and Tang, Xun},
	journal = {Econometrica},
	number = {1},
	pages = {143--172},
	title = {Inference of signs of interaction effects in simultaneous games with incomplete information},
	volume = {80},
	year = {2012}}

@article{sweeting_2009,
	author = {Sweeting, Andrew},
	journal = {The RAND Journal of Economics},
	number = {4},
	pages = {710--742},
	title = {The strategic timing incentives of commercial radio stations: An empirical analysis using multiple equilibria},
	volume = {40},
	year = {2009}}

@article{kalouptsidi_scott_2021,
	author = {Kalouptsidi, Myrto and Scott, Paul and Souza-Rodrigues, Eduardo},
	journal = {Quantitative Economics},
	number = {2},
	pages = {351--403},
	title = {Identification of counterfactuals in dynamic discrete choice models},
	volume = {12},
	year = {2021}}

@article{kalouptsidi_scott_2017_ijio,
	author = {Kalouptsidi, Myrto and Scott, Paul and Souza-Rodrigues, Eduardo},
	journal = {International Journal of Industrial Organization},
	pages = {362--371},
	title = {On the non-identification of counterfactuals in dynamic discrete games},
	volume = {50},
	year = {2017}}

@techreport{kalouptsidi2020partial,
	author = {Kalouptsidi, Myrto and Kitamura, Yuichi and Lima, Lucas and Souza-Rodrigues, Eduardo A},
	institution = {National Bureau of Economic Research},
	title = {Partial identification and inference for dynamic models and counterfactuals},
	year = {2020}}

@article{chevalier_goolsbee_2009,
	author = {Chevalier, Judith and Goolsbee, Austan},
	journal = {The Quarterly Journal of Economics},
	number = {4},
	pages = {1853--1884},
	publisher = {MIT Press},
	title = {Are durable goods consumers forward-looking? Evidence from college textbooks},
	volume = {124},
	year = {2009}}

@article{fang_wang_2015_ier,
	author = {Fang, Hanming and Wang, Yang},
	journal = {International Economic Review},
	number = {2},
	pages = {565--596},
	title = {Estimating dynamic discrete choice models with hyperbolic discounting, with an application to mammography decisions},
	volume = {56},
	year = {2015}}

@article{bayer_mcmillan_2016,
	author = {Bayer, Patrick and McMillan, Robert and Murphy, Alvin and Timmins, Christopher},
	journal = {Econometrica},
	number = {3},
	pages = {893--942},
	title = {A dynamic model of demand for houses and neighborhoods},
	volume = {84},
	year = {2016}}

@article{ching_osborne_2020,
	author = {Ching, Andrew and Osborne, Matthew},
	journal = {Marketing Science},
	number = {4},
	pages = {707--726},
	publisher = {INFORMS},
	title = {Identification and estimation of forward-looking behavior: The case of consumer stockpiling},
	volume = {39},
	year = {2020}}

@article{degroote_verboven_2019,
	author = {De Groote, Olivier and Verboven, Frank},
	journal = {American Economic Review},
	number = {6},
	pages = {2137--72},
	title = {Subsidies and time discounting in new technology adoption: Evidence from solar photovoltaic systems},
	volume = {109},
	year = {2019}}

@article{komarova_sanches_2018,
	author = {Komarova, Tatiana and Sanches, Fabio and Silva Junior, Daniel and Srisuma, Sorawoot},
	journal = {Quantitative Economics},
	number = {3},
	pages = {1153--1194},
	title = {Joint analysis of the discount factor and payoff parameters in dynamic discrete choice models},
	volume = {9},
	year = {2018}}

@techreport{kristensen_nesheim_2015,
	author = {Kristensen, Dennis and Nesheim, Lars and {de Paula}, {\'Aureo}},
	institution = {Mimeo, University College London},
	title = {CCP and the estimation of nonseparable dynamic models},
	year = {2015}}

@inbook{horowitz_hbook_statistics_1993,
	author = {Horowitz, Joel},
	booktitle = {Handbook of Statistics},
	editor = {Maddala, G.S. and Rao, C.R. and Vinod, H.D.},
	pages = {45-72},
	place = {Cambridge},
	publisher = {Elsevier},
	title = {Semiparametric and nonparametric estimation of quantal response models},
	volume = {11},
	year = {1993}}

@article{matzkin_1992_ecma,
	author = {Matzkin, Rosa L},
	journal = {Econometrica},
	number = {2},
	pages = {239--270},
	publisher = {JSTOR},
	title = {Nonparametric and distribution-free estimation of the binary threshold crossing and the binary choice models},
	volume = {60},
	year = {1992}}

@article{aguirregabiria_2010_jbes,
	author = {Aguirregabiria, Victor},
	journal = {Journal of Business \& Economic Statistics},
	number = {2},
	pages = {201--218},
	publisher = {Taylor \& Francis},
	title = {Another look at the identification of dynamic discrete decision processes: An application to retirement behavior},
	volume = {28},
	year = {2010}}

@article{lewbel_1998,
	author = {Lewbel, Arthur},
	journal = {Econometrica},
	number = {1},
	pages = {105--121},
	publisher = {JSTOR},
	title = {Semiparametric latent variable model estimation with endogenous or mismeasured regressors},
	volume = {66},
	year = {1998}}

@article{lewbel_2000,
	author = {Lewbel, Arthur},
	journal = {Journal of Econometrics},
	number = {1},
	pages = {145--177},
	title = {Semiparametric qualitative response model estimation with unknown heteroscedasticity or instrumental variables},
	volume = {97},
	year = {2000}}

@article{norets_tang_2014,
	author = {Norets, Andriy and Tang, Xun},
	journal = {The Review of Economic Studies},
	number = {3},
	pages = {1229--1262},
	publisher = {Oxford University Press},
	title = {Semiparametric inference in dynamic binary choice models},
	volume = {81},
	year = {2014}}

@article{chen_2017_et,
	author = {Chen, Le-Yu},
	journal = {Econometric Theory},
	number = {3},
	pages = {551--577},
	publisher = {Cambridge University Press},
	title = {Identification of discrete choice dynamic programming models with nonparametric distribution of unobservables},
	volume = {33},
	year = {2017}}

@article{buchholz_shum_2021,
	author = {Buchholz, Nicholas and Shum, Matthew and Xu, Haiqing},
	journal = {Journal of Econometrics},
	number = {2},
	pages = {312--327},
	title = {Semiparametric estimation of dynamic discrete choice models},
	volume = {223},
	year = {2021}}

@article{simon_1959_aer,
	author = {Simon, Herbert A},
	journal = {American Economic Review},
	number = {3},
	pages = {253-283},
	title = {Theories of decision-making in economics and behavioral science},
	volume = {49},
	year = {1959}}

@article{iskhakov_rust_2020_ej,
	author = {Iskhakov, Fedor and Rust, John and Schjerning, Bertel},
	journal = {The Econometrics Journal},
	number = {3},
	pages = {S81--S124},
	publisher = {Oxford University Press},
	title = {Machine learning and structural econometrics: contrasts and synergies},
	volume = {23},
	year = {2020}}

@article{aguirregabiria_magesan_2020,
	author = {Aguirregabiria, Victor and Magesan, Arvind},
	journal = {The Review of Economic Studies},
	number = {2},
	pages = {582--625},
	publisher = {Oxford University Press},
	title = {Identification and estimation of dynamic games when players' beliefs are not in equilibrium},
	volume = {87},
	year = {2020}}

@article{goldfarb_xiao_2011,
	author = {Goldfarb, Avi and Xiao, Mo},
	journal = {American Economic Review},
	number = {7},
	pages = {3130--61},
	title = {Who thinks about the competition? Managerial ability and strategic entry in US local telephone markets},
	volume = {101},
	year = {2011}}

@article{hortaccsu_puller_2008,
	author = {Horta{\c{c}}su, Ali and Puller, Steven L},
	journal = {The RAND Journal of Economics},
	number = {1},
	pages = {86--114},
	title = {Understanding strategic bidding in multi-unit auctions: a case study of the Texas electricity spot market},
	volume = {39},
	year = {2008}}

@article{hortaccsu_puller_2019,
	author = {Horta{\c{c}}su, Ali and Luco, Fernando and Puller, Steven L and Zhu, Dongni},
	journal = {American Economic Review},
	number = {12},
	pages = {4302--42},
	title = {Does strategic ability affect efficiency? Evidence from electricity markets},
	volume = {109},
	year = {2019}}

@article{huang_ellickson_2020,
	author = {Huang, Yufeng and Ellickson, Paul B and Lovett, Mitchell J},
	note = {Manuscript, University of Rochester - Simon Business School},
	title = {Learning to set prices},
	year = {2020}}

@article{xie_2021_jbes,
	author = {Xie, Erhao},
	journal = {Journal of Business \& Economic Statistics},
	title = {Inference in games without Nash equilibrium: An application to restaurants' competition in opening hours},
	volume = {Forthcoming},
	year = {2021}}

@article{brown_camerer_2013,
	author = {Brown, Alexander L and Camerer, Colin F and Lovallo, Dan},
	journal = {Management Science},
	number = {3},
	pages = {733--747},
	publisher = {INFORMS},
	title = {Estimating structural models of equilibrium and cognitive hierarchy thinking in the field: The case of withheld movie critic reviews},
	volume = {59},
	year = {2013}}

@article{ho_su_2013_ms,
	author = {Ho, Teck-Hua and Su, Xuanming},
	journal = {Management Science},
	number = {2},
	pages = {452--469},
	publisher = {INFORMS},
	title = {A dynamic level-k model in sequential games},
	volume = {59},
	year = {2013}}

@article{an_hu_2021_joe,
	author = {An, Yonghong and Hu, Yingyao and Xiao, Ruli},
	journal = {Journal of Econometrics},
	number = {1},
	pages = {645--675},
	title = {Dynamic decisions under subjective expectations: A structural analysis},
	volume = {222},
	year = {2021}}

@article{seim_2006,
	author = {Seim, Katja},
	journal = {The RAND Journal of Economics},
	number = {3},
	pages = {619--640},
	title = {An empirical model of firm entry with endogenous product-type choices},
	volume = {37},
	year = {2006}}

@article{das_1992,
	author = {Das, Sanghamitra},
	journal = {The Review of Economic Studies},
	number = {2},
	pages = {277--297},
	publisher = {Wiley-Blackwell},
	title = {A micro-econometric model of capital utilization and retirement: the case of the US cement industry},
	volume = {59},
	year = {1992}}

@article{rust_rothwell_1995,
	author = {Rust, John and Rothwell, Geoffrey},
	journal = {Journal of Applied Econometrics},
	number = {S1},
	pages = {S75--S118},
	title = {Optimal response to a shift in regulatory regime: The case of the US nuclear power industry},
	volume = {10},
	year = {1995}}

@article{hendel_nevo_aer_2013,
	author = {Hendel, Igal and Nevo, Aviv},
	journal = {American Economic Review},
	number = {7},
	pages = {2722--51},
	title = {Intertemporal price discrimination in storable goods markets},
	volume = {103},
	year = {2013}}

@article{rotemberg_saloner_aer_1987,
	author = {Rotemberg, Julio J and Saloner, Garth},
	journal = {American Economic Review},
	pages = {917--926},
	publisher = {JSTOR},
	title = {The relative rigidity of monopoly pricing},
	year = {1987}}

@article{rotemberg_jpe_1982,
	author = {Rotemberg, Julio J},
	journal = {Journal of Political Economy},
	number = {6},
	pages = {1187--1211},
	publisher = {The University of Chicago Press},
	title = {Sticky prices in the United States},
	volume = {90},
	year = {1982}}

@article{gertner_1985,
	author = {Gertner, Robert},
	journal = {PhD Thesis, MIT},
	title = {Dynamic duopoly with price inertia},
	year = {1985}}

@article{cooper_haltiwanger_2006,
	author = {Cooper, Russell and Haltiwanger, John},
	journal = {Review of Economic Studies},
	number = {3},
	pages = {611--633},
	title = {On the nature of capital adjustment costs},
	volume = {73},
	year = {2006}}

@book{evans2012learning,
	author = {Evans, George W and Honkapohja, Seppo},
	publisher = {Princeton University Press},
	title = {Learning and expectations in macroeconomics},
	year = {2012}}

@article{abadie2011bias,
  title={Bias-corrected matching estimators for average treatment effects},
  author={Abadie, Alberto and Imbens, Guido W},
  journal={Journal of Business \& Economic Statistics},
  volume={29},
  number={1},
  pages={1--11},
  year={2011},
  publisher={Taylor \& Francis}
}

\end{document}